\documentclass
[tightenlines,eqsecnum,floats,aps,amsmath,amssymb,nofootinbib,prd,shownopacs,notitlepage]{revtex4-1}
\usepackage{graphicx, wrapfig}
\usepackage{amssymb}
\usepackage{color}
\usepackage{mathrsfs}
\usepackage{amsmath}
\usepackage{subfigure}
\usepackage{afterpage}

\def\f{\frac}

\def\lp{l_{\rm Pl}}

\def\Vpl{V_{\rm Pl}}
\def\rhopl{\rho_{\rm Pl}}

\usepackage{enumerate}

\newcommand{\be}{\nopagebreak[3]\begin{equation}}
\newcommand{\ee}{\end{equation}}
\newcommand{\bfig}{\nopagebreak[3]\begin{figure}}
\newcommand{\efig}{\end{figure}}
\newcommand{\ba}{\nopagebreak[3]\begin{eqnarray}}
\newcommand{\ea}{\end{eqnarray}}

\newcommand{\bmult}{\nopagebreak[3]\begin{multline}}
\newcommand{\emult}{\end{multline}}
\newcommand{\fref}[1]{Fig.\,\ref{#1}}
\newcommand{\eref}[1]{eq.\,(\ref{#1})}
\newcommand{\secref}[1]{Sec.\,\ref{#1}}

\begin{document}
\title[Squeezed]{Numerical evolution of squeezed and non-Gaussian states in loop quantum cosmology}

\author{Peter Diener$^{1,2}$}
\email{diener@cct.lsu.edu}

\author{Brajesh Gupt$^2$}
\email{brajesh@phys.lsu.edu}

\author{Miguel Megevand$^2$}
\email{megevand@phys.lsu.edu}

\author{Parampreet Singh$^2$}
\email{psingh@phys.lsu.edu}

\affiliation{
$^1$ 
Center for Computation and Technology, Louisiana State University, Baton Rouge, LA 70803, U.S.A.
}
\affiliation{
$^2$ 
Department of Physics and Astronomy, Louisiana State University, Baton Rouge, LA 70803, U.S.A.
}
%\email{\mailto{diener@cct.lsu.edu}, \mailto{bgupt1@lsu.edu}, \mailto{psingh@phys.lsu.edu}}

\begin{abstract}
In recent years, numerical simulations with Gaussian initial states have demonstrated the existence 
of a quantum bounce in loop quantum cosmology in various models. 
A key issue pertaining to the robustness of the bounce and the associated 
physics is to understand the quantum evolution for more general initial states which may depart 
significantly from Gaussianity and may have no well 
defined peakedness properties.
The analysis of such states, including squeezed and highly non-Gaussian states,
has been computationally challenging until now. In this manuscript, we overcome these challenges 
by using the Chimera scheme for the spatially flat, homogeneous and isotropic model sourced with 
a massless scalar field. We demonstrate that the quantum bounce in this model 
occurs even for states which are highly squeezed or are non-Gaussian with multiple peaks and with 
little resemblance to 
semi-classical states. The existence of the bounce is found to be robust, being independent of 
the properties of the states. The evolution of squeezed and non-Gaussian states turns out to be 
qualitatively similar to that of Gaussian states, and satisfies strong constraints on the growth of the 
relative fluctuations across the bounce. 
We also compare the results from the effective dynamics and find that, although it captures 
the qualitative aspects of the evolution for squeezed and highly non-Gaussian states, it always 
underestimates the bounce volume. We show that various properties of the evolution, such as the 
energy density at the bounce, are in excellent agreement with the predictions from an exactly 
solvable loop quantum cosmological model for arbitrary states.

\end{abstract}
 
%\submitto{\CQG}
\maketitle

%%%%%%%%%%%%%%%%%%%%%%%%%%%%%%%%%%%%%%%%%%%%%%%%%%
\section{Introduction} \label{sec:intro}
%%%%%%%%%%%%%%%%%%%%%%%%%%%%%%%%%%%%%%%%%%%%%%%%%%
Over the past decade, loop quantum cosmology (LQC) has proved to be a promising 
avenue to address the problem of classical singularities, providing a consistent framework to 
gain insights on the physics at the Planck scale  \cite{as1}. LQC is a symmetry reduced canonical quantization of homogeneous spacetimes using 
the techniques of loop quantum gravity (LQG) where the underlying geometry is discrete. This quantum discreteness of geometry results in an 
upper bound on the curvature of spacetime which leads to the absence of singularities in LQC. The main feature of the singularity resolution in LQC is 
the replacement of the big bang singularity by a quantum bounce. Resulting from the non-perturbative quantum gravitational effects, the quantum bounce   
establishes a non-singular bridge between the disjoint expanding and contracting branches of the classical theory \cite{aps1,aps2,aps3}. 
The robustness of the singularity resolution 
has been studied in great detail for various cosmological models
\cite{aps2,aps3,szulc_open,warsaw_flat,kv,warsaw_closed,bp,kp1,ap,rad},
including in the presence of potentials \cite{aps4,cyclic} and anisotropies
\cite{awe2,awe3,we1,chioub1,b1madrid1}. For all the models where numerical
simulations have been performed, the evolution of  semi-classical states has
been shown to be free of the classical 
singularities, and the occurrence of a quantum bounce has been
established.\footnote{For a review of various numerical methods in LQC, see
Ref. \cite{brizuela_cartin_khanna,ps12}.} The spatially flat homogeneous
Friedmann-Robertson-Walker (FRW) model with a massless scalar field can also
be solved exactly in LQC. This quantization known as solvable LQC (sLQC) \cite{acs}, 
predicts the existence of a universal maximum for the expectation value of the energy density
and a bounce for all the 
states in the physical Hilbert space. This model has been used to obtain
powerful constraints on the growth of the relative fluctuations across the
bounce \cite{recall,kp,montoya_corichi1,montoya_corichi2}, and to compute a
consistent quantum probability for the occurrence of a bounce (which turns out to be unity) \cite{craig_singh_lqc1}.

Due to the underlying quantum geometry, 
the quantum Hamiltonian constraint in isotropic models in LQC is a finite difference equation in volume. This is in contrast to the Wheeler-DeWitt quantization where the corresponding quantum 
Hamiltonian  constraint is a differential equation. In the cases where the matter part is considered to be a massless scalar field, the latter 
serves as a good internal clock to measure the variation in the volume of the
universe and the evolution can be studied following a rigorous quantization.  In
contrast to LQC where the quantum evolution is non-singular and
physical states undergo a quantum bounce, the Wheeler-DeWitt quantization of this model 
does not resolve the classical singularity, and the physical states follow the classical trajectories all the way to the big bang singularity 
in the backward evolution, or the big crunch singularity in the forward evolution.  At curvature scales much smaller than the Planck scale, the discrete quantum geometry leads to 
the classical geometry and the difference equation in LQC is well approximated by the Wheeler-DeWitt differential equation. Therefore, 
in the low curvature regime, the trajectories of LQC, Wheeler-DeWitt theory and 
classical general relativity are in extremely good agreement. Interestingly, there also exist effective descriptions 
of discrete quantum evolution in LQC, which for suitably chosen sharply peaked states can capture the underlying quantum evolution extremely well all the way to the quantum bounce \cite{vt}. Note that for the effective description, the underlying geometry is a continuum geometry but is only classical at scales much smaller than the Planck curvature. The effective dynamics for the spatially flat model results in a modified Friedmann equation where the dominant modification is encoded in the $\rho^2$ correction \cite{ps06,aps3}, and has been extensively used to extract phenomenological predictions in LQC \cite{as1}.

One of the main  goals of earlier work in LQC was to demonstrate the
existence of a quantum bounce for Gaussian 
states which are sharply peaked at classical trajectories in a macroscopic universe at late times. For the massless scalar field model in the spatially flat isotropic and homogeneous spacetime,
this implies that the initial state is peaked at a large value of the field momentum and at a large volume. For such states, 
the quantum bounce was found to occur approximately at an energy density $\rho \approx \rho_{\rm max}$ where $\rho_{\rm max} \approx 0.409 \rho_{\mathrm{Pl}}$ is  the universal maximum of energy density predicted in sLQC \cite{acs}.\footnote{Strictly speaking, the energy density at which the quantum bounce occurs, never saturates 
the upper bound $\rho_{\rm max}$ in sLQC. For the sharply peaked states, the bounce density is extremely close to but less than $\rho_{\rm max}$ \cite{aps3,dgs2}, and for the widely spread states the bounce density can be much smaller \cite{montoya_corichi2,dgs2}.} Numerical simulations of initial states which are sharply peaked also demonstrated an excellent agreement between the quantum evolution and the trajectory obtained from the effective 
Hamiltonian constraint. To establish the robustness of new physics at the Planck scale, it is important to understand the quantum evolution of states which are not sharply peaked and depart from Gaussianity. Such states may be widely spread, squeezed and multi-peaked with no semi-classical properties. Though such states do not lead to 
a classical macroscopic universe at late times, 
they are important for testing the genericity of predictions and also hold significance for some phenomenological reasons, such as the understanding of the role of non-Gaussianities on the 
cosmological perturbations in a loop quantum universe. A first step in this direction was performed recently in Ref. \cite{dgs2}, where quantum evolution of 
widely spread Gaussian states in the spatially flat homogeneous and isotropic model sourced with a massless scalar field were extensively studied. These simulations were
performed using the Chimera scheme \cite{dgs1}, which has been recently proposed to overcome computational difficulties associated with the evolution of widely spread states.  
It was found that the 
qualitative features of the quantum bounce remain true for the widely spread states. 
However, there are quantitative differences between the LQC and 
the corresponding effective trajectories. A general feature of these differences, 
is that the effective theory alway underestimates the bounce volume and 
overestimates the energy density at the bounce.

%zzz
%% In this article, our goal is to test the robustness of the quantum bounce by considering non-Gaussian
%% states. 
In this article we test the robustness of the quantum bounce by considering non-Gaussian 
states in the quantization of homogeneous and isotropic spatially flat spacetime with a massless 
scalar field as given in Ref.~\cite{aps3}.\footnote{For a comparison of different quantizations of this model in LQC, see Ref.~\cite{MenaMarugan:2011me}} Non-Gaussian states present computational challenges far more severe than the corresponding Gaussian states. To understand this, let us take 
an example of a squeezed state.  It turns out that for the same value of field momentum and its absolute dispersion, 
 the squeezed state is more widely spread in volume 
than its Gaussian counterpart.  A typical simulation for a sharply peaked initial state with   
$p_\phi=1000\,\sqrt{G}\hbar$ and relative dispersion in volume observable of $1 \%$  
requires about $30,000$ grid points on the spatial grid. Such a simulation takes 
approximately $240\, {\rm sec}$ on a 2.4 GHz Sandybridge workstation with 
16 cores. On the other hand, a simulation of a squeezed state with the same values of field momentum has larger volume dispersion and requires $4\times10^{11}$ grid points.
Such a simulation, running on a computer equipped with enough memory for the
entire simulation,  
 would take approximately $7\times10^{12}\,{\rm hrs}\approx10^8\, {\rm years}$ 
with pre-Chimera techniques.
As for the widely spread states considered in Ref~ \cite{dgs2}, the 
Chimera scheme dramatically brings down the computational costs and makes such
simulations possible. The Chimera scheme utilizes the fact that in the large volume limit, the quantum Hamiltonian constraint in LQC 
can be very well approximated by the Wheeler-DeWitt equation \cite{dgs1}. 
In this scheme  a hybrid (spatial) grid is introduced. The hybrid grid is composed of an inner grid
corresponding to the small volume regime, where we solve the LQC difference
 equation, and an outer grid where the evolution is governed by the 
Wheeler-DeWitt differential evolution equation. The inner grid, of course,
has to be chosen sufficiently large to capture all the non-trivial LQC 
physics.  This leads to a significant reduction in the computational 
cost of the simulation. A simulation of the squeezed state mentioned above,
that would have taken $10^8\, {\rm years}$, can now be performed in about
10 minutes.

Using the Chimera scheme we perform numerical simulations of a spatially flat
isotropic and homogeneous model in LQC  in the presence of a massless scalar field
with three types of 
non-Gaussian states: (i) squeezed states (ii) a sum of Gaussians in the
momentum space which lead to a multipeaked state in volume, and  
(iii) a state with even more peaks. The two latter types of states will be
denoted multipeaked-1 and multipeaked-2 in this manuscript. 
We find, as in the earlier studies with Gaussian states in this model~\cite{aps2,aps3,dgs2}, 
that the evolution is non-singular, and the classical singularity is
replaced by a quantum bounce. 
This is shown to be a generic feature of all the types of states considered, 
irrespective of the initial parameters. 
The 
relative dispersions across the bounce, for all types of states {\it{always}}
obey the triangle inequalities derived in  
Ref. \cite{kp}. We also find that certain types of squeezed states satisfy a
stronger set of triangle inequalities as derived in
Ref. \cite{montoya_corichi1} for sharply peaked states. If the state is highly
squeezed then this particular triangle inequality is found to be violated.
The latter inequality is also satisfied by multipeaked states studied in this
work, even though they are not sharply peaked. The reason for this lies in the  
way these states are constructed and the values of parameters
considered.\footnote{All the states considered in this manuscript are
  constructed using ``method-3'' of Refs. \cite{aps3,dgs2}. In this method,  
the initial state uses Wheeler-DeWitt eigenfunctions but with an additional
phase factor, carefully chosen to mimic the behavior of eigenfunctions of LQC
at large volumes.  
Depending on the parameters of the state, in this construction, the relative
fluctuations of volume at times much earlier and much later after the bounce
can be very similar.  
If a multipeaked state has this behavior, the stronger form of triangle
inequality (\ref{eq:corichieps}) can be satisfied.} 
These results 
confirm the robustness of the constraints on the growth of relative
fluctuations across the bounce for highly non-Gaussian states. We find that
 the profile   
of the state at early  times before the bounce is not affected at late times
after the bounce. In synergy with the results obtained in Ref. \cite{dgs2},  
our analysis reveals that the effective theory always underestimates the
bounce volume  
and overestimates the energy density at the bounce. Since the effective
dynamics is derived under the assumptions of sharply peaked Gaussian states,  
we find, as expected, significant quantitative departures between the
predictions of the effective theory and the quantum evolution of states with
large volume dispersion. However, the
effective dynamics qualitatively  
captures the  main features of the physics in LQC, 
and it still provides a very good approximation to the underlying quantum dynamics
for states with small dispersion, even when they are highly non-Gaussian.
We
find that the energy density can be much smaller than the maximum density
predicted in sLQC.  
In particular, for squeezed states, we find that the 
behavior of the energy density with increased squeezing agrees extremely well 
with the analytical calculations performed for sLQC in
Ref. \cite{montoya_corichi2}. It is important to note that the quantum
constraint considered in our analysis, which is the same as  
the one analyzed in Ref. \cite{aps3}, does not 
correspond to the one in sLQC. Despite this difference, we find sLQC to
provide important insights on various findings and in agreement with several
results in our analysis.  

%zzz  This paragraph to be reviewed:
Finally, it is worth emphasizing that, although the states studied here do not
necessarily correspond to a classical universe at late times, they are in the physical Hilbert space. Therefore, 
their study is important for testing the robustness of the Planck scale physics, 
in particular the resolution of the classical singularity, the existence of a quantum bounce, 
constraints on the growth of fluctuations and the reliability of the effective dynamics. 
Our results show that many features of the new physics first observed for the Gaussian states also hold true for the states which are highly non-Gaussian. Thus, our analysis provides a strong robustness test of the physics in LQC. 
Furthermore, these results are potentially important  in computing corrections to the
observational signatures of LQC arising from the state fluctuations.

This manuscript is organized as follows. We give a brief overview of the 
loop quantization, relative fluctuations and the effective dynamics of a flat FRW model with a 
massless scalar field in \secref{sec:lqc}. In \secref{sec:initialdata}, we describe 
the construction of the three different types of non-Gaussian initial states. For completeness we briefly discuss the main features of the Chimera scheme 
in \secref{sec:chimera}. In \secref{sec:results} we discuss the results 
of the numerical simulations for the three kinds of states and also compare their 
LQC evolution with the corresponding effective trajectories. We also study the 
variation of the energy density at the bounce and the validity of the triangle equalities  
for the relative fluctuations of various non-Gaussian states. In \secref{sec:disc} we present 
a summary and discussion of the main results. 

%%%%%%%%%%%%%%%%%%%%%%%%%%%%%%%%%%%%%%%%%%%%%%%%%%
\section{Loop quantum cosmology of the spatially flat model: Quantum Constraint, Fluctuations and Effective Dynamics}
\label{sec:lqc}
%%%%%%%%%%%%%%%%%%%%%%%%%%%%%%%%%%%%%%%%%%%%%%%%%%
In this section we briefly describe the loop quantization of a flat FRW spacetime 
with a massless scalar field as the matter source. We start with a summary of the key ideas behind the quantum Hamiltonian 
constraint in LQC and the 
way the difference equation arises. After discussing the relation of the LQC quantum constraint with the Wheeler-DeWitt equation at 
 small spacetime curvatures, we summarize the main results on the 
bounds of relative fluctuations of the Dirac observables across the bounce. We conclude this section with a discussion of the effective 
dynamics derived from the effective Hamiltonian for sharply 
peaked Gaussian states \cite{vt}. For details, we refer the reader to the original works where this quantization was first performed \cite{aps2,aps3} (see also Ref. \cite{acs}) and Ref. \cite{as1} 
for a review.

%%%
\subsection{Quantum Hamiltonian Constraint}
%%%

Loop quantization of the cosmological spacetimes is based on the techniques of LQG,  where the primary variables 
for the quantization of the gravitational sector are the holonomies of the Ashtekar-Barbero connection $A^i_a$ and the fluxes of the triads $E^a_i$.
In this canonical quantization, due to the homogeneity of the cosmological spacetimes, the only non-trivial constraint is the Hamiltonian constraint which 
is expressed in terms of the symmetry reduced version of the canonically conjugate connection-triad pair $(c, p)$
\be
A_a^i = c~V_o^{1/3}~ \mathring{\omega}_a^i, \quad  {\rm and} \quad E_i^a=  p~V_o^{-2/3}\sqrt{q} ~\mathring{e}_i^a,
\ee
where $\mathring{e}_i^a$ are the densitized triads and $\mathring{\omega}_a^i$ are the 
fiducial co-triads compatible with the fiducial metric $\mathring{q}_{ab}$, and $V_o$ denotes the volume of the fiducial cell introduced to 
define the symplectic structure in the canonical quantization. At the kinematical level, the triad is related to the scale factor in the physical metric as 
$|p|=V_o^{1/3} a^2 = V^{2/3}$, where the modulus sign arises due to the two possible orientations of the triads.\footnote{Since the matter sector 
of the Hamiltonian constraint in this model has no fermions, the resulting physics is insensitive to the orientation of the 
triad.} For the classical 
solutions, the connection is proportional to the time derivative of the scale factor $c=\gamma V_o^{1/3}\dot{a}$, 
where $\gamma \approx 0.2375$ is the Barbero-Immirzi parameter and  the derivative is taken with respect to proper time.

The gravitational part of the Hamiltonian constraint 
expressed in terms of the triads and the field strength $F_{ab}^i$ of the connection is
\be
C_{\rm grav} = -\f{1}{\gamma^2} \int d^3x~\varepsilon_{ijk}~ \f{E^{ai}E^{bj}F^k_{ab}}{\sqrt{|\rm{det}(E)|}} ~\;,
\ee
where we have chosen the lapse to be unity. For the massless scalar field model under consideration, the matter part of the Hamiltonian constraint is 
\be
C_{\rm matt}= \f{p_{\phi}^2}{2|p|^3},
\ee
where $p_\phi$ is the momentum of the scalar field, which is a constant of motion in this case.

To quantize the Hamiltonian constraint, the field strength is expressed in terms of the holonomies $h_i^{(\lambda)}$ of the symmetry reduced connection $c$ on 
a square loop $\Box_{ij}$ 
\be
F_{ab}^i = -2 \lim_{Ar_\Box\rightarrow 0} {\rm Tr}\left(\f{h_{\Box_{jk}}^{(\lambda)}-1}{Ar\Box_{ij}}\right)\tau^k \mathring{\omega}^j_a\mathring{\omega}^k_b,
\ee
where $h_{\Box_{ij}}^{(\lambda)} = h_i^{(\lambda)} h_j^{(\lambda)}\left(h_i^{(\lambda)}\right)^{-1} \left(h_j^{(\lambda)}\right)^{-1}$, and $Ar_\Box$ denotes the 
physical area of the loop whose sides are parameterized by $\lambda$. In LQC, this area has a minimum value determined by the underlying 
quantum geometry which fixes $\lambda$ as 
$\lambda = {2} \sqrt{\left(\sqrt{3} \pi \gamma\right)}\, \lp$. The non-local nature of the field strength which encodes the quantum geometry plays an important role in the new physics near 
the Planck scale in LQC. The elements of the holonomy algebra are almost periodic 
functions of the connection, and the corresponding operators act as a translation 
\be
\widehat{\exp(i \lambda b/2)} \,  |v\rangle = | v - 1\rangle, 
\ee
where $b = c/|p|^{1/2}$ and $|v\rangle$ is the eigenstate of the volume operator
\be\label{eq:Vv}
\hat V |v\rangle = \left(\frac{8 \pi \gamma}{6}\right)^{3/2} \frac{|v|}{K} \lp^3 |v\rangle, ~~ \qquad K = \f{2}{3\sqrt{3\sqrt{3}}}.
\ee
The resulting action of the Hamiltonian constraint 
on states $\Psi(v,\phi)$ is
\be
\widehat{C_{\rm grav}} \Psi(v,\, \phi) = C^+(v) \Psi(v+4, \phi) + C^o(v)\Psi(v,\,\phi)+ C^-(v)\Psi(v-4,\,\phi),
\ee
where the coefficients $C^\pm$ and $C^o$ are given by
\ba
  C^-(v) &=& C^+(v-4)=\frac{3\pi K G}{8} |v-2|\,\Big| |v-3|-|v-1| \Big|, \nonumber \\
  C^+(v) &=& \frac{3\pi K G}{8} |v+2|\,\Big | |v+1|^{}-|v+3| \Big |, \nonumber \\
  C^0(v) &=& -C^+(v)-C^-(v).
\ea
The total Hamiltonian constraint can thus be expressed as a Klein-Gordon type
equation with $\phi$ as the internal time:
\be
\label{eq:lqcevol}\f{\partial^2}{\partial \phi^2}\Psi(v,\,\phi)=-\widehat{\Theta}\Psi(v,\,\phi).
\ee
The operator $\widehat{\Theta}$ is the LQC difference operator, which is
positive definite and self-adjoint, and is defined as follows:
\be
\label{eq:lqctheta}\widehat{\Theta}=-\f{1}{B(v)}\Bigg[C^+(v) \Psi(v+4, \phi) + C^o(v)\Psi(v,\,\phi)+ C^-(v)\Psi(v-4,\,\phi)\Bigg],
\ee
where 
\be\label{eq:Bv}
%B(v)= \f{27}{8}K |v|\left(\arrowvert |v+1|^{1/3} - |v-1|^{1/3}\arrowvert\right)^3
B(v)= \f{27}{8}K |v| \,\Big| |v+1|^{1/3} - |v-1|^{1/3}\Big|^3
\ee
denotes the eigenvalues of the inverse volume operator in LQC. If we label the 
eigenvalues of $\hat \Theta$ with $\omega^2$, then the physical states can be
either chosen as the positive frequency $(\omega > 0)$ or the 
negative frequency ($\omega < 0$) solutions. During the evolution there is no mixing between positive and negative frequencies, and it suffices 
to consider only positive frequency states.
These states are normalized using the following inner product 
obtained by using group averaging procedure \cite{Ashtekar:1995zh, Marolf:1995cn}
\be
\label{eq:inprod}\langle \Psi_1|\Psi_2\rangle = \sum_v \overline{\Psi_1}(v,\phi_o) {B(v)} {\Psi_2}(v,\phi_o).
\ee
This inner product can also be obtained by requiring that the action of Dirac observables in this model are self adjoint. 
Two Dirac observables we will be interested in particularly are the volume at a given slice of time $\phi$, and the field momentum, which is a constant of motion
\be
\hat V|_{\phi_o} \Psi(v,\phi) = \left(\frac{8 \pi \gamma}{6}\right)^{3/2} \frac{\lp^3}{K} |v| e^{i \sqrt{\Theta} (\phi - \phi_o)} \Psi(v,\phi_o), ~~\mathrm{and} ~~ \widehat p_\phi \Psi(v,\phi) = - i \hbar \partial_\phi \Psi(v,\phi) = \sqrt{\Theta} \Psi(v,\phi) ~.
\ee
We can  evaluate  
 the expectation values of the Dirac  
observables  as follows
\be
\label{eq:expect}\langle \widehat{\mathcal O} \rangle = \langle \Psi|\widehat{\mathcal O}|\Psi\rangle=||\Psi||^{-1}\sum_v {B(v)}\overline{\Psi}(v,\phi) \widehat{\mathcal O} {\Psi}(v,\phi)
\ee
where $||\Psi||$ is the norm of the wavefunction 
%$||\Psi||=\langle \Psi|\Psi\rangle$, 
and $\widehat{\mathcal O}$ refers to the quantum operator of interest. The quantum evolution and the action of the Dirac observables 
preserves the lattice in volume labeled by $\epsilon$: $v = \pm \epsilon + 4 n$, where $\epsilon \in [0,4)$ and $n \in \mathbb{Z}$. 
Due to this reason, there is a superselection and the physical Hilbert space can be decomposed into separable Hilbert spaces labeled by $\epsilon$, ${\cal H}_\epsilon$. 
In our analysis, we 
will consider the choice $\epsilon = 0$, since it allows the zero volume. Finally, since there are no fermions in our model, the physical states will  be considered to be symmetric under 
the change of sign of the physical triads: $\Psi(v,\,\phi)=\Psi(-v,\,\phi)$.

The evolution operator in LQC (\eref{eq:lqctheta}) is a non-singular difference operator with a uniform
discretization in $v$ which results in a quantum bounce of the physical
states in the Planck regime \cite{aps2,aps3}. The existence of a 
bounce is tied to the underlying 
quantum geometry which is captured by the non-local nature of the field strength in the quantum Hamiltonian constraint. In contrast, in the Wheeler-DeWitt theory, the quantization of this 
model yields the following evolution equation:
\be
\label{eq:wdw} \f{\partial^2}{\partial \phi^2} \underline{\Psi}(v,\,\phi)=-\widehat{\underline{\Theta}}\,\underline{\Psi}(v,\,\phi)= 12 \pi G v\f{\partial }{\partial v}\left(v\f{\partial }{\partial v}\right)\underline{\Psi}(v,\,\phi),
\ee
where the operator $\widehat{\underline{\Theta}}$ is the Wheeler-DeWitt
evolution operator with eigenvalues $\omega^2$. The evolution of the physical states in the Wheeler-DeWitt 
theory does not lead to a singularity resolution. Rather, the states follow the classical trajectory throughout the evolution. 
In the large volume approximation, it is straightforward to show that $\widehat{\Theta}$ can be approximated by $\underline{\widehat\Theta}$. Thus, the continuum classical geometry is recovered from LQC 
 at small spacetime curvature. This property turns out to be extremely useful in our numerical simulations.

%%%%%%%%%%%%%%%%%%%%%%%%%%%%%%%%%%%%%%%%%%%%%%%%%%
\subsection{Properties of the dispersion in $p_\phi$ and $V$.}
\label{sec:triangle}
%%%%%%%%%%%%%%%%%%%%%%%%%%%%%%%%%%%%%%%%%%%%%%%%%%
An important property of the loop quantum evolution is that the relative fluctuations of volume observables are very tightly constrained 
across the bounce by the dispersion in the momentum observable \cite{recall,kp,montoya_corichi1,montoya_corichi2}. This implies that semi-classicality of the initial state is preserved through the evolution.   
This property is most evident in terms of the triangle inequalities, which 
constrain the change in the relative fluctuations in volume in the expanding and contracting phases of LQC with the relative fluctuation in 
the momentum $p_\phi$. Note that the latter and its fluctuation remains constant throughout the evolution. The triangle inequalities are derived using sLQC \cite{acs}, which is an exactly solvable loop quantization of the 
massless scalar model in the spatially flat homogeneous spacetime with the lapse $N = a^3$. 
The most general treatment valid for all the physical states in sLQC is
performed in Ref. \cite{kp}, which reaches the conclusion that 
although the volume dispersions $\Delta {\ln(\hat v|_\phi)}$ could be different across the 
bounce,  the difference between their asymptotic values $(\sigma_\pm)$ is bounded by 
the dispersion in the field momentum $\Delta {\ln(\widehat p_\phi)}$ as follows\footnote{The expectation value of the logarithm of field momentum can be computed analytically in this model using 
$\langle {\ln(\widehat p_\phi/\sqrt{G} \hbar)} \rangle = ||{\Psi}(\omega)||^{-1}\int \overline{{\Psi}}(\omega) \ln(\omega/\sqrt{G}) {\Psi}(\omega) d\omega.$ 
}

%The corresponding 
%inequality is like a triangle inequality given as follows:
\be
\label{eq:triangle} |\sigma_+-\sigma_-| \leq 2 \sigma ~.
\ee
In the numerical simulations,  
\be\label{eq:triangle2}
\sigma_\pm= \Delta {\ln(\hat v|_\phi)_\pm}, ~~ \mathrm{and} ~~ \sigma=\Delta {\ln(\widehat p_\phi/\sqrt{G}\hbar)} ~
\ee
are measured at very large values of $|\phi|$ before and after the bounce. 
Since the derivation of (\ref{eq:triangle}) assumes no particular state, it is valid   
 for all the choices of initial states in the physical Hilbert space, irrespective of whether the state is sharply peaked or 
not. A stronger bound on the fluctuations can be obtained by noting that if a state is sharply peaked, 
then the dispersion in the logarithm of a physical observable $\mathcal O$ can be approximated as 
$\langle\Delta \ln \hat {\mathcal O} \rangle \approx  \langle\Delta \hat {\mathcal O}\rangle/\langle\hat{\mathcal O}\rangle$. Under this approximation we can write
\be\label{eq:approx}
\Delta (\ln(\hat v|_\phi)_\pm\approx\Sigma_\pm:=\left(\f{\langle\Delta \hat v|_\phi\rangle}{\langle\hat v|_\phi\rangle}\right)_\pm ~~~ \mathrm{and} ~~~ \Delta \ln(\widehat p_\phi/\sqrt{G} \hbar)\approx\Sigma:=\f{\langle\Delta \widehat p_\phi\rangle}{\langle\widehat p_\phi\rangle} ~.
\ee
The inequality in \eref{eq:triangle} then takes the following form
\be
|\Sigma_+-\Sigma_-| \leq 2 \Sigma,
\ee
which can also be written as
\be
\label{eq:corichieps}\mathcal E:=\f{\left| \Sigma_+-\Sigma_-\right|}{2\Sigma} \leq 1.
\ee
Unlike (\ref{eq:triangle}), the inequality (\ref{eq:corichieps}) is  valid when a state satisfies (\ref{eq:approx}). When the latter approximation is not satisfied, the inequality (\ref{eq:corichieps}) can be violated, as we show in our analysis for highly squeezed 
states which have large relative fluctuation in volume. However, depending on the 
construction of states and the values of parameters, it is possible that ${\mathcal E}$ turns out be smaller than unity even if the state may not be sharply peaked. Examples of such states are studied 
in Sec. IVB and Sec. IVC, where due to a peculiar behavior of $\Sigma_\pm$ tied to the way states are constructed, the above inequality is satisfied for certain parameters for multipeaked states. It is important to note that in contrast to (\ref{eq:corichieps}), the triangle inequality (\ref{eq:triangle}) is satisfied for all the physical states independent of the choice of parameters and the way they are constructed.

These triangle inequalities imply that the difference between the relative 
dispersions across the bounce for a semi-classical state is tightly constrained, and the 
semi-classicality of a state across the bounce is preserved. That is, if the wavefunction of the 
universe on one side of the bounce is semi-classical then it will remain semi-classical throughout 
the evolution. This is an important argument in support of cosmic recall \cite{recall}. Since the triangle inequalities 
 (\ref{eq:triangle}) are derived in sLQC, at first it may seem surprising that they are found to be satisfied for all the states in the present analysis, where the quantum constraint 
 is different from the exactly solvable model in sLQC. The reason for this agreement lies in the fact that the physical differences between sLQC and the presented quantization 
 can only become important when the bounce volume is close to the Planck volume, which never occurs for any state considered in our analysis.

%%%%%%%%%%%%%%%%%%%%%%%%%%%%%%%%%%%%%%%%%%%%%%%%%%
\subsection{Effective dynamics}
\label{sec:effective}
%%%%%%%%%%%%%%%%%%%%%%%%%%%%%%%%%%%%%%%%%%%%%%%%%%
For suitably chosen physical states, it is possible to derive a continuum effective spacetime description of LQC via the geometric formulation 
of quantum mechanics, where the Hilbert space is treated as a quantum phase space. In this manuscript, we will analyze effective dynamics as obtained in the 
embedding approach, where one finds a faithful embedding of the quantum phase space into the classical phase space \cite{josh,vt}.\footnote{For a discussion of different approaches to 
obtain effective dynamics in LQC, see \cite{as1}.}  The 
effective Hamiltonian constraint can then be derived through an appropriate
choice of a semi-classical state, such as a sharpy peaked Gaussian state. Using Hamilton's equations, modified Friedmann and Raychaudhuri 
equations with quantum geometric corrections can then be obtained in a straightforward way. The effective Hamiltonian constraint for the spatially flat, homogeneous 
and isotropic FRW spacetime sourced with a massless scalar field is given as \cite{vt}\footnote{In the effective Hamiltonian, we ignore the contributions from the 
inverse volume effects (encoded in $B(v)$ (eq.\ref{eq:Bv})). These corrections play little role on the physical implications in LQC unless the bounce occurs close to the Planck volume.}
\be
\label{eq:heff}C_{\mathrm{eff}} = -\f{3V}{8\pi G \gamma^2}\f{\sin^2\left(\lambda b \right)}{\lambda^2} + \f{p_\phi^2}{2V} \approx 0~.
\ee
The variable $b = c/|p|^{1/2}$ is the conjugate variable to $V$ and satisfies the Poisson bracket relation: 
$\{b, V\}=4\pi G\gamma$, where $\lambda^2=4\sqrt{3}\pi\gamma\lp^2$ is the minimum eigenvalue of the area 
operator in loop quantum cosmology. This effective Hamiltonian constraint neglects any state dependent fluctuation terms, which are small for sharply peaked states with small relative fluctuations.
For a more accurate analysis, these terms should be included for a given choice of semi-classical states. For states which are not semi-classical or have large 
relative fluctuations, the underlying assumptions in the derivation of the effective Hamiltonian are violated and the effective dynamics can not be trusted. 
In this manuscript, we will restrict our analysis to the study of the effective dynamics resulting from eq.(\ref{eq:heff}). 

%$\mathcal{H}_\phi$ is the matter Hamiltonian for massless scalar field and 
%
Using Hamilton's equation of motion for the volume variable $V$, one can 
obtain the time derivative of $V$ as follows
\be
\dot{V}=\f{3}{2\gamma\lambda} \sin\left(2\lambda b\right) V.
\ee
This equation, along with the vanishing of the Hamiltonian constraint, yields the 
modified Friedmann equation
\be
\label{eq:fried}  H^2=\left(\f{\dot V}{3V}\right)^2 = \f{8\pi G}{3}\rho\left(1-\f{\rho}{\rho_{\rm b}^{\rm eff}}\right), ~~~\mathrm{where} ~~~ \rho_{\rm b}^{\rm eff} = \f{3}{8\pi G \gamma^2\lambda^2}\approx 0.409\,\rhopl.
\ee
Here $H = \dot V/3V = \dot a/a$ is the Hubble rate and $\rho = p_\phi^2/(2V^2)$ is the energy density  of the scalar field. % $H$ is the Hubble rate, $a$ is the scale factor and 
The modified Friedmann equation predicts a maximum bound for energy density 
$\rho_{\rm b}^{\rm eff}$ where the Hubble rate vanishes and a quantum bounce occurs. It is 
interesting to note that this bound is the same as the maximum bound 
$\rho_{\rm max}$ for the expectation values of the energy density observable in sLQC. However, 
unlike in sLQC, where the bound is valid for all the states in the physical Hilbert space, the bound in 
energy density in the effective dynamics is derived for sharply peaked 
Gaussian states.

%$\rho_{\rm b}^{\rm (eff)}

The modified Raychaudhuri equation can in a similar way be 
derived from the time derivative of $b$, giving
\be
\label{eq:rayc} \f{\ddot a}{a} = \f{-4\pi G}{3}\rho\left(1-4\f{\rho}{\rho_{\rm b}^{\rm eff}}\right) -4\pi G P\left(1-2\f{\rho}{\rho_{\rm b}^{\rm eff}}\right). 
\ee
%where $a=V^{1/3}$ is the scale factor.
The modified Friedmann and Raychaudhuri equations given in equations 
(\ref{eq:fried}) and (\ref{eq:rayc}) respectively lead to the classical Friedmann and Raychaudhuri equation when $\rho \ll \rho_{\rm b}^{\rm eff}$. These equations can 
be used to obtain the effective 
dynamical trajectory of a cosmological model by providing initial conditions at some 
$t=t_0$. In this paper we compute the effective dynamical trajectories corresponding 
to an LQC evolution by giving the initial conditions far from the bounce in the low 
curvature regime in the expanding branch.

For sharply peaked initial states, extensive numerical simulations for different matter models show that 
the effective dynamics provides an excellent  description of the underlying quantum dynamics (see Ref. \cite{as1,ps12} for a review of these results). For 
states which are not sharply peaked, there are additional corrections to the effective Hamiltonian constraint which affect 
the energy density in the effective theory at the bounce. A detailed numerical analysis of widely spread Guassian states has been recently performed in Ref. \cite{dgs2}.  
It was concluded that, for states with significant relative fluctuations, even though the effective dynamical trajectory obtained from the above constraint captures the 
qualitative features of the underlying quantum dynamics, it always overestimates the spacetime curvature at the bounce. In our analysis, we will test the 
validity of the effective Hamiltonian constraint for states which are non-Gaussian. We will show that for states which are highly non-Gaussian, there are large departures between the 
quantum evolution and the above effective dynamics. It is important to note that, though the evolution of such highly non-Gaussian states can not be reliably captured
by the effective dynamics, it 
turns out to be in excellent agreement with predictions from sLQC. This is not surprising because the predictions extracted from sLQC are valid for arbitrary states in the 
physical Hilbert space, whereas the effective description assumes that the states have some peakedness properties.

%%%%%%%%%%%%%%%%%%%%%%%%%%%%%%%%%%%%%%%%%%%%%%%%%%
\section{Initial data and numerical scheme} 
%\label{sec:initialdata}
%%%%%%%%%%%%%%%%%%%%%%%%%%%%%%%%%%%%%%%%%%%%%%%%%%
In the previous section we discussed an important relation between  the quantum difference equation in LQC and the differential WDW equation in the regime of 
small spacetime curvature. In the spatially flat model under consideration, for a given value of the field momentum $p_\phi$, 
the spacetime curvature becomes smaller as the 
volume increases. Therefore, at large volume, the eigenfunctions of LQC can be approximated by 
a linear superposition of the eigenfunctions of $\widehat{\underline{\Theta}}$ (eq.(\ref{eq:wdw})) in the Wheeler-DeWitt theory, which are given by %The eigenfunctions  are %denoted by $\omega^2$ (as in LQC), and its 
%eigenfunctions are % ${\underline{e}}_k$ as its eigenfunctions
\be
\underline{e}_{k}(v)=\f{1}{2\pi}e^{i k \ln |v|},
\ee
with $\omega = \sqrt{12 \pi G} |k|$.  A general solution of the Wheeler-DeWitt equation can be written as a superposition of the outgoing $(k < 0)$ and the incoming $(k > 0)$
solutions. The outgoing part corresponds to the expanding branch, and the ingoing part corresponds to the contracting branch in the FRW spacetime. In the Wheeler-DeWitt  theory, the 
contracting and expanding trajectories are disjoint, and the initial data can be chosen on any of the trajectories.
If an initial state is chosen peaked on an expanding branch, i.e. 
 a wavefunction with no support on the positive axis of $k$, its evolution in the Wheeler-DeWitt theory yields a trajectory which agrees with the classical GR trajectory all the way
 to the big bang singularity at $v = 0$ in its past evolution. Similarly, if the intial state is chosen on a contracting branch, 
 it encounters a big crunch singularity in its future evolution.\footnote{It may seem that a general superposition of incoming and outgoing solutions can potentially resolve 
 the big bang singularity in the Wheeler-DeWitt theory. This expectation turns out to be wrong. It can be analytically shown that for this matter model, even with the states 
 constructed from a superposition of expanding and contracting solutions, the  probability for a singularity to occur is unity in the Wheeler-DeWitt quantization  \cite{craig_singh_wdw1,craig_singh_wdw2}.} 

 In our numerical simulations we will consider initial states
 which are peaked on the expanding trajectory at late times (i.e. large $\phi$) and have $\omega > 0$. A general positive frequency Wheeler-DeWitt initial state at time $\phi = \phi_o$ is of the form  
\be
\label{eq:sqstate}\underline{\Psi}(v,\,\phi) = \int \Psi(k)
\underline{e}_k(v)e^{i\omega(\phi-\phi_o)} e^{-i\alpha} \, dk, %e^{i\alpha}\,dk,
\ee
where $k=-p_\phi/\sqrt{12\pi G \hbar^2}$. % $\omega^2$ is the eigenvalue of the Wheeler-DeWitt 
The time derivative of $\underline\Psi(v,\phi)$ can be computed by evaluating the following integral
\be
\label{eq:psidot}\f{\partial }{\partial \phi} \Psi(v,\,\phi)=\int i\omega~\Psi(k) \underline{e}_k(v) e^{i\omega(\phi-\phi_o)} e^{-i\alpha} \, dk .%e^{i\alpha} \,dk.
\ee
Given the form of a wavepacket $\Psi(k)$ (or equivalently $\Psi(\omega)$), the initial state and its derivative
can be computed by numerically evaluating the integrals in \eref{eq:sqstate} 
and (\ref{eq:psidot}). Since the physical states are required to be symmetric under the change in the orientation of the
triads, i.e.\ $\Psi(v, \phi) = \Psi(-v, \phi)$, we construct the initial states
using the Wheeler-DeWitt eigenfunctions that approximate the symmetric eigenfunctions of the LQC operator at large volumes. This is achieved by multiplying the integrand in eq.(\ref{eq:sqstate}) by a phase factor $e^{-i \alpha}$ where $\alpha = k (\ln |k| - 1)$ \cite{aps3,dgs2}.\footnote{The initial states used for the simulations in this manuscript are based on ``method-3'' in the convention used in earlier works \cite{aps2,dgs2}.}

Before going into more details about our prescription for the initial data, it is useful to point out an important property of the eigenfunctions 
of the $\hat \Theta$ operator in the small volume regime. Earlier studies show that the eigenfunctions of $\hat \Theta$, unlike those of ${\underline{\hat \Theta}}$, decay almost exponentially near zero volume \cite{aps2,aps3,Craig:2012gw}. The decay starts near a cutoff volume $V_c$, which can be computed from the analytical results in Ref. \cite{Craig:2012gw} as\footnote{In Ref. \cite{Craig:2012gw}, the cutoff  was obtained on $k$  (see eq.(3.36a) of Ref. \cite{Craig:2012gw}). Since our goal will be to understand this cutoff in volume and compare it to the bounce volume, we have written the same equation in a different form.}
\be \label{eq:vcut}
 V_{\rm c} = \sqrt{\f{4 \pi G}{3}}\, \gamma\lambda \, \omega \hbar . 
\ee 
Although  $V_{\rm c}$ is computed for eigenfunctions in sLQC, one
may expect that, for states with a well defined peak, a good estimation
 of the volume at which  the wavefunction amplitude decays almost exponentially will be 
 given by replacing  $\omega$  with the eigenvalue at which the state is peaked.
The reason for this is that the main contribution to the wavefunction comes from
eigenstates with eigevalues around this value. In fact, we find that for 
states with a well defined peak and $\langle \widehat p_\phi \rangle = \omega^* \hbar$ 
(with $\omega^*=-\sqrt{12\pi G}k^*$, see \eref{eq:sqpsi}), the bounce volume is very close to the 
cutoff volume. For states with a well defined peak for which the expectation
value of the field momentum is different from $\omega^*$, 
the cutoff volume for the amplitude of the wavefunction can be estimated by using $\langle \widehat p_\phi \rangle$ in eq.(\ref{eq:vcut}).
We will see in~\secref{sec:results} that the
squeezed states  do show a decay of the amplitude at the bounce, starting at the value of $V_{\rm c}$ estimated in this way.

An initial state $\Psi(v,\phi)|_{\phi=\phi_o}$ and its time derivative 
$\partial_\phi\Psi(v,\phi)|_{\phi=\phi_o}$  provide the initial data for
evolution using the Chimera method, which uses the Wheeler-DeWitt theory for evolution in an 
outer grid corresponding to very large volumes, and the LQC quantum difference equation in an 
inner grid  (discussed in Sec.\ref{sec:chimera}). The wave profile $\Psi(k)$ corresponds to a choice 
of $p_\phi = p_\phi^*$, and the initial state is constructed at $v|_{\phi_o} = v^*$ with $v^* \gg 1$. 
In our simulations we consider three types of initial states: squeezed states, double peaked states 
in the $k$ space (which in general have more than two peaks in the $v$ space), and multipeaked 
states with several peaks. We will refer to the latter two types  as  multipeaked-1 and multipeaked-2 
in the following. Note that all of these states are far more general than the Gaussian states 
considered in the previous numerical simulations in LQC. In the following we first discuss the 
construction of the initial states and then briefly summarize the 
numerical algorithm used in the simulations.

%%%%%%%%%%%%%%%%%%%%%%%%%%%%%%%%%%%%%%%%%%%%%%%%%%
\subsection{Types of initial states} \label{sec:initialdata}
%%%%%%%%%%%%%%%%%%%%%%%%%%%%%%%%%%%%%%%%%%%%%%%%%%

({\em i}) {\em Squeezed states:} 
These states can be written as a generalization of 
the Gaussian states considered in previous numerical works 
\cite{aps2,aps3,dgs2}. A Gaussian state is characterized by two parameters: 
the field momentum where the state is peaked ($p_\phi^*$) for which $k^*=-p_\phi^*/\sqrt{12\pi G}\hbar$ and
the spread of the Gaussian quantified by a real quantity $\eta$. Here 
we generalize the Gaussian to a squeezed Gaussian state by allowing 
$\eta$ to be a complex quantity. As yet another variation we include a factor $|k|^n$, which leads to the following expression for the wavepacket in $k$ space
\be
\label{eq:sqpsi}\Psi(k) = |k|^n e^{-\eta(k-k^*)^2},
\ee 
where $n \in \mathbb{R}$ and  $\eta \in \mathbb{C}$. \\

({\em ii}) {\em Multipeaked-1 states:}
For this type of initial data  we use the sum of two separated 
Gaussian waveforms in $k$ space as
\be
\label{eq:avg}\Psi(k) = \frac{1}{2}\left ( e^{-\eta\left(k-\left(k^*+\f{\delta k}{\sqrt{24\pi|\eta|}}\right)\right)^2}+
                           e^{-\eta\left(k-\left(k^*-\f{\delta k}{\sqrt{24\pi|\eta|}}\right)\right)^2} \right ),
\ee
where $k^*=-p_\phi^*/\sqrt{12\pi G}\hbar$ and $\delta k$ is a free parameter which parameterizes the separation between the two Gaussian components in $k$ space.
We will later see that the above sum of Gaussians, when expressed as a function of $v$ can have multiple peaks. Note that unlike the case of squeezed states, $\eta$ 
is assumed to be real for multipeaked-1 states.\\
%In this sense, the average Gaussian state may not be peaked at a single volume at the initial time.\\

({\em iii}) {\em Multipeaked-2 states:}
For our second multipeaked state we choose the following waveform in the $k$ space
\be
\label{eq:multi}\Psi(k) = e^{-\eta^2{\left(k-\left(k^*+\f{\delta k}{\sqrt{24\pi|\eta|}}\right)\right)^2}{\left(k-\left(k^*-\f{\delta k}{\sqrt{24\pi|\eta|}}\right)\right)^2}},
\ee
where $k^*=-p_\phi^*/\sqrt{12\pi G}\hbar$, $\delta k$ is a free parameter and $\eta$ is a real parameter.
We will see later that the above state, when expressed in $v$ space, has 
several local maxima. 
%Thus, these states have almost no peakedness property.
Thus, these states have no well defined peakedness property.

%%%%%%%%%%%%%%%%%%%%%%%%%%%%%%%%%%%%%%%%%%%%%%%%%%
\subsection{Numerical scheme}\label{sec:chimera}
%%%%%%%%%%%%%%%%%%%%%%%%%%%%%%%%%%%%%%%%%%%%%%%%%%
Using the numerical techniques of Ref.\ \cite{aps2,aps3} for the quantum
Hamiltonian constraint~\eref{eq:lqcevol} it would be computationally very 
expensive to evolve widely spread non-Gaussian states. 
On one hand, for states which are not 
sharply peaked, large computational domains are needed. And, since the quantum difference equation in LQC has a fixed discretization,  a large domain translates into 
a large number of grid points. On the other hand, since
the characteristic mode speeds increases linearly with volume, the use of a large
domain also demands a shorter time step in order to ensure a stable
evolution \cite{dgs1}. These two factors combine to require unfeasibly long numerical 
simulations for the states with large spread in volume.
Fortunately, this issue can be resolved by implementing a hybrid spatial grid.
This is the Chimera scheme that was presented in~\cite{dgs1}. 
As shown in that work, the computation time can be reduced significantly. As an example, simulations which would take billions of years using pre-Chimera 
techniques can be completed in a couple of hours. Here we present a brief summary of the method, and for more details refer the
reader to Ref. \cite{dgs1}.

The main idea of the Chimera scheme is to first identify the volume at which the Wheeler-DeWitt
equation is a good approximation to the LQC difference 
equation and then divide the entire domain into two parts: an inner grid where
the LQC difference equation is solved and an outer grid where the 
Wheeler-DeWitt equation is solved. 
On the outer grid we are not restricted by the limitations imposed by LQC on the discretization as the Wheeler-DeWitt equation has a continuum limit. Therefore
we can choose any suitable discretization. In particular, it is convenient to perform a change
of variables $x=\ln(v)$ and discretize in $x$ instead of $v$. A uniform
discretization in $x$ then translates into larger intervals $\Delta v$ for
larger $v$, reducing the number of grid points needed for large
domains. Furthermore, since the characteristic speed is constant in the new
coordinates, larger spatial domains no longer demand shorter time
steps. Finally, we need to choose an adequate numerical scheme for the solution
of the partial differential equations. Of the two implementations presented 
in~\cite{dgs1}, we chose, based on efficiency considerations, the Discontinuous
Galerkin (DG) method with uniformly sized discrete elements in $x$.

After setting up the initial state peaked on an expanding classical trajectory, 
as explained in~\secref{sec:initialdata}, we
evolve the state backward in time ($\phi$). All the simulations show the
existence of a quantum bounce, and the states are evolved through the bounce until the 
evolution trajectory meets a contracting classical solution corresponding to a mean volume similar to that of the initial state. We compute the expectation values of various observables using eq.~\eqref{eq:expect}, with the summation in $v$ converted to an integral on the outer grid and the 
upper limit of the integral set to a finite value $v_{\rm int}$. 
The eigenfunctions of the quantum constraint in LQC are superpositions of incoming and outgoing eigenfunctions of the Wheeler-DeWitt quantum constraint. Thus,
though the initial state is constructed as a purely expanding solution of the
Wheeler-DeWitt equation, it is not a purely expanding solution on the LQC
grid. Thus, a small part of the state will separate and evolve as a contracting mode, that
is, an outgoing mode in the backward evolution. The amplitude of this mode
decreases as we move the interface between the inner and outer grid to larger
values of $v$. In our simulations we choose $v_{\rm int}$ small enough so that
this mode is not included in the calculation of expectation values, but large
enough to accurately obtain expectation values for the ingoing mode.
This is achieved by performing careful convergence studies where both 
$v_{\rm int}$ and the location of the interface between the inner and outer 
grid are varied and convergence of the expectation values are established.

The resources used for the simulations performed in this work varied
greatly from case to case due to the variety of states studied, some with
much larger spread than others. Furthermore, each simulation was repeated using
different resolutions in order to check convergence of the solutions. The number of grid points used in the LQC grid varied form $7500$ to
$60000$, whereas the number of  DG elements in the Wheeler-DeWitt grid 
varied from $40$ to $653$, resulting in a total domain in $v$ that varied from $10^{9}$ to 
$10^{15}$. The computation time in this  case ranged from a few minutes to at most two 
hours when running on a 2.4 GHz Sandybridge workstation with 16 processors.

%%%%%%%%%%%%%%%%%%%%%%%%%%%%%%%%%%%%%%%%%%%%%%%%%%
\section{Results}\label{sec:results}
%%%%%%%%%%%%%%%%%%%%%%%%%%%%%%%%%%%%%%%%%%%%%%%%%%
Numerical simulations of all the non-Gaussian states considered in our analysis demonstrate
the existence of a quantum bounce. When the spacetime curvature is 
very small, the quantum trajectory (defined by the expectation value of the volume and the internal time) 
agrees quite well with a classical trajectory. This agreement persists as long 
as the energy density $\rho$ is very small compared to the density at the bounce $\rho_{\rm b}$. The 
departures between the LQC and the classical trajectory (or the Wheeler-DeWitt trajectory) becomes significant once the energy density becomes greater than about one percent of the bounce density. As the 
backward evolution is continued, due to the quantum geometric effects, the evolution in 
LQC is  non-singular, whereas the classical trajectory encounters a big bang singularity. The 
occurrence of a quantum bounce  for all the states 
considered in this paper provides a strong numerical evidence in favor of 
the generic occurrence of the quantum bounce and is in agreement with the 
predictions of the exactly solvable model~\cite{acs}, 

We present the representative  results of over 100 simulations performed by studying 
the numerical evolution of the three different classes of non-Gaussian states described 
in \secref{sec:initialdata}.We have considered a wide variety of initial parameters for 
squeezed states with various values of $n$ (where $n$ is the exponent of the wavenumber 
$k$ in \eref{eq:sqpsi}) and other non-Gaussian states with various $\omega^*=-\sqrt{12\pi G}\, k^*=p_\phi^*/\hbar$. 
A summary of some of the representative simulations 
performed for each class of initial states is presented in Table \ref{tab:summary}. In this table, 
we show the expectation value of the field momentum ($\langle \widehat{p}_\phi\rangle$), 
the time of bounce ($\phi_{\rm b}$), the bounce volume in LQC ($V_{\rm b}$) and the 
bounce volume in the effective theory ($V_{\rm b}^{\rm (eff)}$) for several values of $\omega$
and the complex valued $\eta$ (with the real part denoted as $\eta_r$ and the imaginary part denoted as $\eta_i$) for the three types of 
non-Gaussian states discussed in the previous section.\footnote{In this section, results are discussed in terms of the physical volume of the fiducial cell which is related to $v$ 
in the quantum difference equation via eq.(\ref{eq:Vv})).} 
It should be noted that the values of $\omega^*$ and $\langle \widehat{p}_\phi\rangle$ are 
the same for the squeezed states with $n=0$, multipeaked-1 and multipeaked-2 states, while the 
value of $\langle \widehat{p}_\phi\rangle$ is greater than that of $\omega^*$ for $n=50$. This can be easily 
understood by noting that it is only the latter kind of states which are 
asymmetric around $k^*$ in $k$ space. Numerical simulations show that $\langle \widehat{p}_\phi\rangle$ is a constant of 
motion 
in all cases. An important feature of all the simulations is that the bounce volume in the 
effective theory is smaller than in the LQC evolution. A similar feature was also 
observed in the  recent numerical study in Ref.~\cite{dgs2} of Gaussian states. 

\begin{table}
\caption{Summary of the representative simulations shown 
              in various figures in this paper. All the values are in  Planck units.}
\begin{tabular}{cccccccc}
\hline
 $\omega^*$ & $\eta_{\rm r}$  &$\eta_{\rm i}$ & $\langle \widehat{p}_\phi\rangle$  &$\phi_b$  & $V_{\rm b}$ & 
$V_{\rm b}^{\rm (eff)}$ & Fig. \\
\hline
\hline
\textbf{Squeezed, $n=0$}    & \text{}       & \text{}       &\text{}   & \text{}     & \text{}  & \text{} & \\
 1000     & 0.0001          & 0.0001        & 1000          &-0.7830   & 1109.3      & 1105.2    & \ref{f:eta1000b}\\
 1000     & 0.0001          & 0.001         & 1000          &-0.7785   & 1336.3      & 1105.2    & \ref{f:eta1000}\\
 500      & 0.0001          & 0.001         & 500           &-0.8865   & 667.22      & 552.58    & -- \\

%%  500      & 0.01            & 0.01          & 500                      &-0.8954   & 805.61        & 
%% 552.83    & \ref{f:eta500}\\
1000      & 0.01            & 0.01          & 1000          &-0.7831   & 1611.2     & 1105.2    & \ref{f:eta1000c}\\

 60       & 0.01            & 0.01          & 60            &-1.233    & 96.729      & 66.310    & \ref{f:eta60}\\
 \hline
 \textbf{Squeezed, $n=50$}  & \text{}       & \text{}       &\text{}   & \text{}    & \text{} & \text{}\\
 1000     & 0.0003          & 0.0003        & 1077.4       &-0.7366   & 1204.3    & 1190.7    & \ref{f:kn1000}\\
 500      & 0.0003          & 0.0003        & 632.07        &-0.8164   & 706.63     & 698.54  & -- \\
 50       & 0.03            & 0.03          & 63.219         &-2.611    & 221.22     & 69.867  & \ref{f:kn50}\\
 \hline
 \textbf{Multipeaked-1}  & \text{}       & \text{}          &\text{}   & \text{}     & \text{}& \text{}\\
 1000     & 0.0002          & 0             & 1000          &-0.7835   & 1108.7    & 1105.2    & \ref{f:avg1000}\\
% 500      & 0.0002          & 0             & 500           &-0.8964   & 554.38     & 552.58    & -- \\
 50       & 0.02            & 0             & 50            &-1.271    & 78.231     & 55.258    & \ref{f:avg50}\\
 \hline
 \textbf{Multipeaked-2}       & \text{}       & \text{}                  &\text{}   & \text{}      & \text{} & \text{}\\
 1000     & 0.0001          & 0             & 1000          &-0.7835   & 1121.1    & 1105.2    & \ref{f:multi1000}\\
 200      & 0.0025          & 0             & 200           &-1.796    & 318.85     & 221.03    & \ref{f:multi200}\\
   \hline
   \label{tab:summary}
\end{tabular}
\end{table}

In the following, we will study the way the profile of a given 
initial wavefunction evolves across the quantum bounce, compare the trajectory 
given by the expectation value of the volume observable in LQC with the 
corresponding effective trajectory obtained from the effective Hamiltonian and investigate the properties of the dispersion in 
the matter and volume observables for various choices of parameters and types of initial 
state. The properties of the dispersion are directly 
related to the validity of the triangle inequalities which strongly constrain  the growth of the relative fluctuations of volume and 
momentum observables across the bounce. 
We analyze our numerical results in the light of two different 
formulations of the triangle inequalities, as described in \secref{sec:triangle}. 
We also study the way the energy density behaves 
during the evolution and compare our numerical results with the analytical results 
obtained in the Refs.\ \cite{montoya_corichi1,montoya_corichi2} for the squeezed 
states. Note that in the discussion of results from our analysis, data points in all the figures are plotted in Planck units and
we have chosen the notation $V=\langle \widehat{V}|_\phi \rangle$ and 
$\Delta V=\langle \Delta \widehat{V}|_\phi\rangle$, and will continue to do 
so in the rest of the paper when discussing the evolution trajectories of the
expectation value of the volume observable.

%%%%%%%%%%%%%%%%%%%%%%%%%%%%%%%%%%%%%%%%%%%%%%%%%%
\subsection{Squeezed States} \label{sec:squeezed}
%%%%%%%%%%%%%%%%%%%%%%%%%%%%%%%%%%%%%%%%%%%%%%%%%%
The squeezed states considered in this paper are obtained by evaluating the 
integral in \eref{eq:sqstate} with $\Psi(k)$ given by \eref{eq:sqpsi}. The squeezedness of the wavefunction is controlled by the real and 
the imaginary parts of the complex parameter $\eta$: $\eta_{\rm r}$ and $\eta_{\rm i}$. In the case of Guassian states, the 
parameter $\eta$ is real, hence the %relative 
dispersion of the state can be varied relative to that of a Gaussian state for the same 
$\eta_{\rm r}$ by varying $\eta_{\rm i}$. For example, for a given value of 
$\eta_{\rm r}$ for which the corresponding Gaussian states is sharply peaked, the 
squeezed state can be made widely spread by varying $\eta_{\rm i}$. 
In the following we present a detailed study of the numerical evolution of 
squeezed states for various initial parameters $\omega^*$ and 
$\eta$. In our numerical analysis, several values of $n$ were chosen and the results are qualitatively similar. Here 
we discuss only two representative cases: $n=0$ and $n=50$. In the following subsections, we first highlight the qualitative 
features of the evolution of the squeezed state wavepacket, which is followed by the analysis of the relative fluctuations in the volume 
observable $\hat V|_\phi$ using the triangle inequalities (eqs.(\ref{eq:triangle}) and (\ref{eq:corichieps})). The key results of this analysis are: (i) the confirmation of the validity of 
the triangle inequality (\ref{eq:triangle}) derived  in Ref.\ \cite{kp} for states with arbitrary parameters and (ii) finding the 
range of validity of a stronger version of inequalities (\ref{eq:corichieps}) derived in Ref. \cite{montoya_corichi1} with assumptions on the semi-classicality of the initial state. 
We also study the behavior of the
energy density at the bounce as the imaginary part of $\eta$ is varied while keeping the 
real part fixed and compare our results with the analytical results obtained in Ref. \cite{montoya_corichi1} using sLQC \cite{acs}. After 
that, we  illustrate the dynamical trajectories for 
various  parameters by studying the behavior of the expectation value of the 
physical observables with respect to the internal time and compare them with the 
corresponding effective trajectories obtained from the effective Hamiltonian in Ref. \cite{vt}. Since the derivation of the 
effective Hamiltonian assumes Gaussian states we find significant departures between the quantum evolution and the effective 
trajectory (Sec.~\ref{sec:effective}). 

%%%
\subsubsection{Evolution}
%%%

At a qualitative level, the evolution of all states constructed using (\ref{eq:sqpsi}) is very similar. They all undergo a quantum bounce. However, depending on the values 
of $\eta_r$ and $\eta_i$,  the evolution of the initial states demonstrate rich distinct features. 
In particular, the profile of the state and dispersions can vary significantly during evolution for different initial states. Larger values of field momentum leads to  larger 
bounce volumes and lower bounce densities. 
To demonstrate the way the choice of squeezedness parameters effect the properties of the state 
during quantum evolution 
we discuss three representative cases:
(i) a sharply peaked initial state with 
$n=0$, $\omega^*=1000\,\sqrt{G}$, $\eta=(1+i)\times 10^{-4}$, 
(ii) a very widely spread initial state with 
$n=0$, $\omega^*=1000\,\sqrt{G}$, $\eta=(1+10i)\times 10^{-4}$, and  
(iii) a sharply peaked initial state with 
$n=50$,  $\omega^*=1000\,\sqrt{G}$, $\eta=(1+i)\times 10^{-4}$. In the latter case, $\langle \widehat{p}_\phi \rangle \neq \omega^* \hbar$ unlike the first two cases. 
These three cases are also summarized in Table \ref{tab:summary}.

\begin{figure}[tbh!]
  \subfigure[]{
    \includegraphics[angle=0,width=0.62\textwidth,height=!,clip]{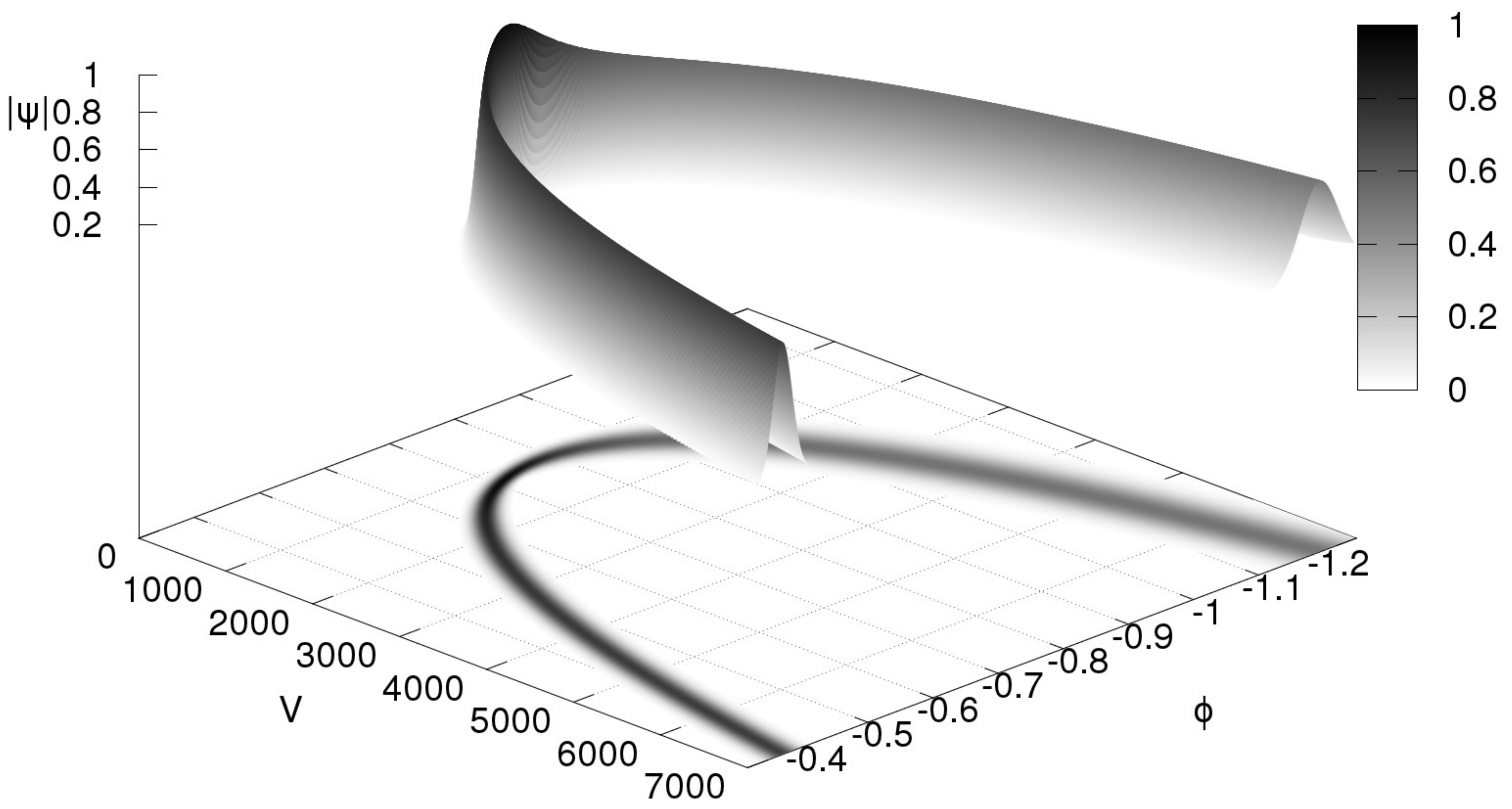}
    \label{f:squeezed_3Da}
  }
  \subfigure[]{
    \includegraphics[angle=0,width=0.35\textwidth,height=!,clip]{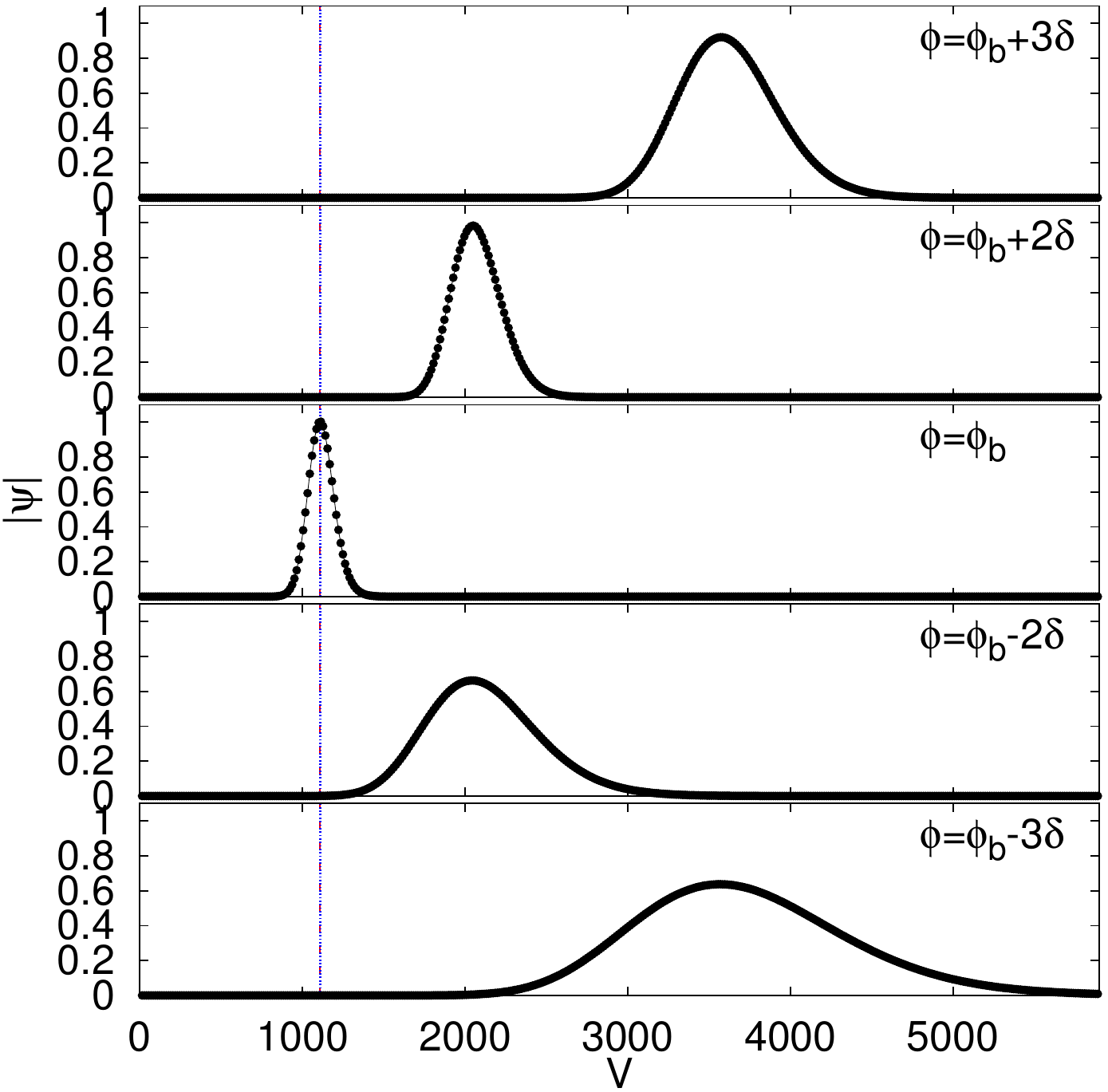}
    \label{f:squeezed_3Db}
  }
  \subfigure[]{
    \includegraphics[angle=0,width=0.47\textwidth,height=!,clip]{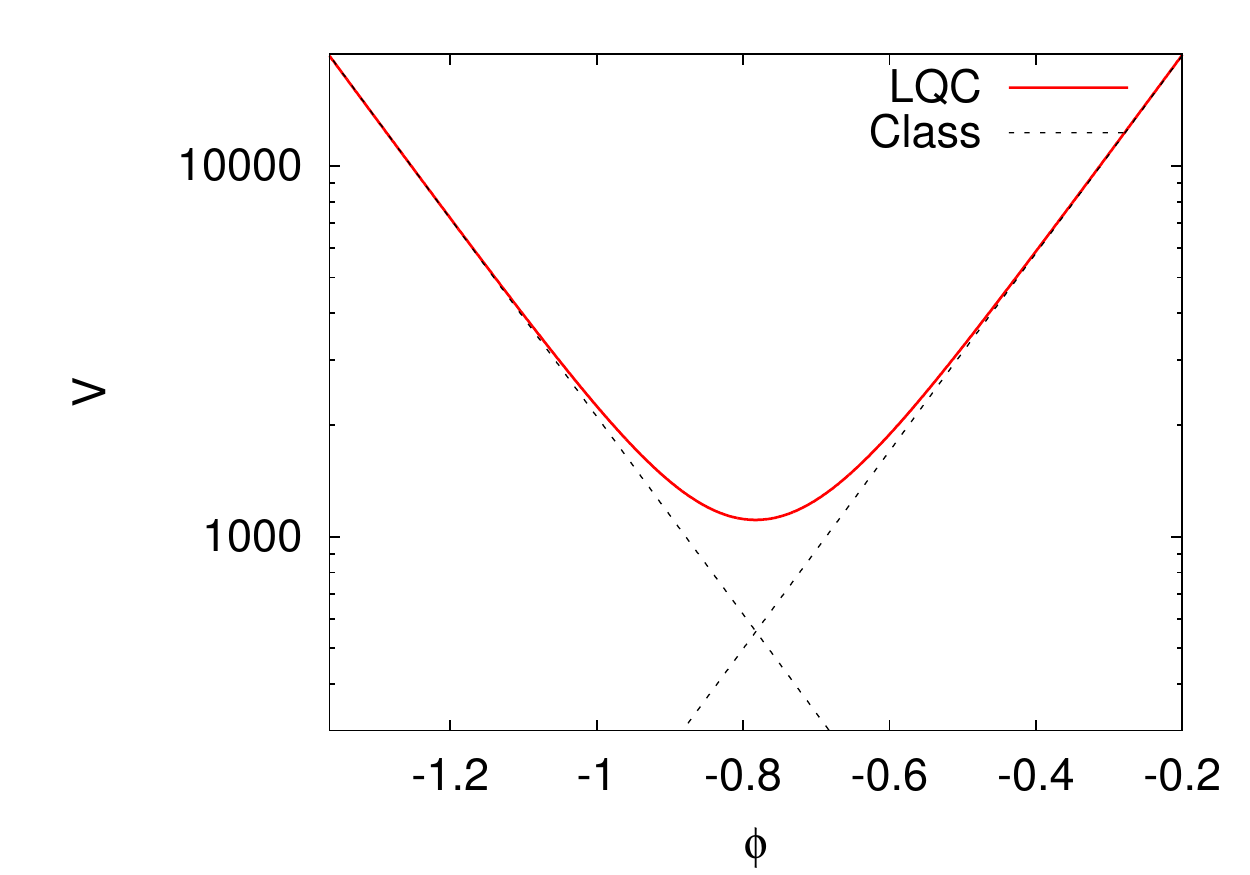}
    \label{f:squeezed_3Dc}
  }
  \subfigure[]{
    \includegraphics[angle=0,width=0.47\textwidth,height=!,clip]{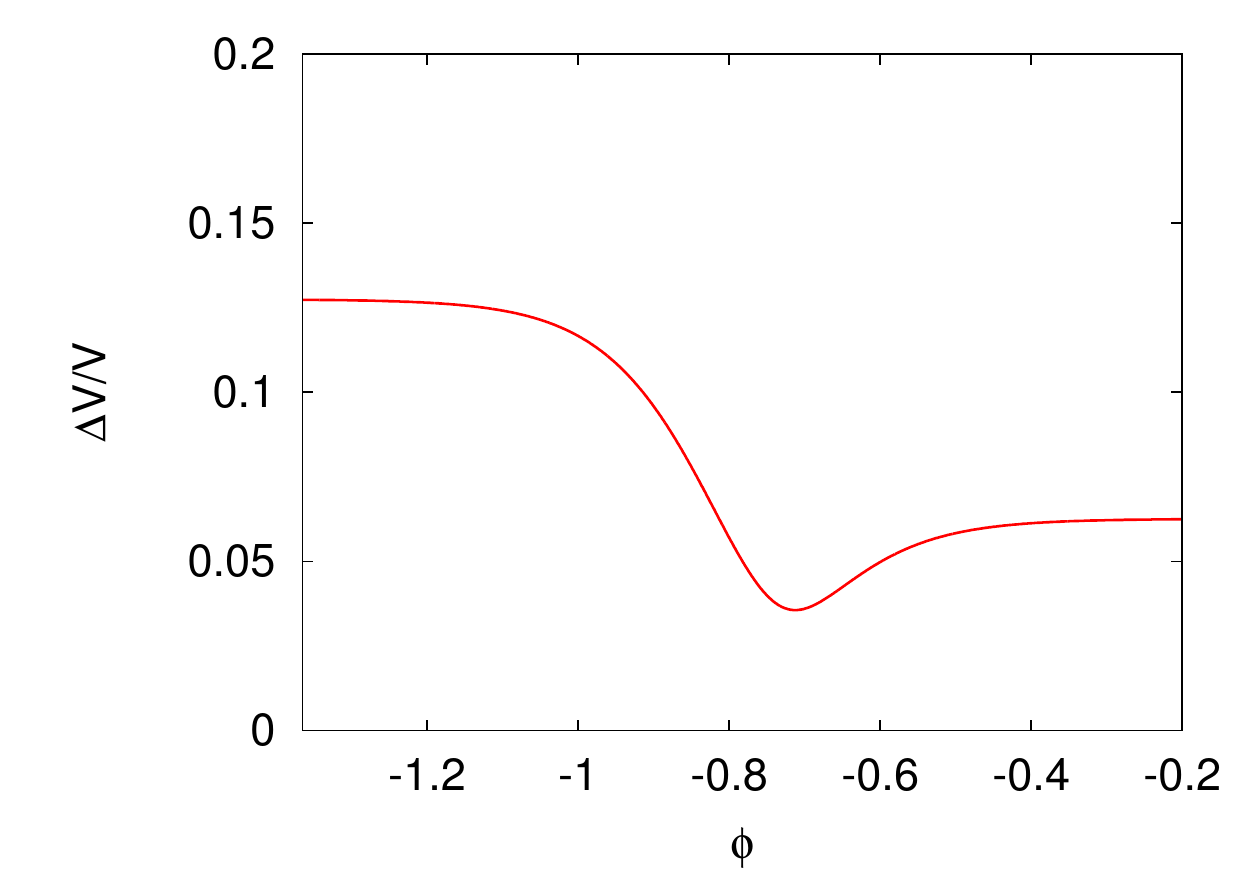}
    \label{f:squeezed_3Dd}
  }
  \caption{Evolution of a squeezed state with $n=0$, $\eta=(1+i)\times
    10^{-4}$ and $\omega^*=1000\,\sqrt{G}$. Panel~\subref{f:squeezed_3Da} shows  
    the amplitude of the wavefunction, $|\Psi|$, as a function of the volume $V$ and the  
    internal time $\phi$, including a projection onto the $V-\phi$ plane to
    help visualizing the 3D graph. Panel~\subref{f:squeezed_3Db} shows
    $|\Psi|$ at different values of $\phi$ around the bounce time,
    $\phi_{\rm b}=-0.7830$,
    where $\delta=0.1$. Note that the wavefunction has support only on the discrete lattice, but for a better visualization we use continuous curves for the snapshots of $|\Psi|$ in this and the following figures. 
    The blue dotted line indicates the bounce volume, $V_{\rm b}=1109.3\,\Vpl$,
    and the red dashed one indicates the cutoff volume, $V_{\rm c}=1105.7\,\Vpl$ estimated from sLQC. In this case
    the values are so close that one cannot easily distinguish the two vertical curves in
    the figure.
Panel~\subref{f:squeezed_3Dc} presents a comparison
    between the LQC and classical GR
    trajectories. Panel~\subref{f:squeezed_3Dd} shows the relative dispersion
    $\Delta V/V$.
All values are given in Planck units, and
 we have chosen the notation $V=\langle \widehat{V}|_\phi \rangle$ and $\Delta V=\langle \Delta \widehat{V}|_\phi\rangle$ in the trajectory figures.
    The same convention is used in all the figures in this paper.
  }
  \label{f:squeezed_3D}
\end{figure}

Results from the quantum evolution of case (i) are presented in \fref{f:squeezed_3D}.  
The evolution of $|\Psi|$ with respect to volume and time $\phi$ is shown in \fref{f:squeezed_3Da}. \fref{f:squeezed_3Db} shows snapshots of $|\Psi|$ at 
different times around the bounce time $\phi_b$. Note that, in all 3D figures
and figures showing snapshots of the state, we do not include the whole
computational domain, but rather only show the region relevant for the bounce.
It is evident from these figures that the wavefunction is 
peaked on a non-zero finite value of the spatial volume during the entire evolution. This feature is more 
apparent in the projection of the amplitude of the wavefunction in the $V-\phi$ plane in \fref{f:squeezed_3Da}. From the plot corresponding to $\phi = \phi_b$ in 
\fref{f:squeezed_3Db}, we also find that the amplitude of the wavefunction decays rapidly for $V<V_b$ (shown by the dotted blue vertical curve).
This decay is almost exponential, in agreement with the earlier 
results on the decay of the eigenfunctions in this model \cite{aps2,aps3} and in 
sLQC \cite{Craig:2012gw}. In the same plot we also show, via a dashed red
vertical line, 
the estimation of the cutoff volume $V_{\rm c}$ obtained by replacing
$\omega$ with the peak value $\omega^*$ (which in this case equals  $\langle \widehat{p}_\phi \rangle$) in \eref{eq:vcut}. 
Note that for this initial state $V_{\rm c}$ is almost identical to the bounce volume, since the state is sharply peaked at the bounce. The expectation 
value of the volume are plotted against $\phi$ in \fref{f:squeezed_3Dc} and a comparison is made with the classical trajectories. 
As expected from the evolution of the 
wavefunction, the expectation value of the volume always remains non-zero. Since the initial data is provided when the spatial curvature is very small, the LQC 
trajectory coincides with the classical theory in the region far from the
bounce. Under the backward evolution, the wavepacket travels inwards 
and the expectation value of the volume decreases. This in turn leads to an increment in 
the spacetime curvature. Quantum geometric effects 
become prominent when the spacetime curvature approaches Planck scale, which in turn leads to a notable difference between the LQC and classical 
trajectories. The LQC trajectory undergoes a non-singular bounce while the  
classical trajectory corresponding to the expanding solution encounters a big-bang singularity in the past evolution. In the subsequent evolution, 
the expectation values of the volume observable in LQC 
turn out to be in excellent agreement with the contracting classical solution (for the same value of field momentum) when spacetime curvature becomes much smaller than the 
Planck scale. The disjoint classical trajectories hence get connected by a smooth non-singular quantum gravitational bridge in LQC. 
Although the shape of the 
wavefunction before and after the bounce is similar, the spread of the state,
shown in \fref{f:squeezed_3Dd}, 
is not symmetric. The asymmetry
is also apparent in~\fref{f:squeezed_3Db}. This feature is in contrast 
with the behavior of the spread of a Gaussian counterpart ($\eta_{\rm i}=0$) of the squeezed state \cite{dgs2}.

\begin{figure}[tbh!]
  \subfigure[]{
    \includegraphics[angle=0,width=0.62\textwidth,height=!,clip]{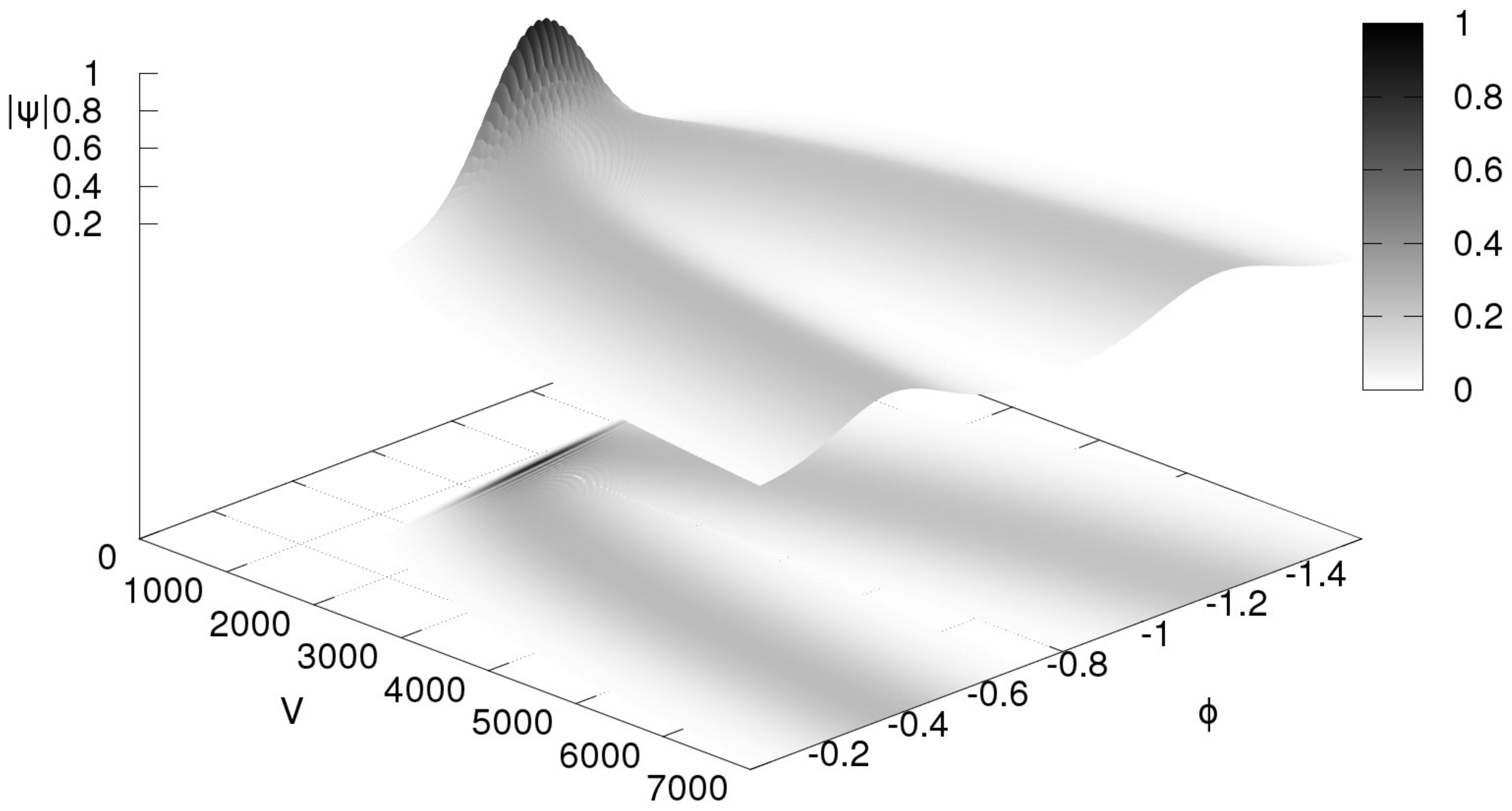}
    \label{f:squeezed2_3Da}
  }
  \subfigure[]{
    \includegraphics[angle=0,width=0.35\textwidth,height=!,clip]{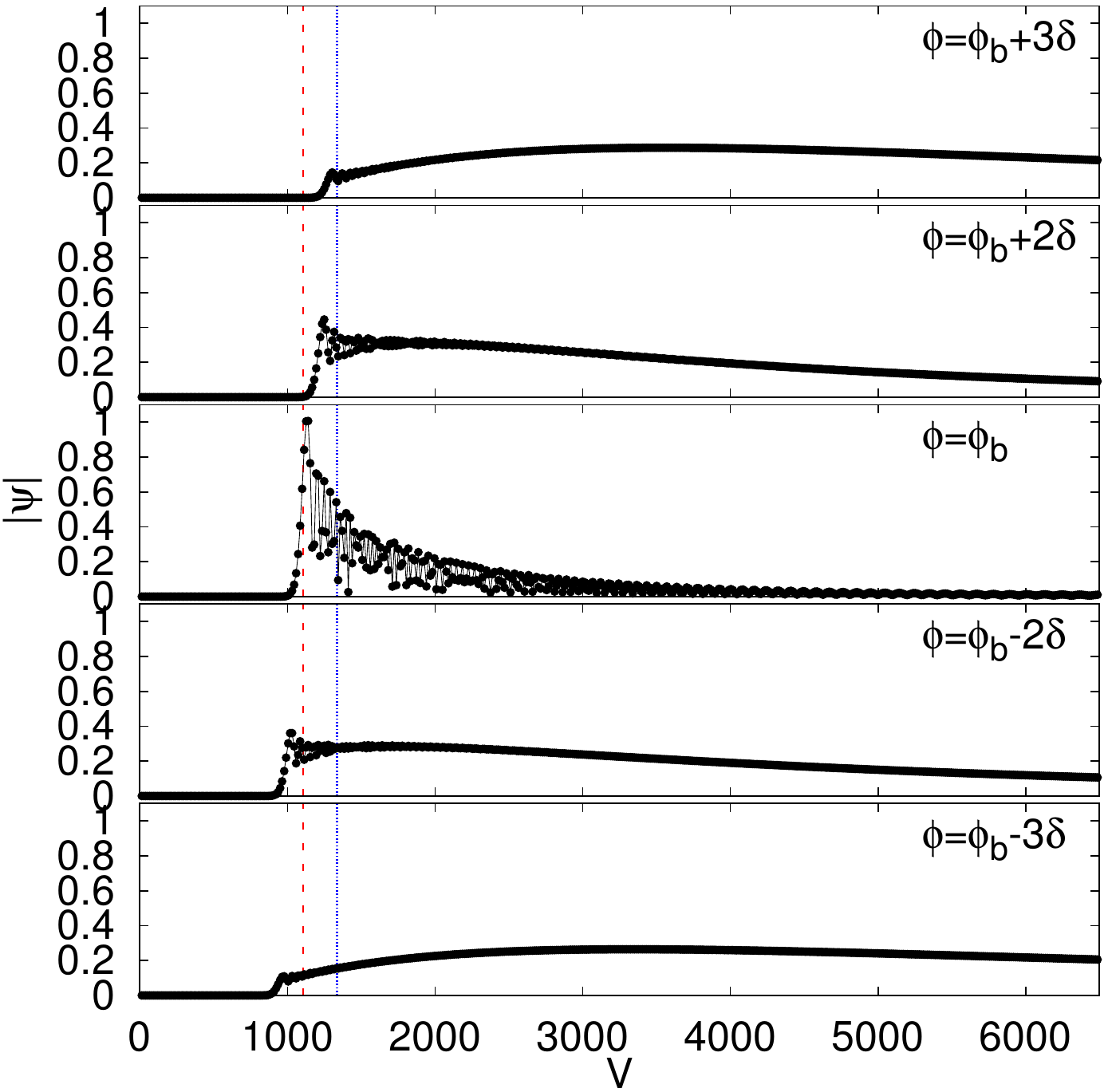}
    \label{f:squeezed2_3Db}
  }
  \caption{Evolution of a squeezed state with $n=0$, 
    $\eta=(1+10i)\times 10^{-4}$ and $\omega^*=1000\,\sqrt{G}$. 
    Panel~\subref{f:squeezed2_3Da}: $|\Psi|$, including the
    projection onto the $V$-$\phi$ plane to help
    visualize the 3D plot. Panel~\subref{f:squeezed2_3Db}: $|\Psi|$ at different values of $\phi$ close to
    the bounce, as indicated in the figure, where $\phi_{\rm b}=-0.7785$ and
    $\delta=0.1$. The blue dotted line indicates the bounce volume, $V_{\rm b}=1336.3\,\Vpl$,
    and the red dashed one indicates the cutoff volume, $V_{\rm c}=1105.7\,\Vpl$. 
  }
  \label{f:squeezed2}
\end{figure}

Case (ii), a very widely spread squeezed state with parameters
$\omega^*=1000\,\sqrt{G}$, $n=0$ and
$\eta=(1+10i)\times 10^{-4}$ is shown in~\fref{f:squeezed2}.
It is noteworthy that the shape of the state in \fref{f:squeezed2_3Da} is
quite different from that in \fref{f:squeezed_3Da},  
even far from the bounce. Moreover, in the vicinity of the bounce  the shape
of the wavefunction has prominent non-Gaussian features (see also
\fref{f:squeezed2_3Db}). It is noticeable, however, 
 that despite the non-trivial features close to the bounce, the original shape of 
the wavefunction is recovered  on the other side of the bounce. That is, the 
profile of the wavefunction on both  sides of the bounce (at early and late times) 
is similar.
Such non-Gaussian
features are also seen in the simulation of very widely  
spread Gaussian states, as shown in Fig.~10 of Ref.\ \cite{dgs1}. It is worth 
mentioning here that the non-Gaussian features observed in the case of widely 
spread states is a phenomena which was not seen in the early numerical 
simulations of \cite{aps2,aps3} as they were limited to sharply peaked states only. Also, these features 
are {\it not} artifacts of numerical techniques as confirmed by several robustness
tests of these simulations. 
It is apparent from these figures that, despite  some small differences in the evolution in the high curvature regime, squeezed states also undergo a smooth quantum bounce.
Unlike case (i), since this squeezed state has more significant quantum features at the bounce time, the bounce volume ($V_b$) computed from the 
expectation value of volume observable and the cutoff volume ($V_c$) are significantly different.  For this reason, the almost exponential decay of $|\Psi|$ does not occur below $V_b$, but occurs  below $V_c$ as is evident from the plot for $\phi = \phi_b$. 

\begin{figure}[tbh!]
  \subfigure[]{
    \includegraphics[angle=0,width=0.62\textwidth,height=!,clip]{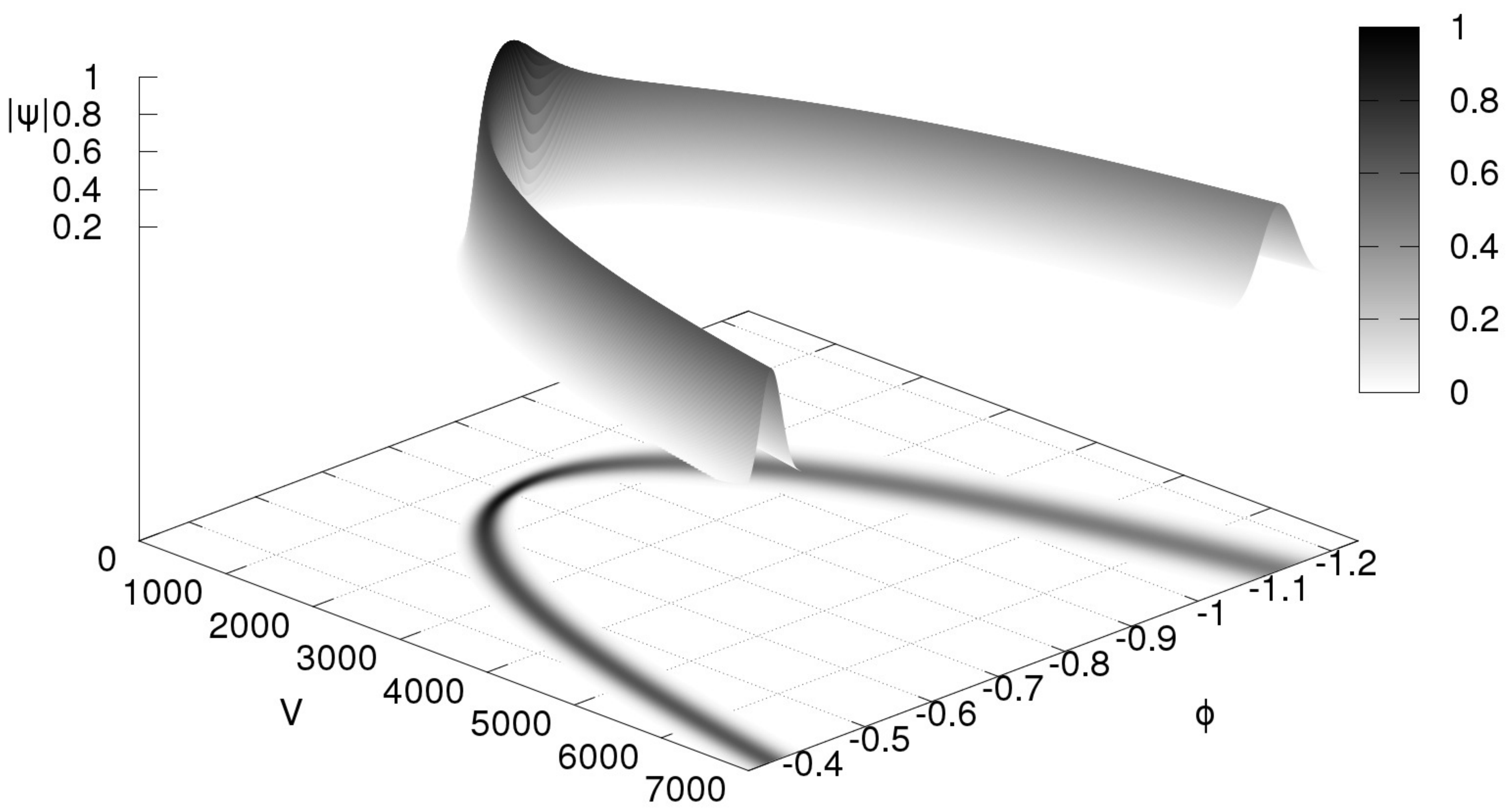}
    \label{f:kn_squeezed_3Da}
  }
  \subfigure[]{
    \includegraphics[angle=0,width=0.35\textwidth,height=!,clip]{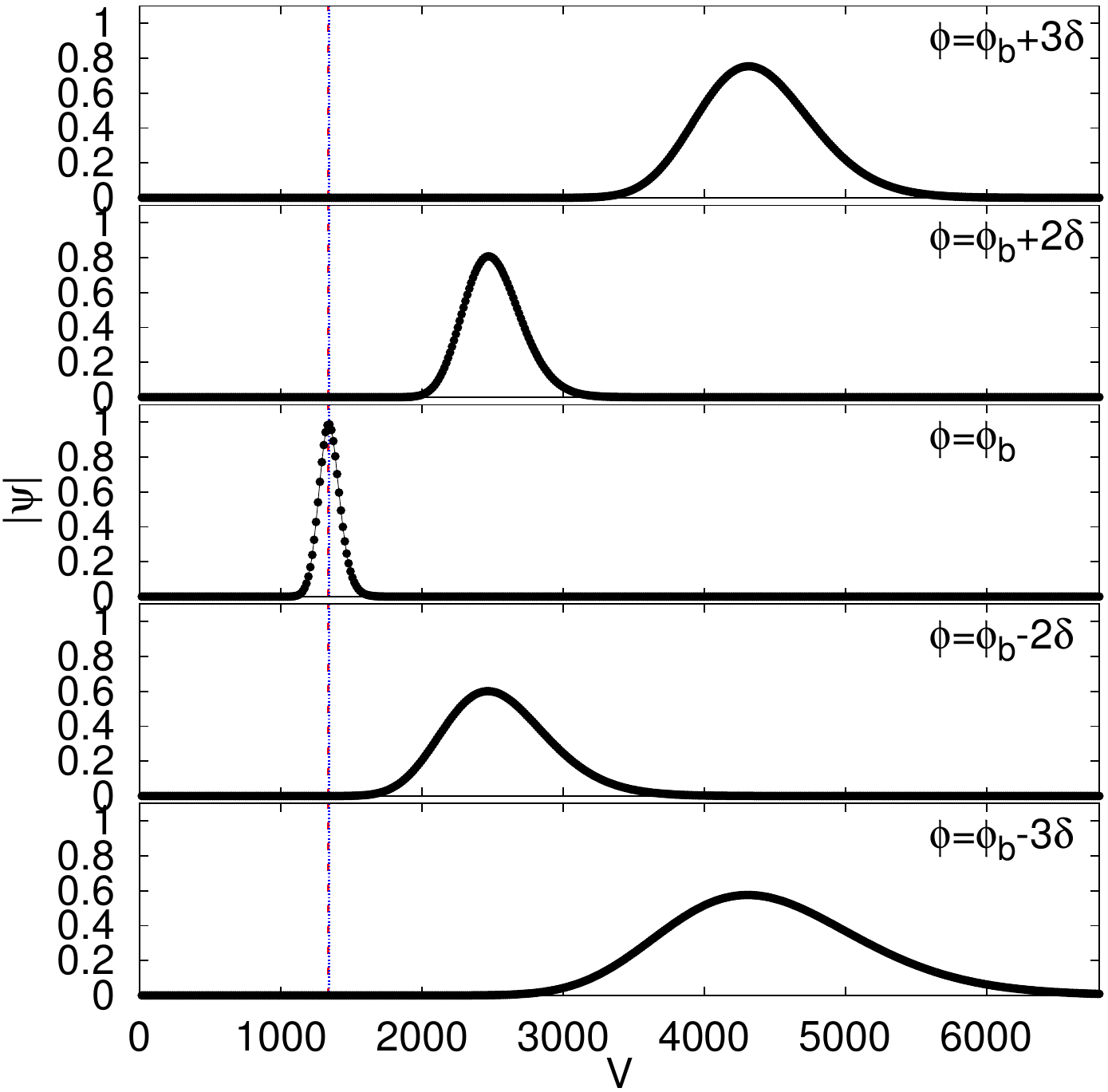}
    \label{f:kn_squeezed_3Db}
  }
  \caption{
    Evolution of a squeezed state with $n=50$, $\eta=(1+i)\times
    10^{-4}$ and $\omega^*=1000\,\sqrt{G}$. The expectation value of the field 
    momentum is $\langle \widehat p_\phi\rangle=1207\,\sqrt{G}\hbar$.
    Panel~\subref{f:kn_squeezed_3Da}: $|\Psi|$, including the projection onto
    the $V$-$\phi$ plane.
    Panel~\subref{f:kn_squeezed_3Db}: $|\Psi|$ at different values of $\phi$ close to
    the bounce, as indicated in each sub-figure, where $\phi_{\rm b}=-0.7417$ and
    $\delta=0.1$. 
    The blue dotted line indicates the bounce volume, $V_{\rm b}=1339.4\,\Vpl$,
    and the red dashed curve indicates the cutoff volume, $V_{\rm c}=1334.6\,\Vpl$. Note that
    the values are so close that the two lines are indistinguishable from
    each other.
  }
\end{figure}

Finally, figures~\ref{f:kn_squeezed_3Da} and~\ref{f:kn_squeezed_3Db} show the 
evolution of the squeezed state with parameters $\omega^*=1000\,\sqrt{G}$, $n=50$ and 
$\eta=(1+i)\times 10^{-4}$ (case (iii)). 
As in the cases with $n=0$, the state is always peaked on a non-zero finite 
volume, while undergoing a non-singular bounce. Further, as expected, the expectation value of the field momentum remains constant throughout the 
evolution. 
%at $\phi\approx-0.783$. 
Note that, even though cases (i) and (iii) correspond to the same value of
$\omega^*=1000\,\sqrt{G}$, the expectation value of the field momentum for $n=0$ is 
$\langle \widehat p_\phi \rangle=1000\,\sqrt{G}\hbar$, while for $n=50$ it is $\langle \widehat p_\phi \rangle =1207\,\sqrt{G}\hbar$. If one naively computes the cutoff volume for this state using 
$\omega=\omega^*$, one underestimates the volume below which the almost exponential decay of the amplitude of wavefunction occurs to be $V_c \approx 1106\, \Vpl$. Instead, 
if we estimate the cutoff volume $V_c$ by using $\omega= \langle \widehat{p}_\phi \rangle/\hbar$ in \eref{eq:vcut}, we find that this provides the correct volume below which $|\Psi|$
decays almost exponentially. Similarly to case (i), the bounce volume and the cutoff volume are approximately the same in this case. 

%%%
\subsubsection{Dispersions and triangle inequalities}
%%%
Let us now discuss the behavior of the dispersion of the state as a function of the internal time. 
The relation between  the dispersions on the two sides of the bounce in the 
presence of a massless scalar field has been analytically understood in  Ref.\ 
\cite{kp}, where it was shown that the relative fluctuations in the field momentum 
and the volume observable follow triangle inequalities which constrain the growth of 
the dispersion of the state during the evolution. As discussed in \secref{sec:triangle}, 
a stronger form of the triangle inequality (\ref{eq:corichieps}) can be derived for states which
satisfy
$\langle\Delta {\ln(\widehat{\mathcal O})}\rangle \approx \langle\Delta \mathcal{\hat O}\rangle/\langle\mathcal {\hat O}\rangle$. Here, we will study
how the relative dispersions in the volume observable and in the momentum of the matter field change
across the bounce and the validity of triangle inequalities (\ref{eq:triangle}) and (\ref{eq:corichieps}). 

\begin{figure}[tbh!]
  \includegraphics[angle=0,width=0.6\textwidth,height=!,clip]{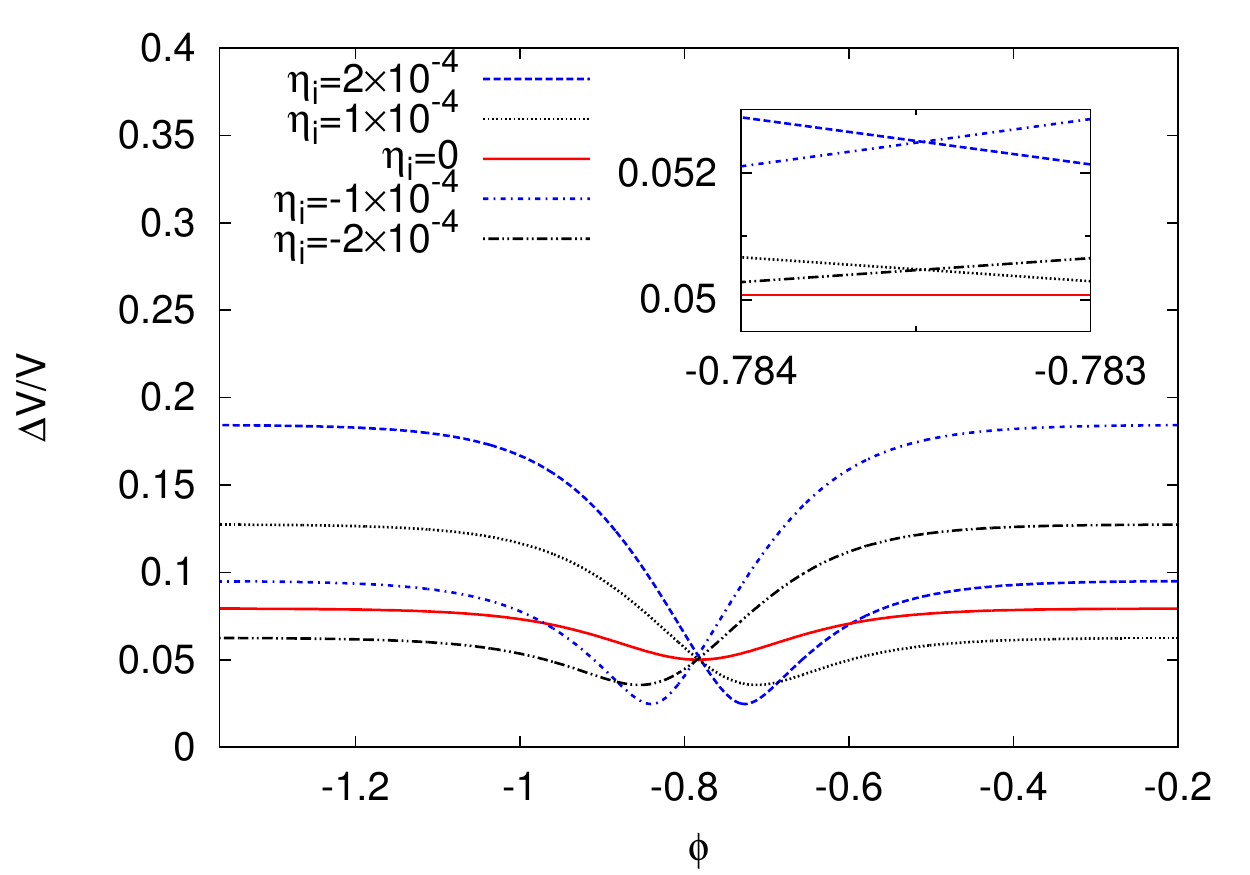}
  \caption{Relative volume dispersion for various values of $\eta_{\rm i}$ 
    and fixed $\eta_{\rm r}=1\times10^{-4}$ are shown. The expectation
    value of the field momentum for all curves is  
    $\langle \widehat p_\phi\rangle=1000 \sqrt{G} \hbar$  and 
    $\omega^*=1000 \sqrt{G}$. It is evident that all 
    curves intersect each other at the same value of $\phi$, 
    and the curves corresponding to $\eta_{\rm i}$ are mirror 
    images of those with $-\eta_{\rm i}$.}
  \label{fig:dvv}
\end{figure}

As an example, let us consider different squeezed states with $n=0$, 
$\eta_{\rm r}=1\times10^{-4}$, $\omega^*=1000\,\sqrt{G}$.
The evolution of $\Delta V/V$ 
for such states are shown in \fref{fig:dvv} for various values of
$\eta_{\rm i}$. Compared to their Gaussian counterparts, 
%that is, a state with the same $\eta_{\rm r}$, but $\eta_{\rm i}=0$, 
squeezed states show some similarities and some important differences. The solid 
(red) curve in~\fref{fig:dvv} shows 
$\Delta V/V$ for $\eta_{\rm i}=0$, that is a pure Gaussian, for 
which the relative volume dispersion takes the same value on either side of the 
bounce. On the other hand, it is evident from the figure that for 
$\eta_{\rm i}\neq 0$, the asymptotic values of $\Delta V/V$ on the two sides of the 
bounce are different. Moreover, it shows a mirror symmetry with respect to the bounce point for opposite signs of $\eta_{\rm i}$ for a fixed $\eta_{\rm r}$. That is, the 
plot of $\Delta V/V$ for $\eta_{\rm i}$ is a mirror image of the one for $-\eta_{\rm i}$, 
with the center of the mirror being the bounce point. As mentioned in \secref{sec:initialdata},  
the initial data is constructed with a phase factor $e^{-i \alpha}$ 
(which corresponds to method-3 of Refs.\ \cite{aps2,dgs2}). Due to this phase factor, the relative 
dispersion in volume of the corresponding Gaussian state has a symmetric
behavior, and the eigenfunctions of the evolution operator in Wheeler-DeWitt theory match those in LQC. 
If the phase factor is not included in the initial data, the above mirror symmetry will not 
be present. We also find that all pairs of mirror symmetric curves intersect at the bounce. %This is a 

It was shown analytically in Ref.~\cite{kp} that,
irrespective of the particular  state, the dispersions $\Delta \ln(V)$ 
and $\Delta \ln{p_\phi}$ for a flat FRW model in the presence of a massless
scalar field obey the inequality 
%which looks very much like a triangle inequality, 
given in \eref{eq:triangle}. For the sharply peaked Gaussian
states, the dispersion in an observable ${\ln(\widehat{\mathcal O})}$ can be 
approximated by $\langle\Delta {\ln(\widehat{\mathcal O})}\rangle \approx \langle\Delta \widehat{\mathcal{O}}\rangle/\langle\widehat{\mathcal O}\rangle$, 
which for the simplicity of the notation is denoted by $\Delta \mathcal O/\mathcal O$ in the rest of the paper. 
Based on this approximation the triangle inequalities can be expressed in terms of the 
quantity $\mathcal E$ as described in \eref{eq:corichieps}, where $\mathcal E$ remains smaller 
than unity in the regime where the above approximation is valid. Whether or not this inequality holds is determined by the 
squeezdness of the state, as is shown below.

\fref{fig:trianglea} shows the difference between the 
asymptotic values of the dispersions $\sigma_\pm$ (computed from (\ref{eq:triangle})) on the two sides of the bounce 
compared with $2\sigma$ for
$\omega^*=1000\,\sqrt{G}$ and two values of $\eta_{\rm r}$ with varying $\eta_{\rm i}$. 
The red (light) curve in the figure corresponds to $\eta_{\rm r}=1\times10^{-4}$
and the black (dark) one to $\eta_{\rm r}=5\times10^{-5}$. 
This figure shows clearly that 
the difference between the asymptotic values of the dispersions $\sigma_\pm$ 
on the two sides of the bounce is always less than 
$2\sigma$. That is, the triangle inequality in \eref{eq:triangle} 
remains valid for all the states considered. \fref{fig:triangleb}, on the other hand, 
shows the variation of $\mathcal E$ with varying $\eta_{\rm i}$. It is apparent 
that $\mathcal E$ is smaller than unity for small $|\eta_{\rm i}|$. As $|\eta_{\rm i}|$ 
increases, $\mathcal E$ also increases. As a result $\mathcal E$ becomes greater 
than unity for some value of $|\eta_{\rm i}|$ (dependent on $\eta_{\rm r}$), hence violating the inequality in 
\eref{eq:corichieps} for large $\eta_{\rm i}$. 
\begin{figure}[tbh!]
  \subfigure[]{
    \includegraphics[angle=0,width=0.45\textwidth,height=!,clip]{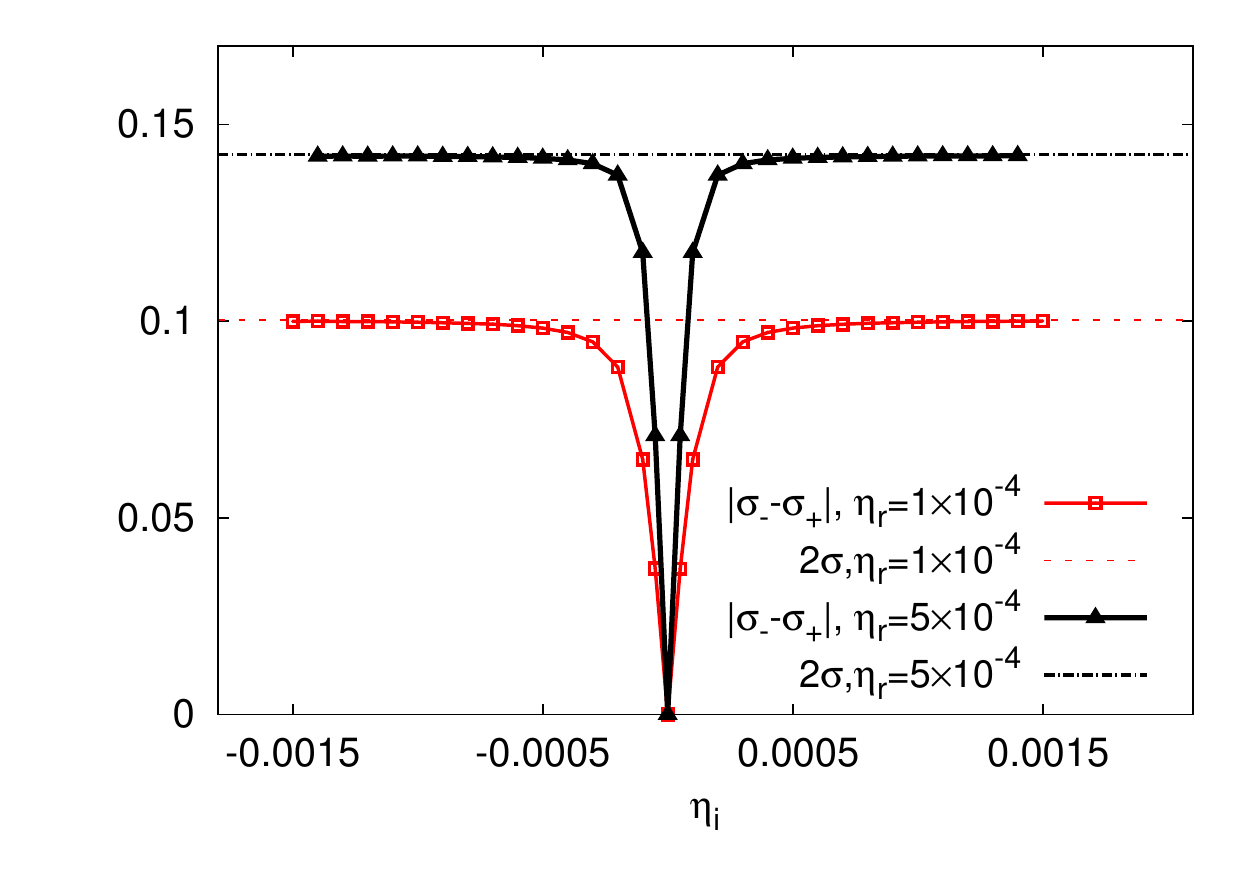}
    \label{fig:trianglea}
  }
  \subfigure[]{
    \includegraphics[angle=0,width=0.45\textwidth,height=!,clip]{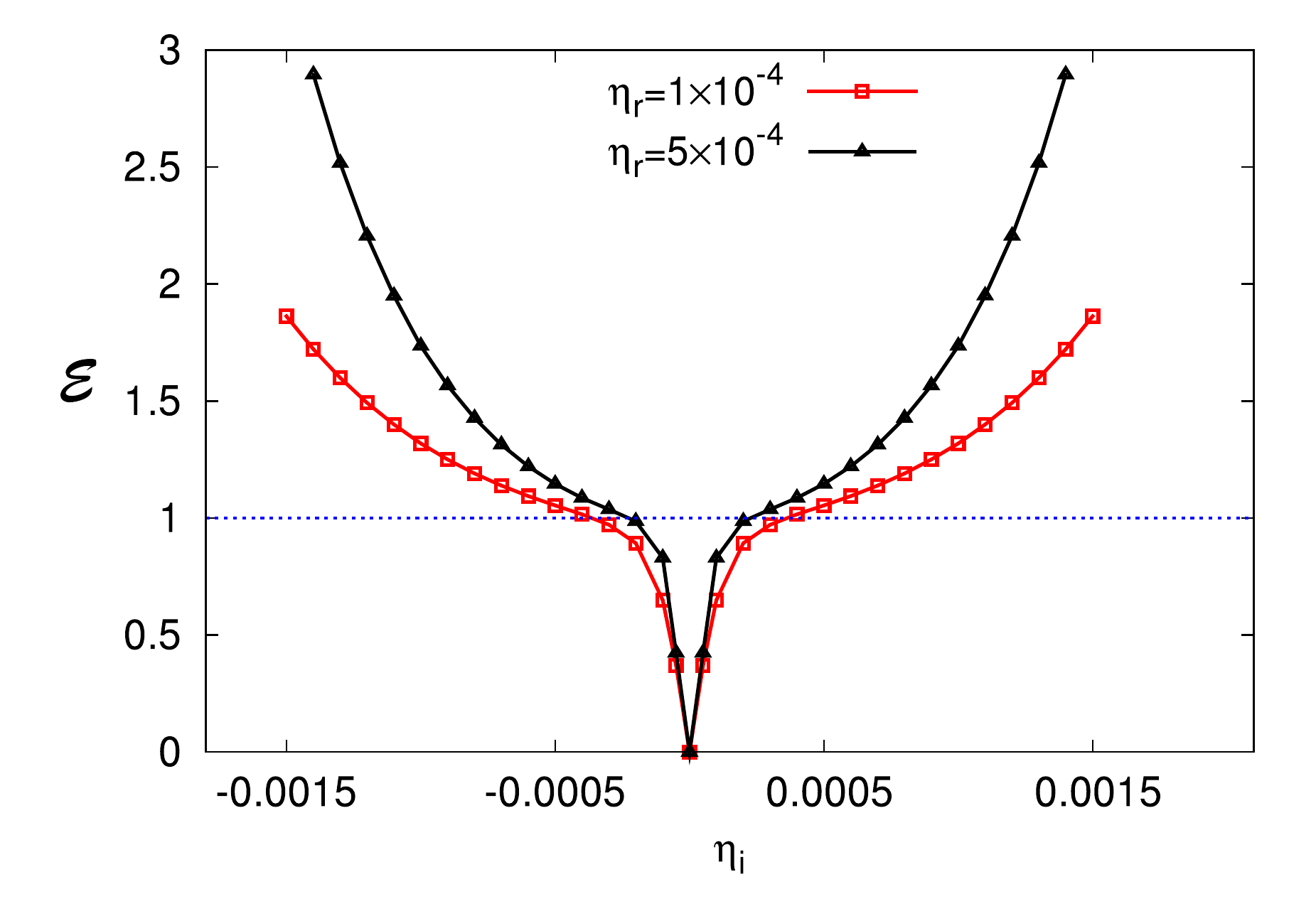}
    \label{fig:triangleb}
  }
  \caption{Panel (a) demonstrates the validity of the triangle inequality as 
    described by \eref{eq:triangle} for the $n=0$ squeezed states. 
    Panel (b) shows the violation of the triangle 
    inequality (\ref{eq:corichieps}) derived for semi-classical states, and the variation of $\mathcal E$ with varying $\eta_{\rm i}$. 
    $\mathcal E$ remains smaller than $1$ as long as $|\eta_{\rm i}|$ 
    is close to zero. The two curves correspond to: 
    $\eta_{\rm r}=(1\times10^{-4}, 5\times10^{-5})$ 
    and $\omega^*=1000\,\sqrt{G}$.}
  \label{fig:triangle}
\end{figure}

%%%
\subsubsection{Energy density}
%%%
As discussed previously, numerical simulations for sharply peaked Guassian states show that the quantum bounce occurs 
close to the maximum value the matter energy density in sLQC, given by 
$\rho_{\rm max}\approx 0.409\,\rho_{\rm Pl}$ \cite{aps3,dgs2}.
For widely spread Gaussian states, however, 
$\rho_{\rm b}$ can be  significantly smaller than $\rho_{\rm max}$ \cite{dgs2}.
On the other hand, the energy density at the bounce in the effective 
description of a flat FRW model is always equal to its 
absolute upper bound, i.e.\ $\rho_{\rm b}^{\rm eff}=\rho_{\rm max}$. Thus, the 
effective theory always overestimates the energy density at the bounce for the Gaussian initial 
states \cite{dgs2}. We see the same qualitative behavior of $\rho_{\rm b}$ to hold for 
squeezed states. 

\begin{figure}[tbh!]
  \subfigure[]{
    \includegraphics[angle=0,width=0.47\textwidth,height=!,clip]{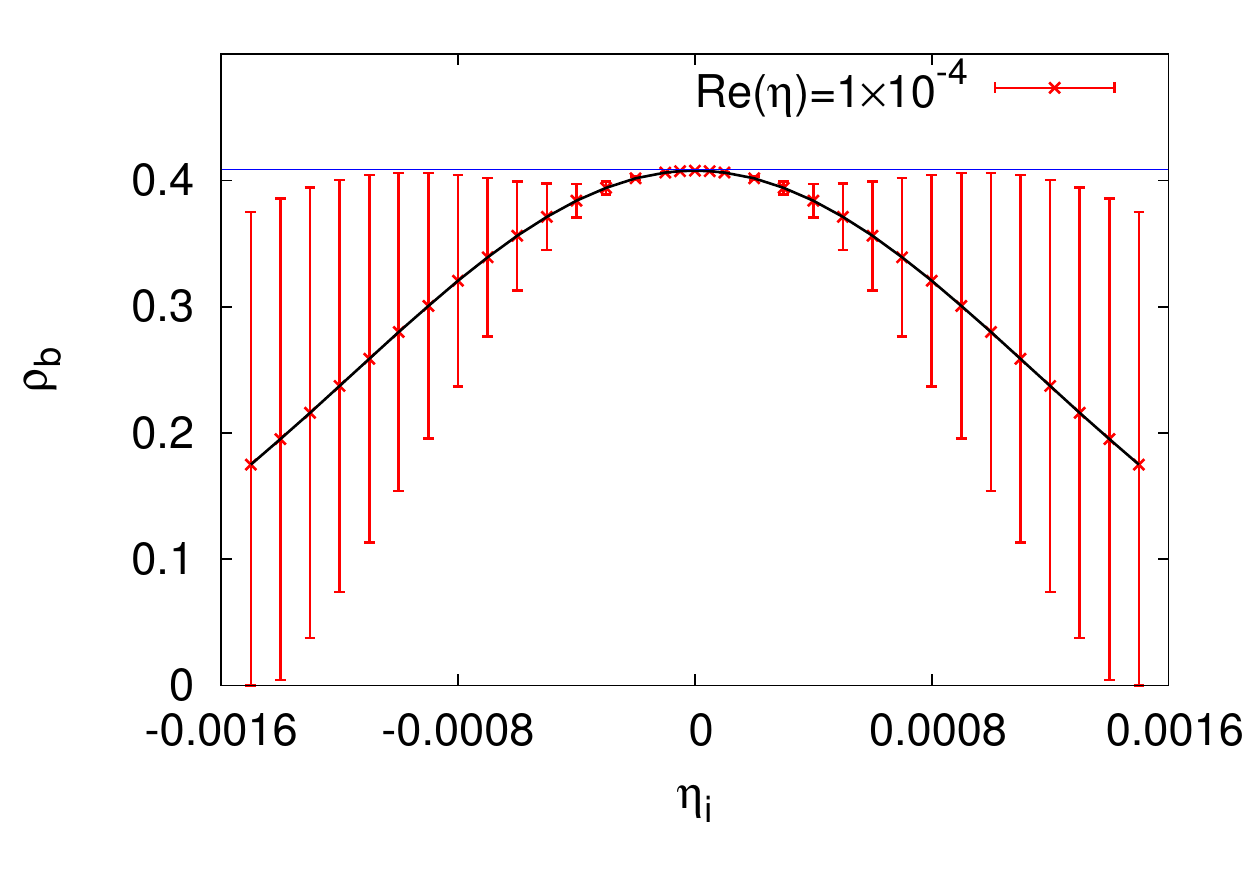}
    \label{fig:rhobounce1}
  }
  \subfigure[]{
    \includegraphics[angle=0,width=0.47\textwidth,height=!,clip]{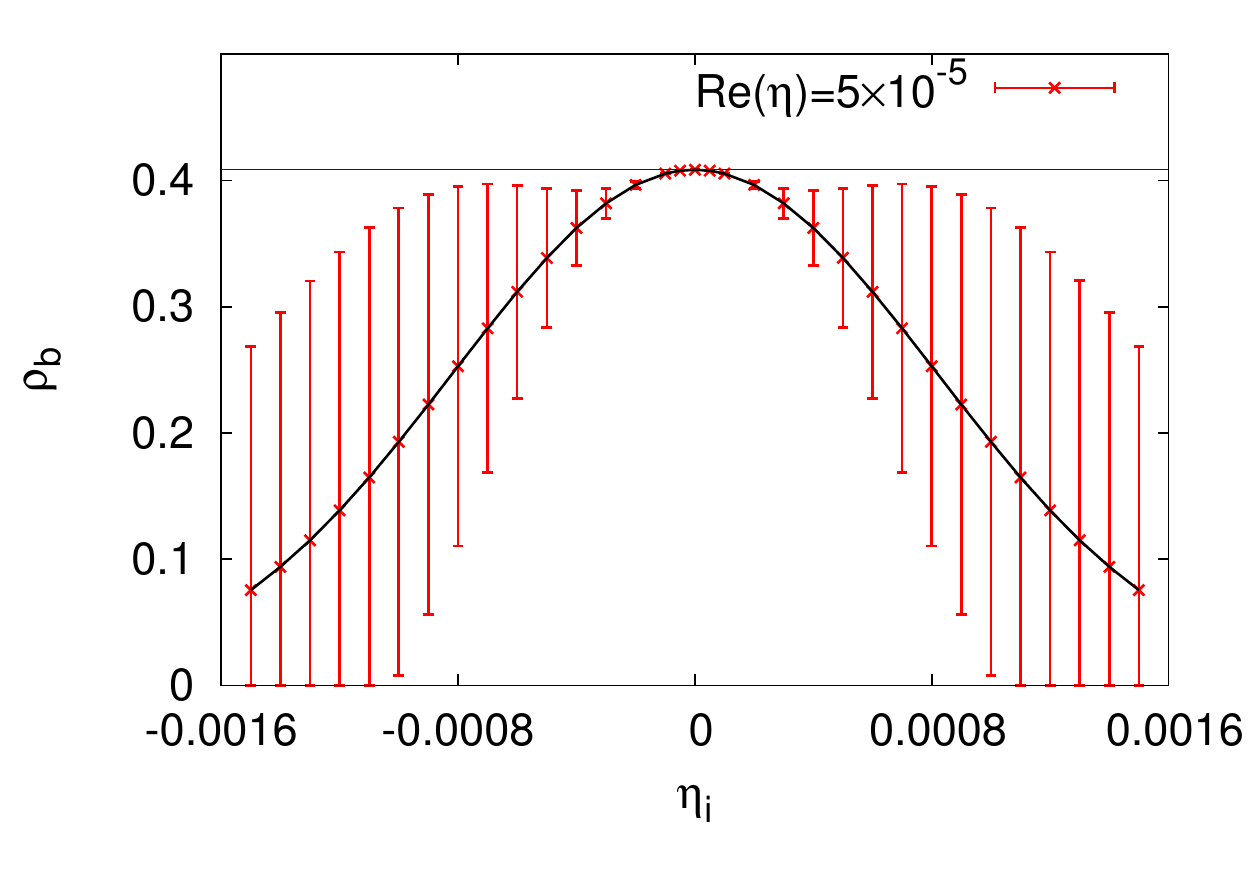}
    \label{fig:rhobounce2}
  }
  \caption{The energy density at the bounce computed for various values of 
     $\eta_{\rm i}$ and fixed $\eta_{\rm r}=1\times10^{-4}$ in 
     \fref{fig:rhobounce1} and $\eta_{\rm r}=5\times10^{-5}$ in 
%   \fref{fig:rhobounce2}. It is clear that the energy density at the bounce is 
    \fref{fig:rhobounce2}. The error bars represent the spread in $\rho$. It is clear that the energy density at the bounce is 
     always bounded above by $\rho_{\rm max} \approx 0.409 \rho_{\rm Pl}$ predicted by sLQC (shown by horizontal line), and the energy 
     density at the bounce decreases with increased squeezing.}
  \label{fig:rhobounce}
\end{figure}

We computed the energy density at the bounce for various squeezed states and
found that it varies monotonically with $|\eta_{\rm i}|$, always remaining
below $\rho_{\rm max}$ and reaching its maximum value for $\eta_{\rm i}=0$.
The variation of the energy density at the bounce for squeezed states was studied 
analytically in the context of the exactly solvable model in Ref.\ 
\cite{montoya_corichi2}. \fref{fig:rhobounce} shows 
the values of the energy density at the bounce for two values of $\eta_{\rm r}$ and
varying $\eta_{\rm i}$. The solid (black) curves show the 
bounce energy density obtained via analytical calculation following
Ref.~\cite{montoya_corichi2}, 
while the (red) symbols correspond to the bounce density 
computed from the numerical simulations. The centers of the error bars show $\rho_{\rm b}$ and the error bars themselves denote the spread 
in the energy density. 
It is remarkable to see that there is an excellent agreement between the 
analytical predictions from sLQC and our numerical simulations even though the Hamiltonian constraint in the present analysis is different from sLQC. 

%%%
\subsubsection{Comparison with the effective theory}
%%%

So far we have discussed the evolution of the squeezed state wavefunctions, their 
dispersion and the energy density at the bounce for various choices of parameters. 
Let us now consider the trajectories, as described by the expectation 
value of the volume observable as a function of the internal time. A detailed 
comparison of such trajectories with the corresponding effective ones for Gaussian 
states were performed in Ref.\ \cite{dgs2}, with the conclusion that for all types 
of Gaussian initial data considered, the effective trajectory obtained from eq.~(\ref{eq:heff}) always underestimates the bounce 
volume. 
We obtain the same qualitative behavior for 
squeezed states in the numerical simulations performed in this paper.

\begin{figure}[tbh!]
  \subfigure[\; $\omega^*=1000\,\sqrt{G},\,\eta=(1+i)\times10^{-4}$]{
    \includegraphics[angle=0,width=0.45\textwidth,height=!,clip]{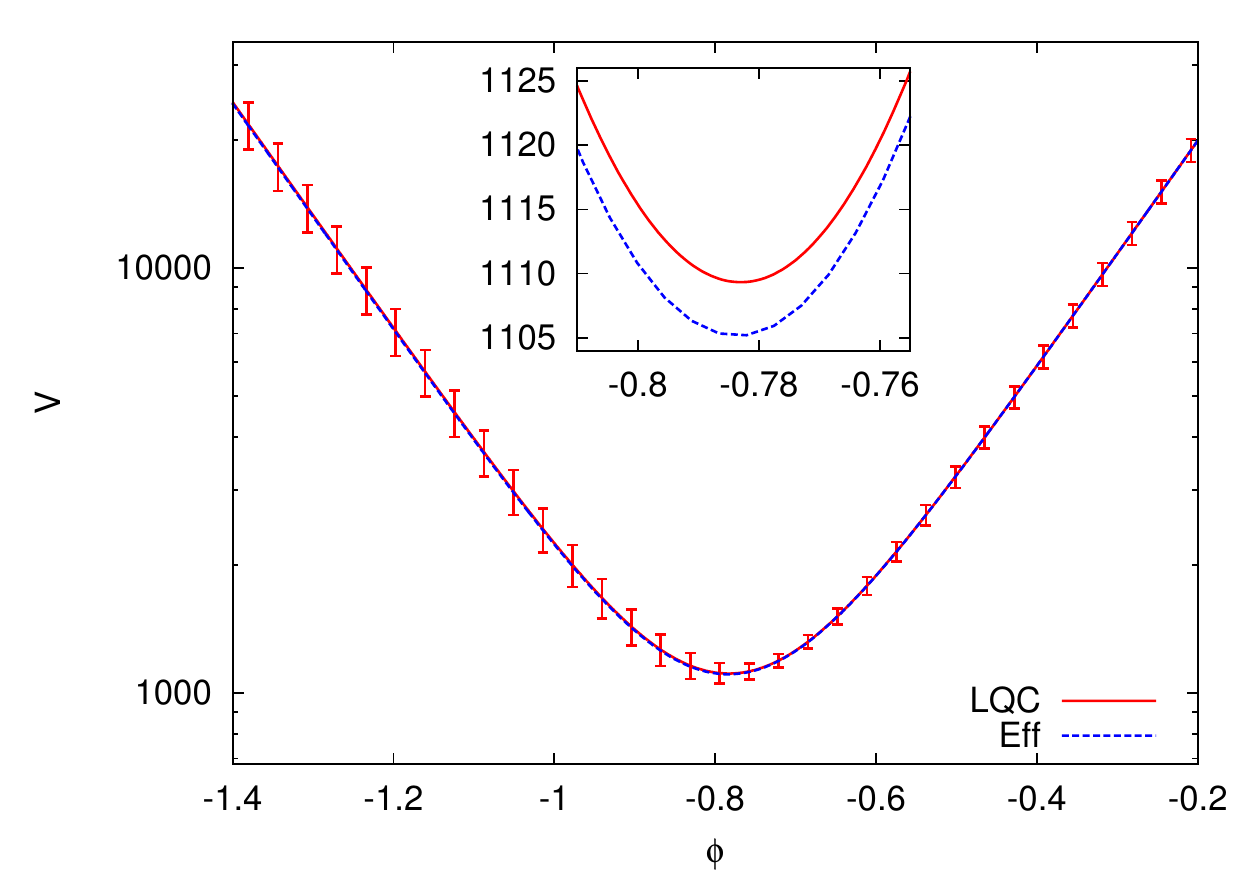}
    \label{f:eta1000b}
  }
  \subfigure[\; $\omega^*=1000\,\sqrt{G},\,\eta=(1+10i)\times10^{-4}$]{
    \includegraphics[angle=0,width=0.45\textwidth,height=!,clip]{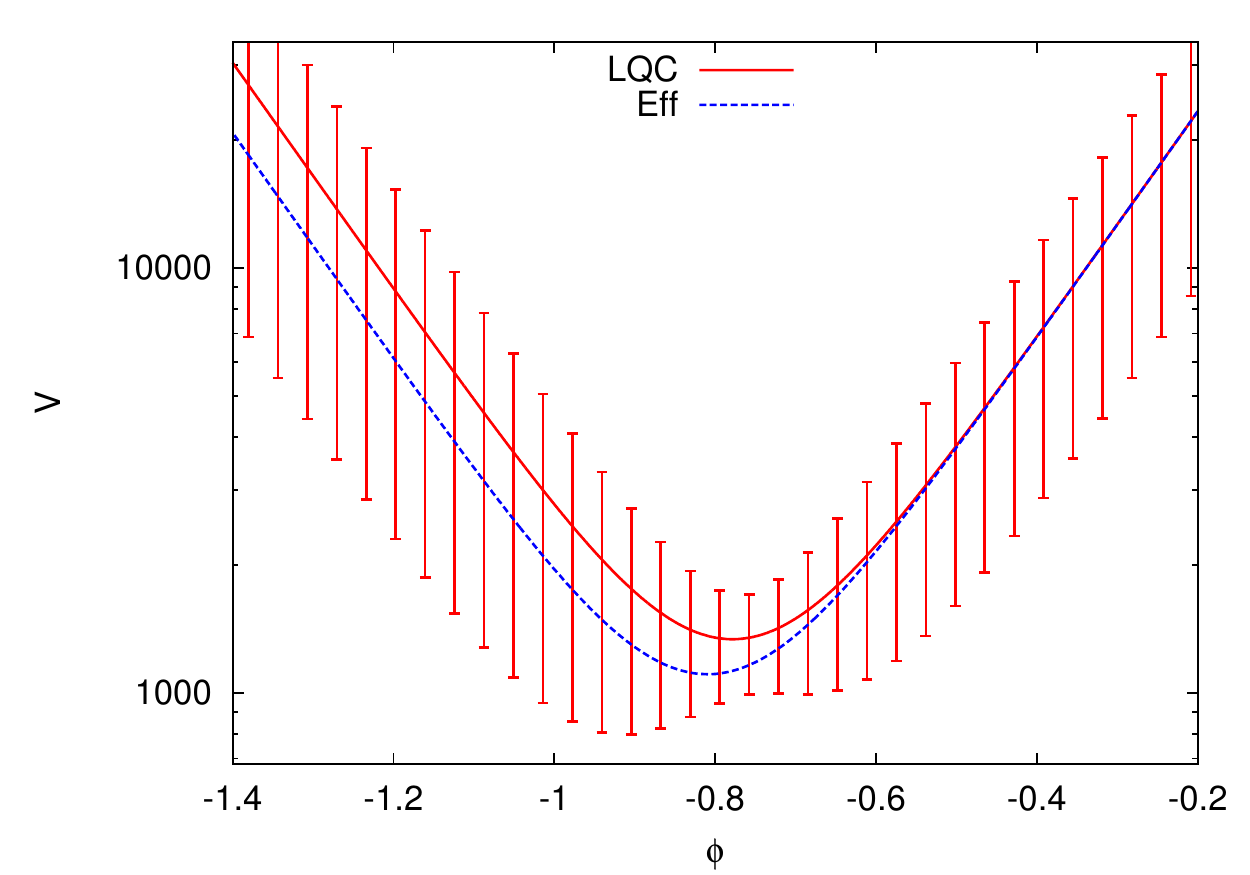}
    \label{f:eta1000}
  }
  \subfigure[\; $\omega^*=1000\,\sqrt{G},\,\eta=(1+i)\times10^{-2}$]{
    \includegraphics[angle=0,width=0.45\textwidth,height=!,clip]{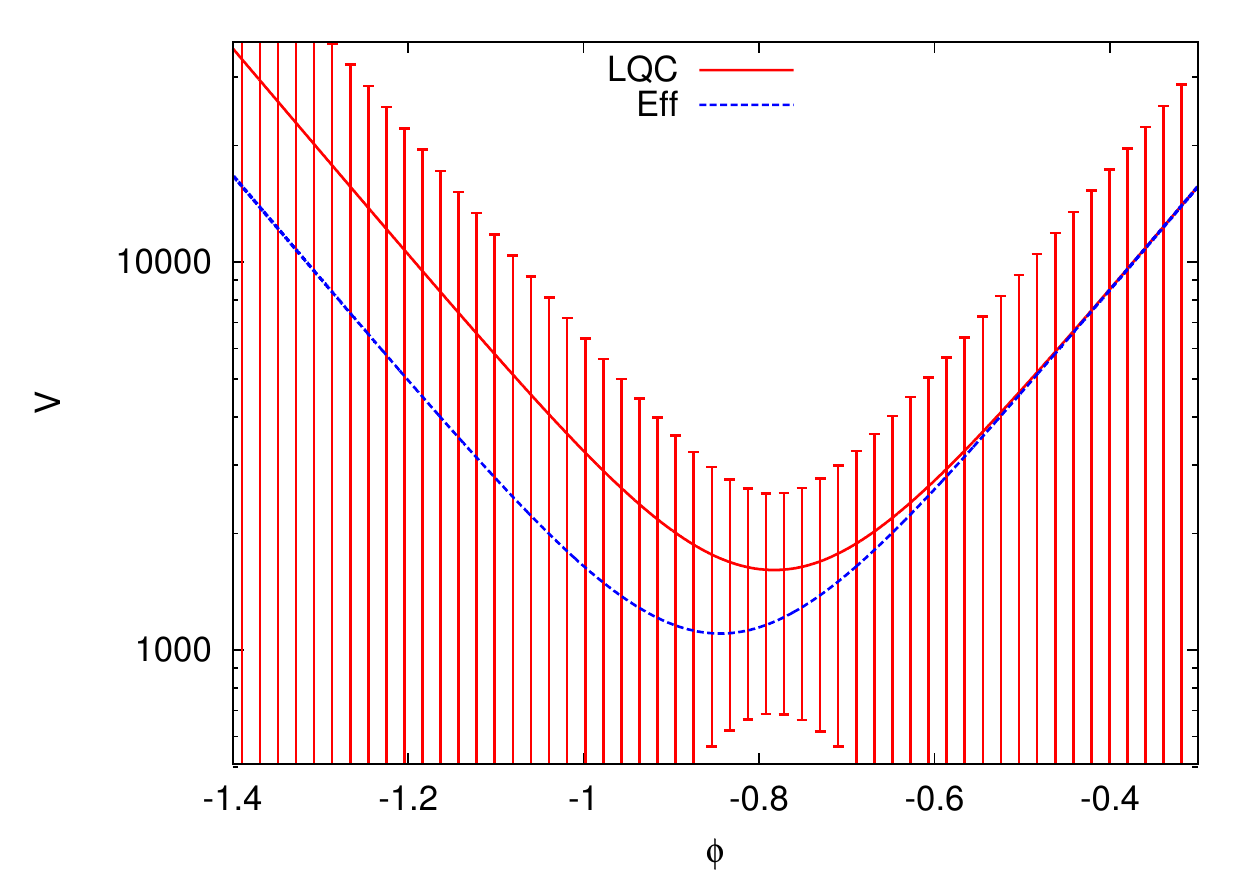}
    \label{f:eta1000c}
  }
  \subfigure[\; $\omega^*=60\,\sqrt{G},\,\eta=(1+i)\times10^{-2}$]{
    \includegraphics[angle=0,width=0.45\textwidth,height=!,clip]{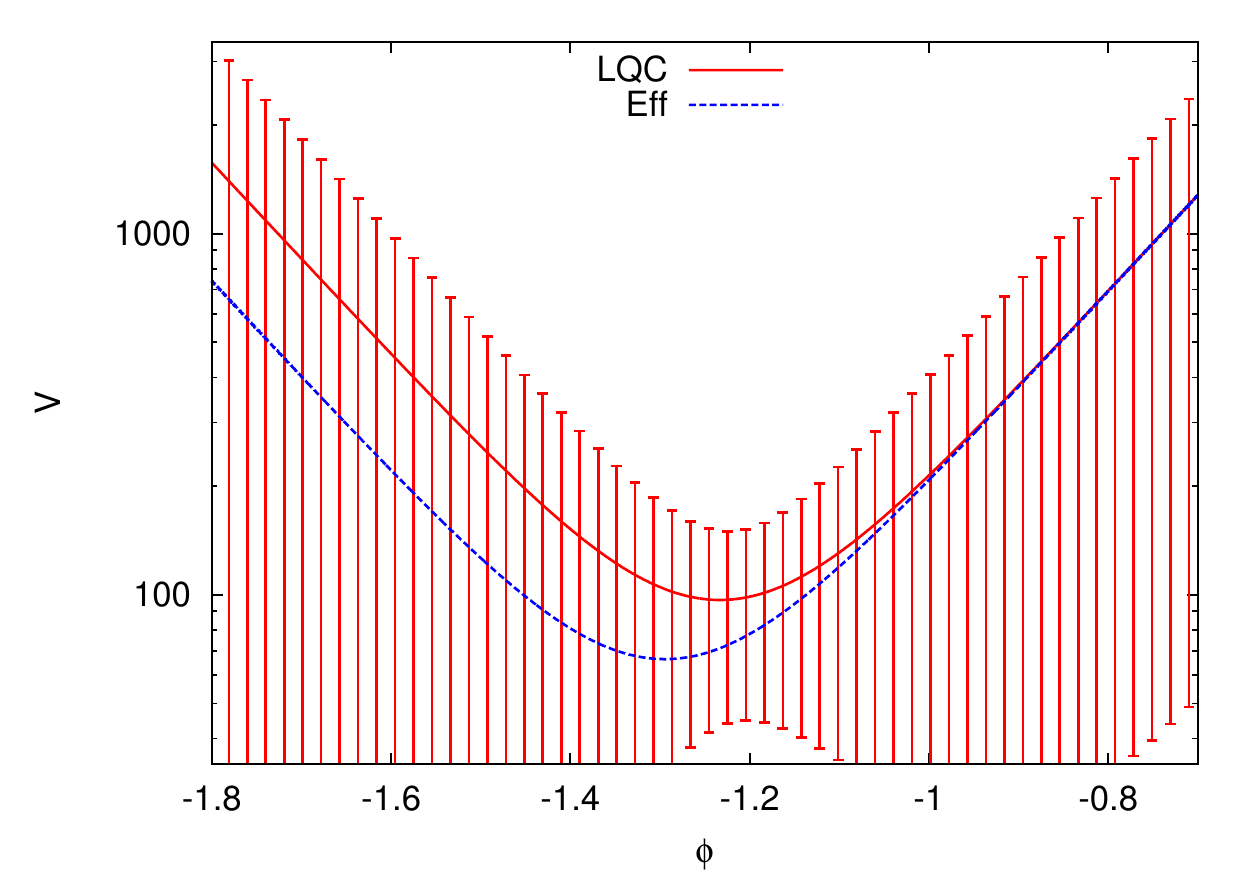}
    \label{f:eta60}
  }
  \caption{Evolution trajectories for squeezed states with $n=0$. 
    Panel~\subref{f:eta1000b}: $\omega^*=1000\,\sqrt{G}$ and $\eta=(1+i)\times10^{-4}$.
    Panel~\subref{f:eta1000}: $\omega^*=1000\,\sqrt{G}$ and $\eta=(1+10i)\times 10^{-4}$.
    Panel~\subref{f:eta1000c}: $\omega^*=1000\,\sqrt{G}$ and $\eta=(1+i)\times 10^{-2}$.
    Panel~\subref{f:eta60}: $\omega^*=60\,\sqrt{G}$ and $\eta=(1+i)\times 10^{-2}$.
    The solid (red) curves, with the error bars representing the
    dispersion, show the LQC trajectories, while the dashed (blue) curves
    show the corresponding effective trajectories. 
  }
  \label{f:w1000}
\end{figure}

In~\fref{f:w1000}, we compare the trajectories of four different squeezed
states with $n=0$. The particular cases shown are chosen to illustrate the effects of varying one
property of the initial state at a time. Specifically,  \fref{f:eta1000b}
and~\fref{f:eta1000} correspond to states that differ only in $\eta_{\rm i}$, whereas
 \fref{f:eta1000b} and~\fref{f:eta1000c} correspond to initial states with different $\eta$ but with
the same ratio $\eta_{i}/\eta_{r}$, all of them with the same $\omega^*$. 
Finally, \fref{f:eta1000c} and~\fref{f:eta60} correspond to states with different
$\omega^*$, but with the same $\eta$.
The particular values of $\omega^*$ and $\eta$ are indicated in each sub-figure.
We see that 
increasing $|\eta|$, either by keeping the same ratio $\eta_{i}/\eta_{r}$ or
by increasing $\eta_{i}$ alone, leads to a wider state, resulting in an
increased deviation between the effective and the LQC trajectories close to the 
bounce as well as an increased bounce volume. 
Comparing \fref{f:eta1000c} and~\fref{f:eta60}  we see that decreasing
$\omega^*$, while resulting in a smaller bounce volume, may nonetheless result
in a similar relative deviation from the effective theory. This is evident
from the qualitative similarity between these two figures (although the scales
differ by an order of magnitude).

As is customary in the literature, we  represent the dispersion $\Delta V$ in the 
figures by means of error bars. Note, however, that for non Gaussian states the error
bars %location (that is, centered around the mean value V) have no particular
%meaning. For instance, the error bars
may fall below $V=0$ in some
cases. This, of course, does not mean $|\Psi|>0$ for $V\leq 0$, but is an artifact of the way error bars are defined (centered around the mean value of volume observable). In fact, as discussed earlier, the amplitude of the wavefunctions decay almost exponentially in the regime $V<V_{\rm c}$, and $|\Psi|$ is 
zero when volume is zero. 
% as shown in \fref{f:psiw60}.

\begin{figure}[tbh!]
  \subfigure[\; $\omega^*=1000\,\sqrt{G}$]{
    \includegraphics[angle=0,width=0.45\textwidth,height=!,clip]{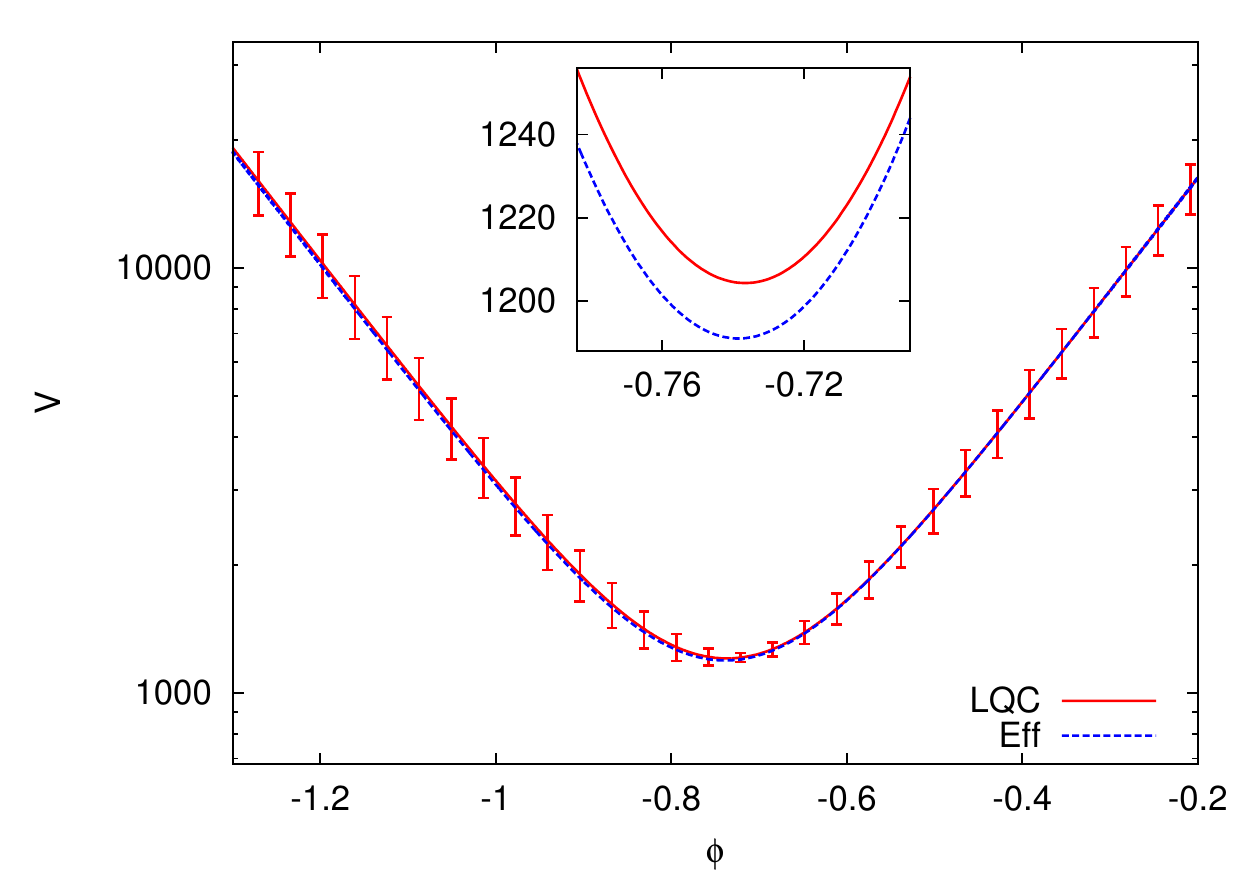}
    \label{f:kn1000}
  }
  \subfigure[\; $\omega^*=50\,\sqrt{G}$]{
    \includegraphics[angle=0,width=0.45\textwidth,height=!,clip]{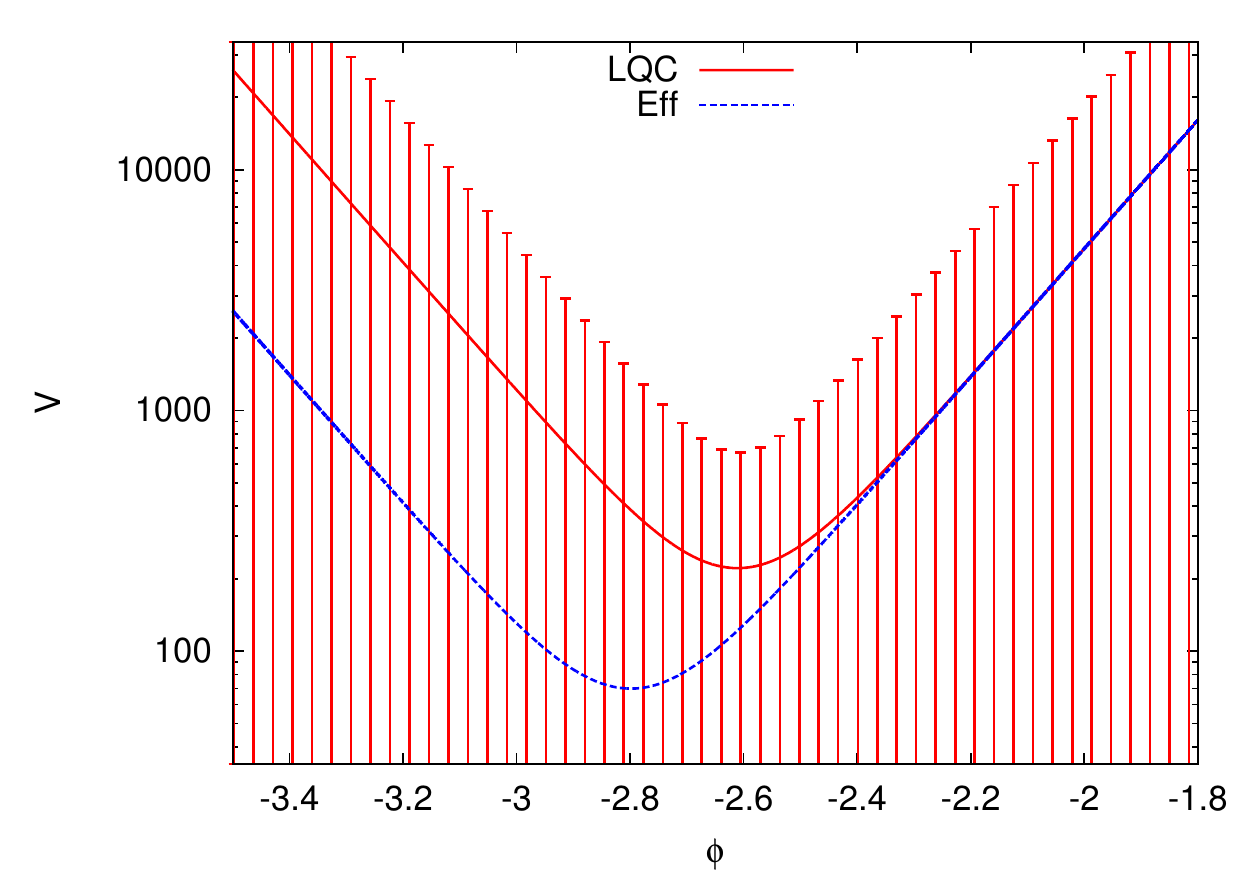}
    \label{f:kn50}
  }
  \caption{Expectation value of $V$ as a function of the emergent time $\phi$ for 
    squeezed states with $n=50$, $\eta=3\times 10^{-4}+ 3\times10^{-4}i$ and
    different values of $\omega^*$, indicated in each panel.
    The solid (red) curves, with the error bars representing the dispersion, show
    the LQC trajectories, while the dashed (blue) curves show the corresponding 
    effective trajectories. There is an excellent agreement between the two 
    trajectories for $\omega^*=1000\,\sqrt{G}$ (panel~\subref{f:kn1000}), whereas the differences are prominent for 
    $\omega^*=50\,\sqrt{G}$ (panel~\subref{f:kn50}).
  }
  \label{f:sqkn50}
\end{figure}

A similar trend can also be observed for the deviation between the effective and the LQC trajectories for the squeezed states with $n=50$. \fref{f:sqkn50} shows the 
comparison of the effective and LQC trajectories for $\omega^*=1000\,\sqrt{G}$ in panel(a) and 
for $\omega^*=50\,\sqrt{G}$ in panel (b), both for $\eta=3\times 10^{-4}+ 3\times10^{-4}i$. It is 
evident from these figures 
that in the case of $\omega^*=1000\,\sqrt{G}$, where the dispersion in the state is small, 
there are small differences between the LQC and effective theory. These differences 
are much more prominent for $\omega^*=50\,\sqrt{G}$, for which the states have large 
dispersion. Once again, the effective theory predicts a smaller bounce volume in comparison to the quantum evolution. 
%in all 
%the cases of the squeezed state considered in this paper. 

\vskip0.5cm
So far we have discussed the  evolution of squeezed states 
for various values of the initial parameters. In all the cases discussed here, the 
evolutions are non-singular and undergo a quantum bounce, showing similar qualitative features as 
the evolution of Gaussian states. We have also seen that the numerical results regarding the variation 
of the energy density with respect to the imaginary part of $\eta$ are in very good 
agreement with the analytical results obtained in Ref.\ \cite{montoya_corichi2}, and 
the triangle inequality derived in the Ref.\ \cite{kp} is obeyed irrespective of the initial 
data considered. In the following we now study the evolution of two more non-Gaussian states, 
which have quite different features compared to a Gaussian state. For these states 
we will discuss the evolution of the wavepacket and compare the LQC trajectories with 
the corresponding effective one for both small and large dispersions.

%\clearpage

%%%%%%%%%%%%%%%%%%%%%%%%%%%%%%%%%%%%%%%%%%%%%%%%%%
\subsection{Multipeaked-1 states}  \label{s:multi1}
%%%%%%%%%%%%%%%%%%%%%%%%%%%%%%%%%%%%%%%%%%%%%%%%%%
%\todo{Begin with some physical motivation ra}
We now discuss results from our analysis for initial states which are constructed as a sum of Gaussians in $k$ space (eq.(\ref{eq:avg})).
The two Gaussians are separated by the parameter $\delta k$. The initial state for the evolution is then obtained
by evaluating the integral that transforms the state into the volume representation. The
resulting state  may have several local peaks, depending on the value of the
parameters. Thus, such states, unlike the
squeezed states considered earlier, are therefore not peaked at any classical volume at the
initial time. 
The state is still chosen such that the expectation value of the
volume at the initial time is very large and the corresponding energy density
is very small compared to the Planck density. Irrespective of the choice of parameters, we find that loop quantum evolution is non-singular for all such states 
and the existence of a quantum bounce is robust. Whether or not the effective trajectory obtained from (\ref{eq:heff}) captures the underlying quantum evolution 
depends on the relative dispersion in volume. As long as the relative dispersion in volume is small, the effective dynamics is an excellent approximation -- a  remarkable 
result, since these states are very different from the Gaussian states used in the derivation of the effective Hamiltonian constraint \cite{vt}.
For multipeaked-1 states with a large dispersion in volume, on the other hand, we find significant 
deviations between the effective and LQC trajectories. 
As with the squeezed states with large dispersions, the energy density at the bounce is much smaller than the universal maximum $\rho_{\rm max}$ in sLQC.
%Let us now go into more detail of the numerical simulations.

\begin{figure}[tbh!]
  \subfigure[]{
    \includegraphics[angle=0,width=0.62\textwidth,height=!,clip]{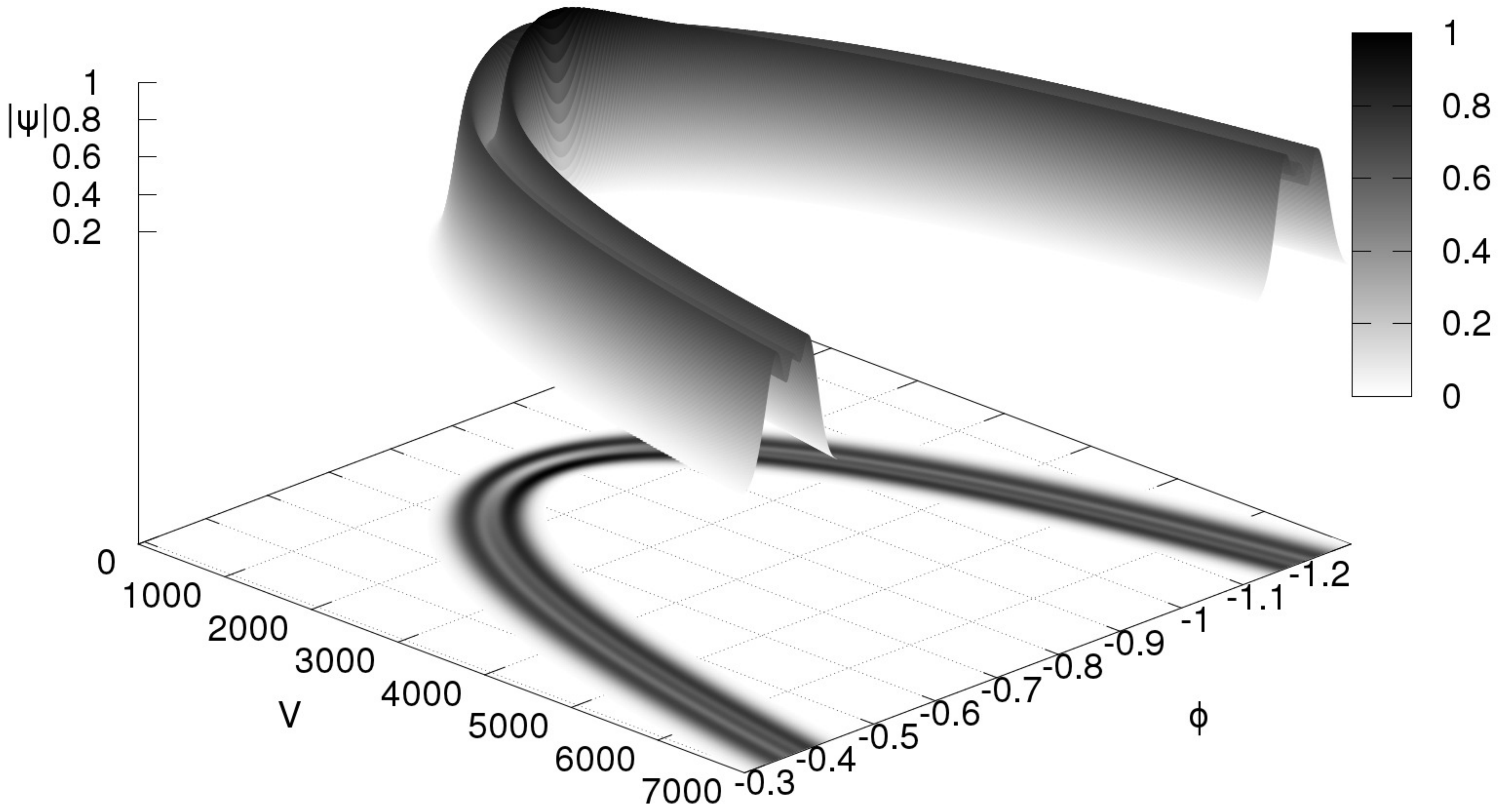}
    \label{f:avg_gauss_3Da}
  }
  \subfigure[]{
    \includegraphics[angle=0,width=0.35\textwidth,height=!,clip]{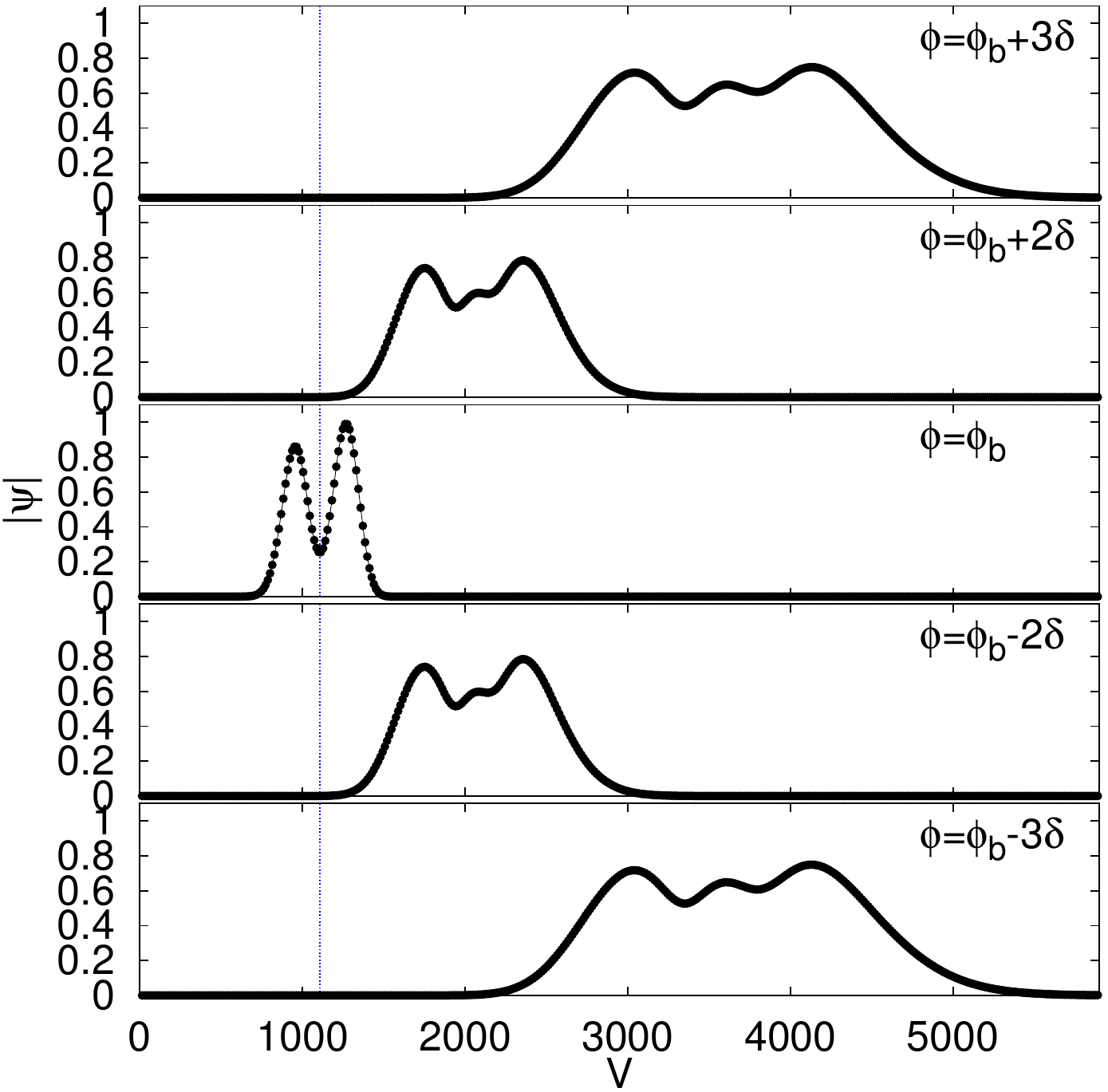}
    \label{f:avg_gauss_3Db}
  }
  \caption{Evolution of a multipeaked-1 initial state with
    $\omega^*=1000\,\sqrt{G}$, $\eta=2\times10^{-4}$ and $\delta k=2$. 
    Panel~\subref{f:avg_gauss_3Da}: $|\Psi|$, including the projection onto the $V$-$\phi$ plane, shown
    to help visualizing the 3D graph. 
    Panel~\subref{f:avg_gauss_3Db}: $|\Psi|$ at different values of $\phi$
    around the bounce time, $\phi_{\rm b}=-0.7835$, as indicated, where
    $\delta=0.1$. 
The blue dotted line indicates the bounce volume $V_{\rm b}=1108.7\, \Vpl$.
  }
  \label{f:avg_gauss_3D}
\end{figure}

As a representative case of simulations of multipeaked-1 states, Fig.~\ref{f:avg_gauss_3D} shows the evolution of a
 state with $\omega^*=1000\,\sqrt{G}$, $\eta=2\times10^{-4}$ and $\delta k=2$. 
\fref{f:avg_gauss_3Da} shows the 3D evolution 
of the wavefunction plotted against the volume $V$ and the emergent time $\phi$. 
\fref{f:avg_gauss_3Db} shows snapshots of the wavefunction around the bounce
time $\phi_b$.
That the shape of the wavepacket is highly non-Gaussian is seen most clearly in~\fref{f:avg_gauss_3Db}. During the entire evolution, the state has support on non-zero finite  
volume and undergoes a non-singular bounce.  
Note that in this case since the state has no well defined single peak, it is not
possible to single out a value of $\omega$ at which the state may be
considered peaked and therefore it is not possible to evaluate the cutoff volume using eq.(\ref{eq:vcut}). 
Another interesting  feature of the evolution is that the shape of the wavepacket is 
recovered on the other side of the bounce. For example, at $\phi_b+3\delta$ and $\phi_b-3\delta$
the amplitude of the wavefunction has almost the same profile, as shown in the first and the last panel of \fref{f:avg_gauss_3Db}. 

\begin{figure}[tbh!]
\subfigure[]
   {
    \includegraphics[angle=0,width=0.62\textwidth,height=!,clip]{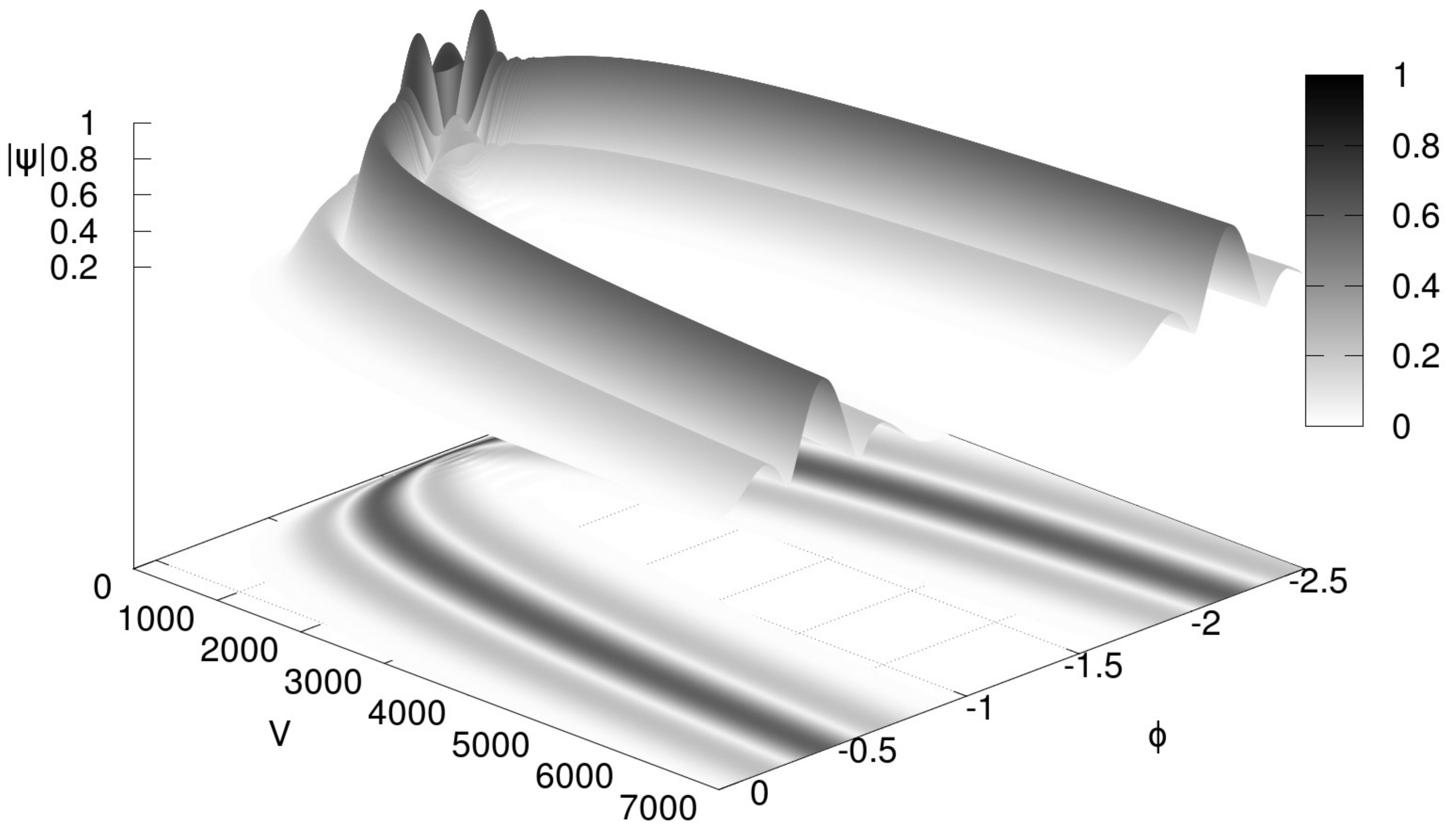}
  \label{f:avg_gauss2_3Da}
  }
\subfigure[]
   {
    \includegraphics[angle=0,width=0.35\textwidth,height=!,clip]{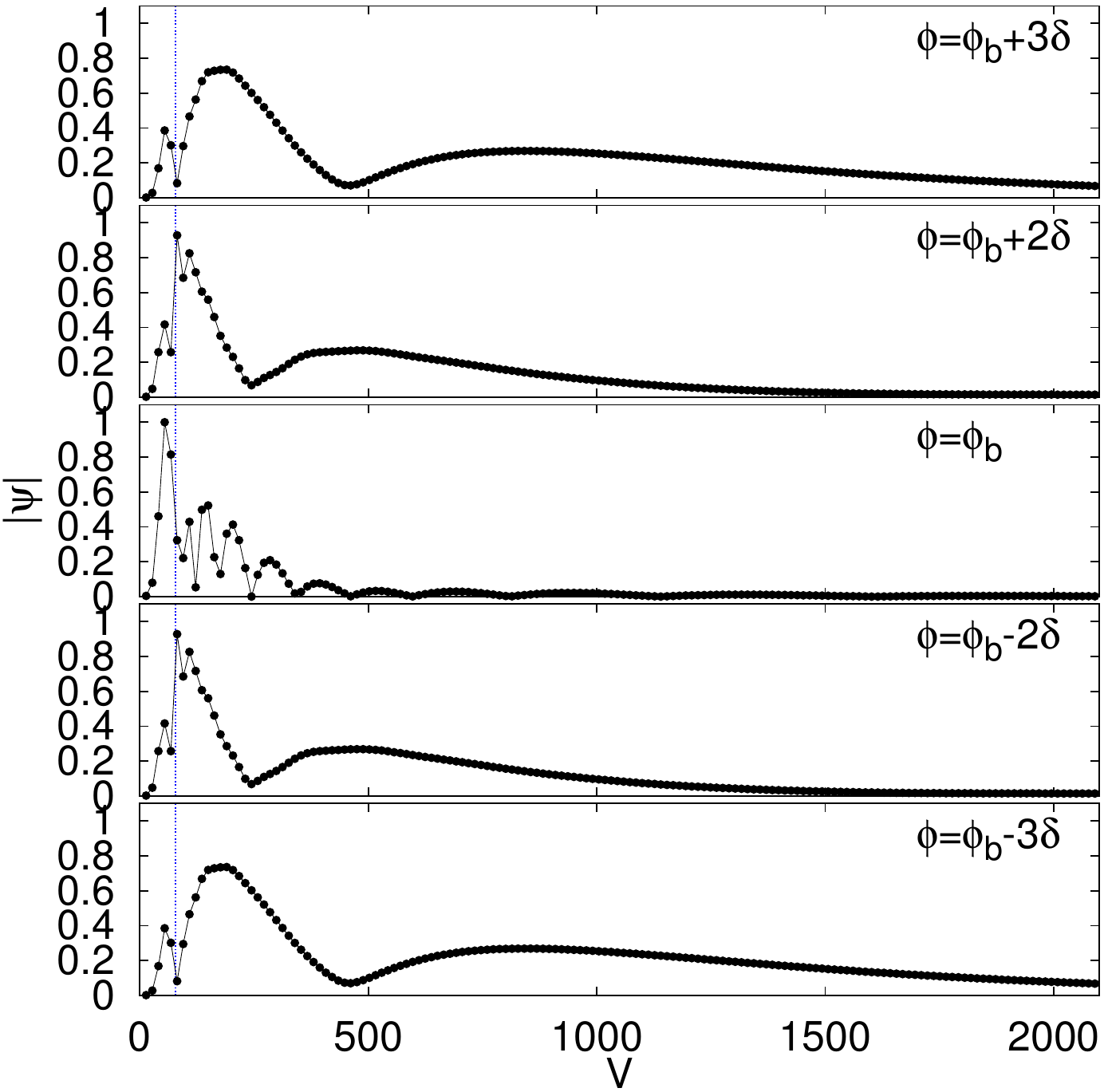}
  \label{f:avg_gauss2_3Db}
  }
\caption{Evolution of an multipeaked-1 initial state with
  $\omega^*=50\,\sqrt{G}$, $\eta=2\times10^{-2}$ and $\delta k=2$. 
  Panel~\subref{f:avg_gauss_3Da}: $|\Psi|$, including the projection onto the $V$-$\phi$ plane. 
  Panel~\subref{f:avg_gauss_3Db}: $|\Psi|$ at different values of $\phi$ close to
  the bounce time, $\phi_{\rm b}=-1.271$, as indicated in the figure, where
  $\delta=0.1$. 
The blue dotted line indicates the bounce volume, $V_{\rm b}=78.231\,\Vpl$.}
\label{f:avg_gauss2_3D}
\end{figure}

In figure~\ref{f:avg_gauss2_3D} we present results for another multipeaked-1 state, with $\eta=2\times 10^{-2}$, $\omega^*=50\,\sqrt{G}$ and $\delta k=2$. 
This state has a quite different profile and a much larger spread, showing
features at small volume that are similar to those in  case (ii) of the squeezed
states presented in Sec. IVA (see \fref{f:squeezed2} for a comparison), 
 which also had a very large spread. Despite being highly quantum in nature, with a very large spread, the shape of the 
 wavefunction is preserved through the non singular bounce. 
We will see, however, that in this case the trajectory obtained from the
effective Hamiltonian  deviates significantly more than the case discussed in Fig.~\ref{f:avg_gauss_3D}.

\begin{figure}[tbh!]
  \subfigure[\; $\omega^*=1000\,\sqrt{G}$]{
    \includegraphics[angle=0,width=0.45\textwidth,height=!,clip]{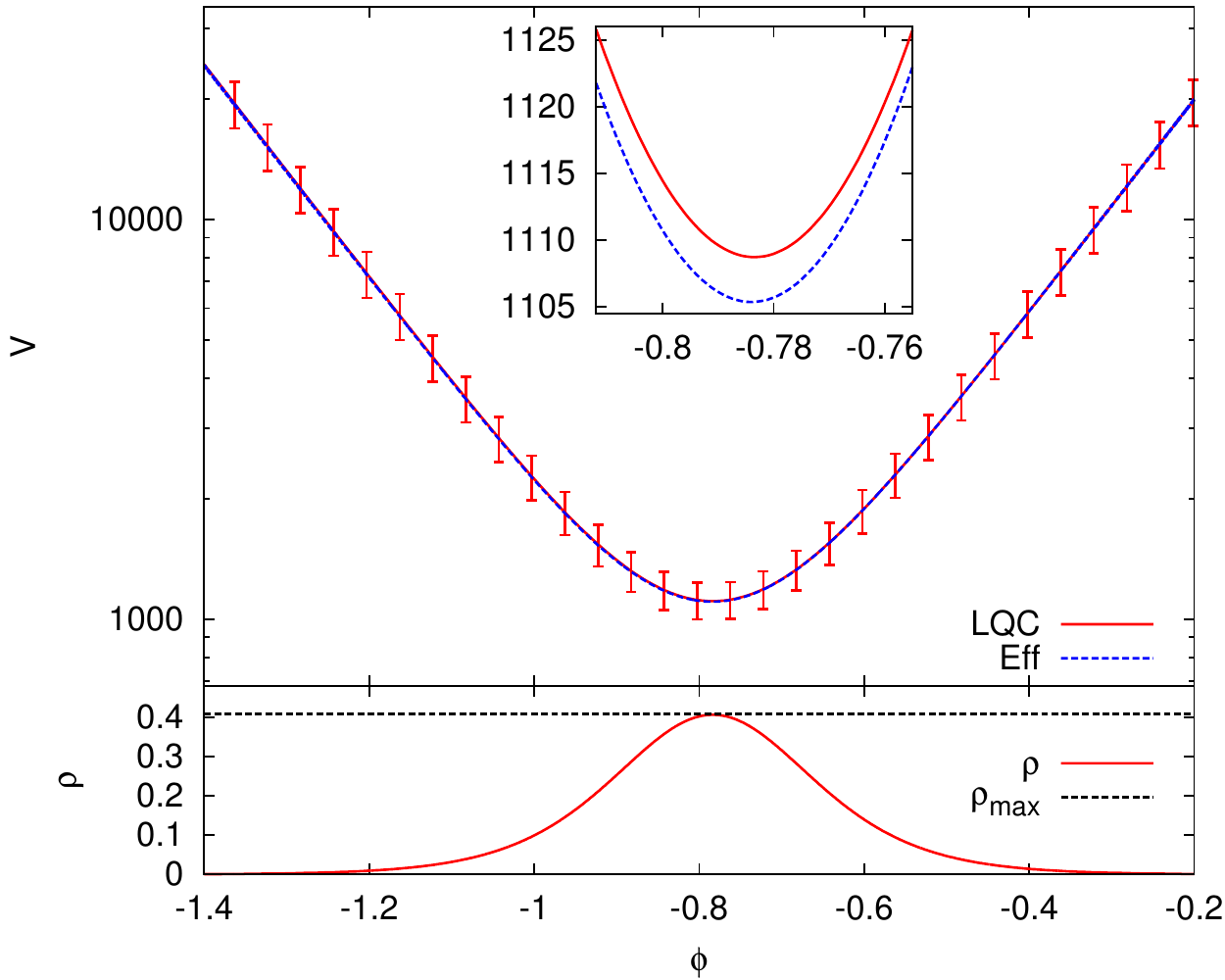}
    \label{f:avg1000}
  }
  \subfigure[\; $\omega^*=50\,\sqrt{G}$]{
    \includegraphics[angle=0,width=0.45\textwidth,height=!,clip]{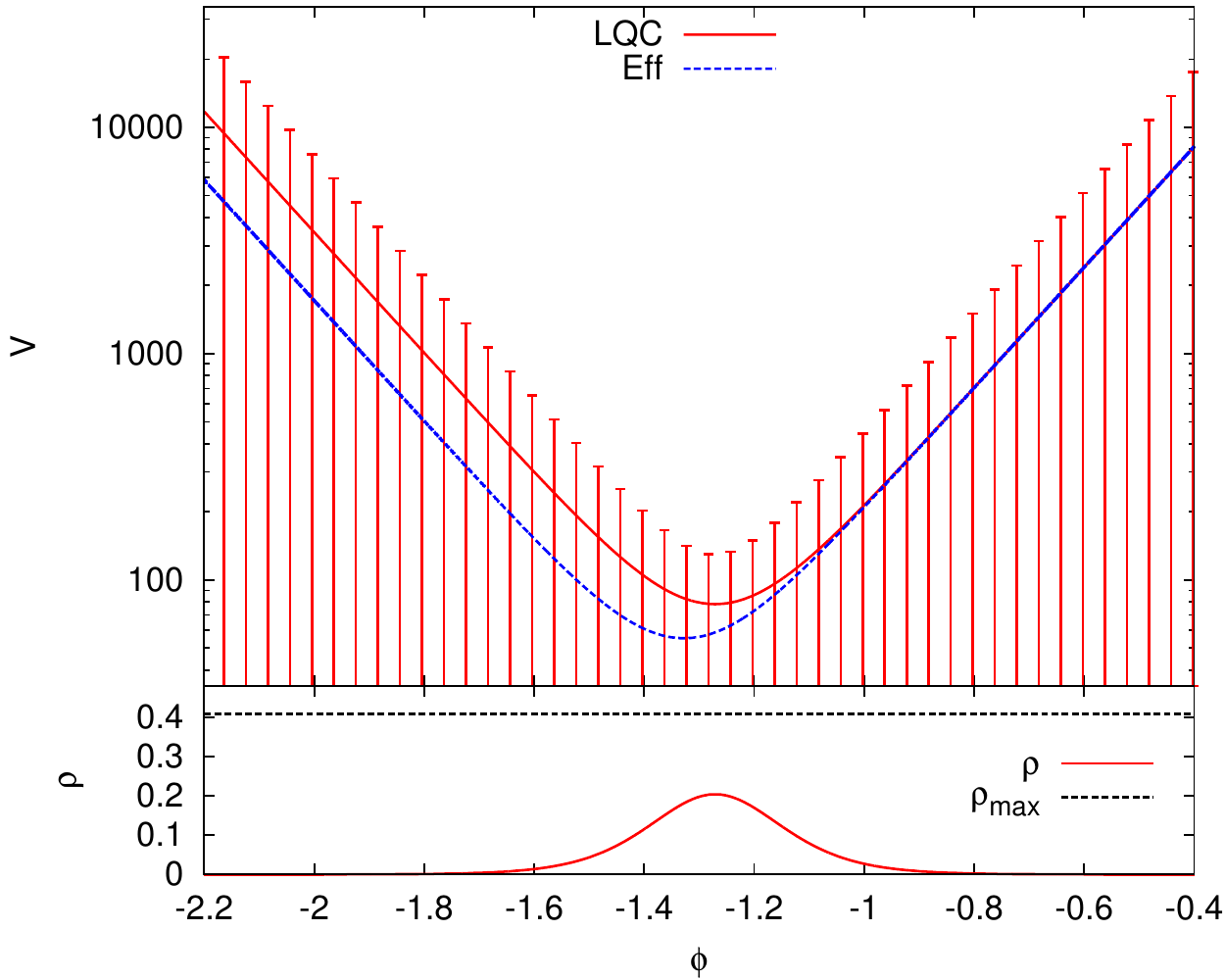}
    \label{f:avg50}
  }
%\subfigure[\; $\omega^*=50\,\sqrt{G}0$]
%   {
%    \includegraphics[angle=0,width=0.45\textwidth,height=!,clip]{./figures/avg500.pdf}
%  \label{f:avg500}
%  }
  \caption{ Comparison of the LQC and effective trajectories and evolution of energy densities for a multipeaked-1 
    state with $\omega^*=1000\,\sqrt{G}$ and $\eta=2\times 10^{-4}$ (panel (a)) and for $\omega^*=50\,\sqrt{G}$ and $\eta=2\times 10^{-2}$ (panel (b)). In both cases
    $\delta k=2$. The 
    solid (red) curve, with the error bars showing the dispersion in volume, 
    corresponds to the LQC trajectory and the dashed (blue) curve shows the 
    corresponding effective trajectory.  
    For large $\omega^*$ the relative volume dispersion is small and the 
    effective theory is in good agreement with the LQC one, whereas 
    $\Delta V/V$ is large for small $\omega$ and the difference between 
    the LQC and the effective theory is more prominent. We see that for the
    multipeaked-1 state with the larger fluctuation the energy density at the bounce is only half of the
    maximum  value of the energy density ($\rho_{\rm max}$) predicted in sLQC. 
    }
  \label{f:avgvphi}
\end{figure}

\fref{f:avgvphi} shows the 
comparison of the LQC and the corresponding effective trajectory for the multipeaked-1 states presented in this subsection.
%$\omega^*=(1000,\, 50)$, $\eta_{\rm r}=2\times 10^{-4}$, $\eta_{\rm i}=0$. 
For the state with $\omega^*=1000\,\sqrt{G}$ and $\eta=2\times 10^{-4}$, shown in \fref{f:avg1000}, the relative volume 
dispersion of the initial state is small: $\Delta V/V=0.13$ and the bounce happens 
at 
%$V_{\rm b}\approx1204\, V_{\rm Pl}$. 
$V_{\rm b}\approx1109\, \Vpl$. 
This results in quite good agreement 
between the two trajectories. On the other hand, for the state with
$\omega^*=50\,\sqrt{G}$ and $\eta=2\times 10^{-2}$, shown in \fref{f:avg50}, the initial dispersion in volume is $\Delta V/V=1.16$ and 
%$V_{\rm b}\approx220\, V_{\rm Pl}$. 
$V_{\rm b}\approx78.23\, \Vpl$ and consequently
there is a significant difference 
between the effective and the corresponding LQC trajectory. 
It is also worth noticing that the energy density at the bounce $\rho_{\rm b}$ satisfies 
the upper bound limit in both cases. Similarly to the case of squeezed states,
$\rho_{\rm b}$ is closer to the absolute 
maximum $\rho_{\rm max}$ for smaller $\Delta V/V$ in these simulations. 

\begin{figure}[tbh!]
  \subfigure[\; $\omega^*=1000\,\sqrt{G}$]{
    \includegraphics[angle=0,width=0.45\textwidth,height=!,clip]{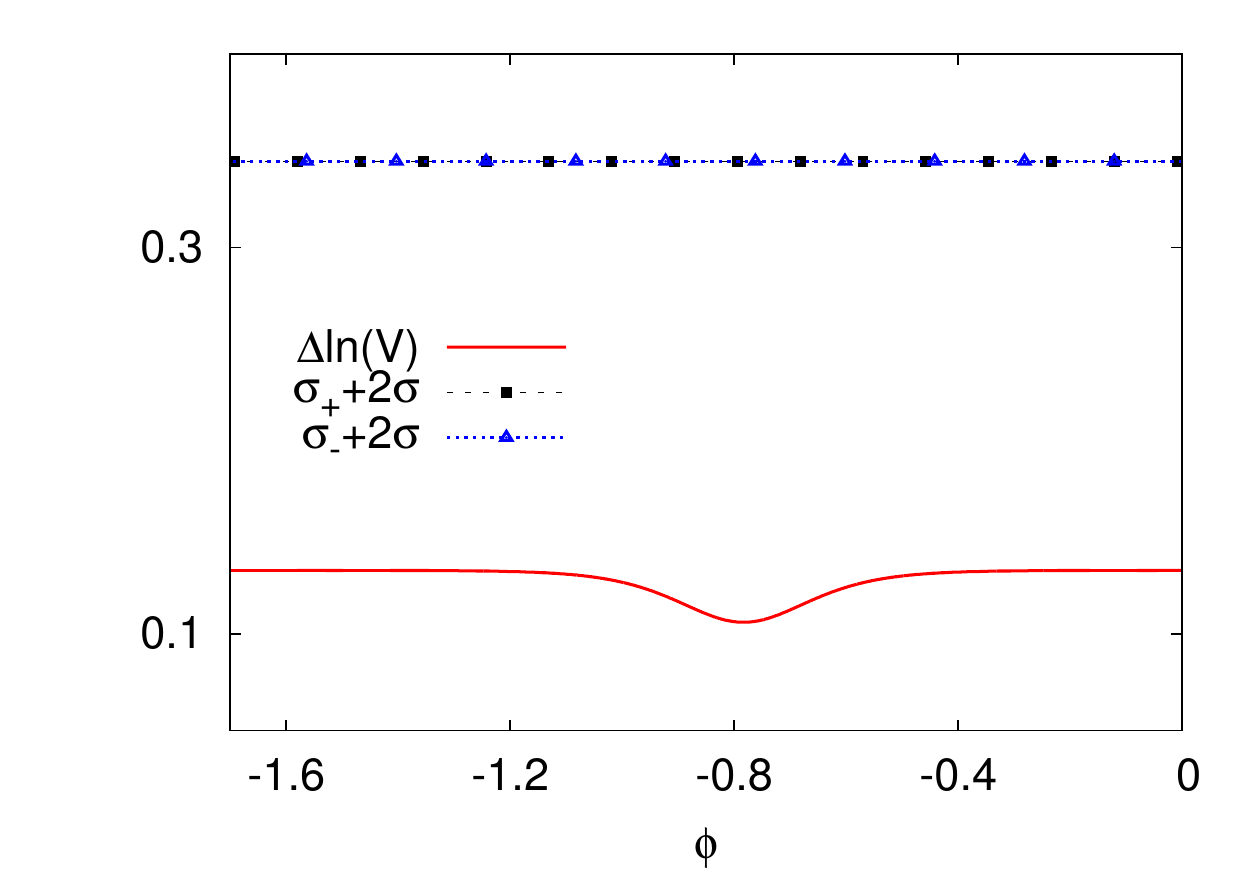}
    \label{f:triangleavg1000}
  }
  \subfigure[\; $\omega^*=50\,\sqrt{G}$]{
    \includegraphics[angle=0,width=0.45\textwidth,height=!,clip]{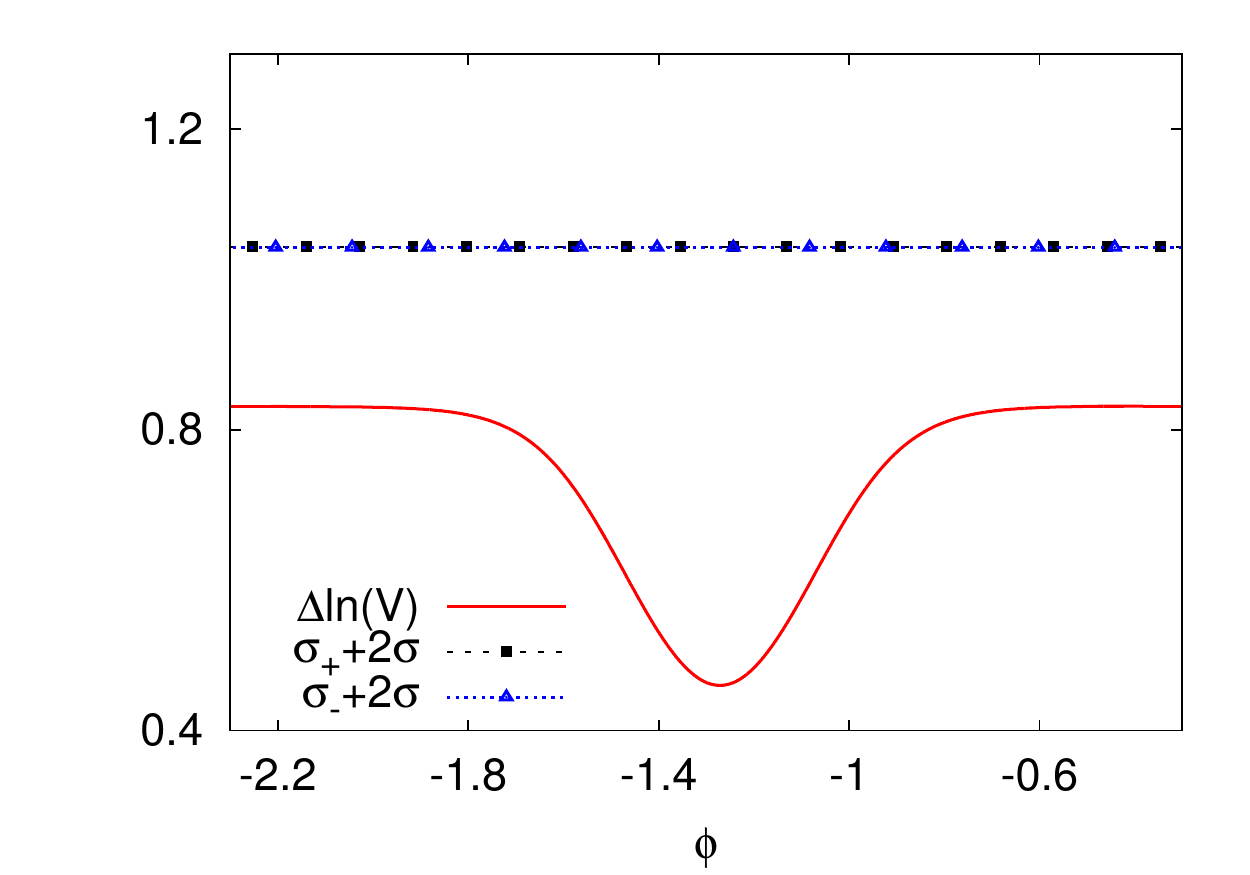}
    \label{f:triangleavg50}
  }
  \caption{The triangle inequality is shown to be valid for multipeaked-1 states with
    $\omega^*=1000 \sqrt{G}\,$ and $\eta=2\times10^{-4}$ (panel(a)); and with $\omega^*=50 \sqrt{G}\,$
    and $\eta=2\times10^{-2}$ (panel(b)). In both cases $\delta k=2$.
   It is evident that $\sigma_\pm+2\sigma$ (shown by horizontal curves) remains greater than $\Delta \ln(\widehat V)$ (shown by solid red curve)
    throughout the evolution. Since, $\sigma$ remains constant 
    throughout the evolution and the volume dispersions on both sides are
    practically the same in both cases, the horizontal curves for $\sigma_++2\sigma$ and 
    $\sigma_-+2\sigma$ overlap.}
  \label{f:traingleavgvphi}
\end{figure}

Let us now examine the validity of the triangle inequalities. \fref{f:traingleavgvphi}  
shows the dispersion $\Delta \ln(\widehat V)$ and the quantity 
$\sigma_\pm+2\sigma$ (computed using (\ref{eq:triangle2})), plotted for 
$\omega^*=1000\,\sqrt{G}$ and $\omega^*=50\,\sqrt{G}$. In this figure $2\sigma$ is a constant of motion 
and the asymptotic values of the volume dispersions on both sides of the bounce are 
the same. Therefore, the horizontal curves for $\sigma_++2\sigma$ and 
$\sigma_-+2\sigma$ overlap.
It is apparent from the figure that $\sigma_\pm+2\sigma$ remains larger than 
$\Delta \ln(\widehat V)$ throughout the evolution, which implies that the triangle inequality 
given in \eref{eq:triangle} is satisfied. A similar analysis shows that the
triangle inequality in \eref{eq:corichieps} is satisfied both for 
the case $\omega^*=1000\,\sqrt{G}$ and $\omega^*=50\,\sqrt{G}$. Note that for
both of these cases, the state is not sharply peaked. In fact, for the case $\omega^* = 50 \sqrt{G}$ the state is highly quantum. Yet, the stronger form of triangle inequality (\ref{eq:corichieps}) is satisfied. The reason for this lies in the fact that 
 the underlying construction of states is based 
on  ``method-3'' in Refs. \cite{aps3,dgs2} (see eq.(\ref{eq:sqstate})) due to which the difference in relative fluctuations in volume at large $|\phi|$ much earlier and after the bounce remains much smaller than unity for the values of parameters considered for these multipeaked-1 states 
(Table \ref{tab:multirelative}).
This is the primary reason why states, which have little in common with sharply
peaked states (in particular the one with $\omega^*=50\,\sqrt{G}$), satisfy
the stronger triangle inequality.\footnote{It can be expected that if the 
same initial state was constructed in a  different way, say by using ``method-2'' of Ref. \cite{aps3,dgs2}, where  
Wheeler-DeWitt eigenfunctions are not multiplied by a phase factor $e^{-i \alpha}$ as in ``method-3'' or the parameter $\eta$ was chosen to be complex as in the case of squeezed states, the inequality (\ref{eq:corichieps}) may not be satisfied.} This observation suggests that it is difficult to associate (\ref{eq:corichieps}) with the semi-classicality of the 
state. As we see here, it is possible to construct highly quantum states which satisfy the inequality (\ref{eq:corichieps}).

\begin{table}
\caption{%Relative dispersions for multipeaked states. 
Relative dispersion in the field momentum ($\Sigma$) and the asymptotic relative volume dispersions ($\Sigma_\pm$) for multipeaked-1 and 2 states used to evaluate the triangle inequality given in \eref{eq:corichieps}. This inequality ($\mathcal E<1$) is satisfied in all cases considered.
}
\begin{tabular}{cccccc}
\hline
 $\omega^*$ & $\eta$  & $\Sigma_+$  &$\Sigma_-$  & $\Sigma$ & $\mathcal E$ \\
\hline
%\hline
 \textbf{Multipeaked-1}        & \text{}          &\text{}   & \text{}     & \text{} & \text{}\\
 1000     & 0.0002               & 0.132983          &0.132982   & 0.1052    & $0.000005$\\
 50       & 0.02                    & 1.15910            &1.15813    & 0.2104     & 0.002068    \\
 \hline
 \textbf{Multipeaked-2}       & \text{}                  &\text{}   & \text{}      & \text{} & \text{}\\
 1000     & 0.0001               & 0.221726          & 0.221724   & 0.1388    & 0.000007  \\
 200      & 0.0025                & 1.17373           &1.17376    & 0.1389     & 0.000108    \\
   \hline
   \label{tab:multirelative}
\end{tabular}
\end{table}

\begin{figure}[tbh!]
\subfigure[]
   {
    \includegraphics[angle=0,width=0.62\textwidth,height=!,clip]{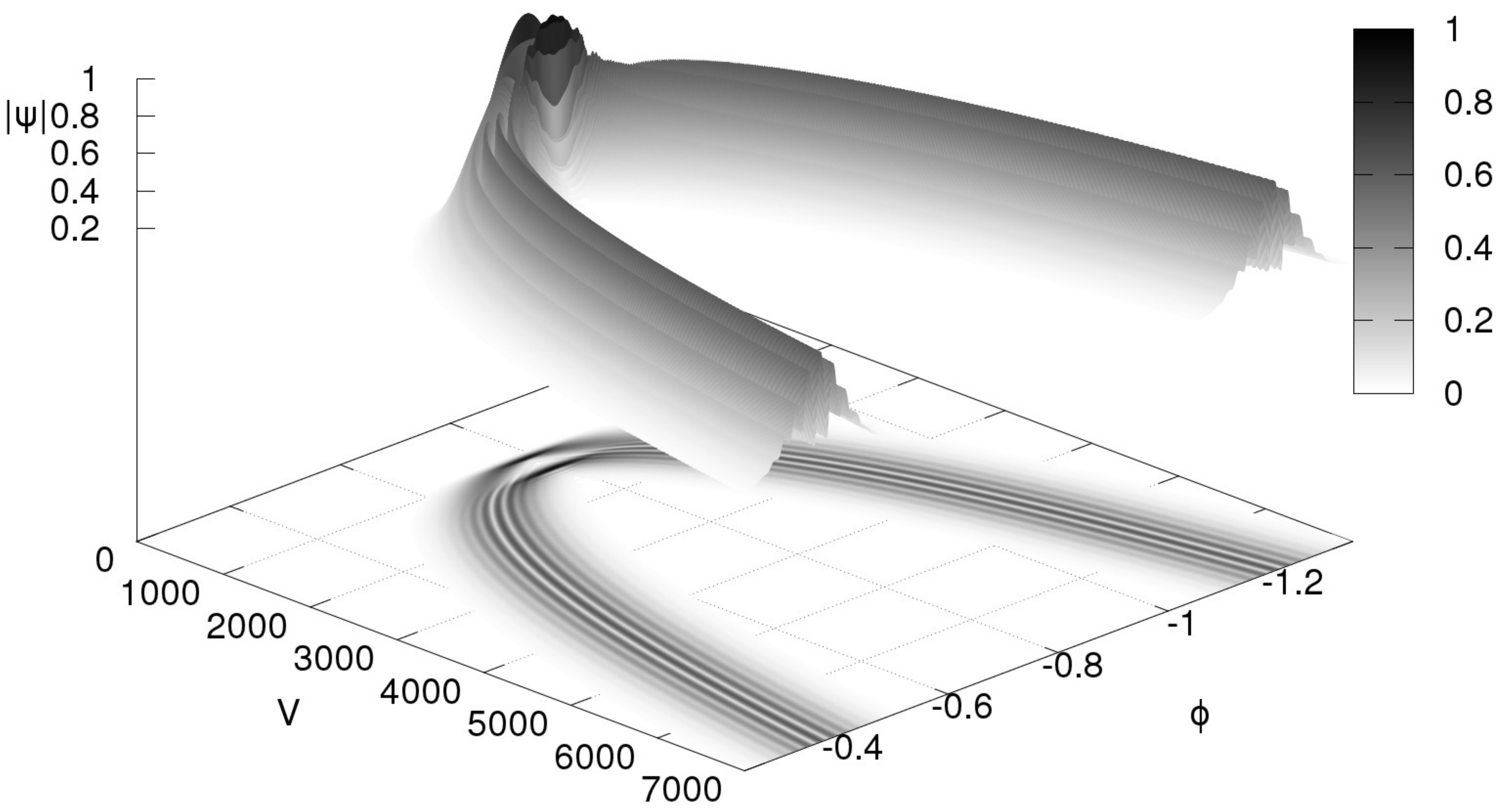}
  \label{f:diener_3Da}
  }
\subfigure[]
  {
    \includegraphics[angle=0,width=0.35\textwidth,height=!,clip]{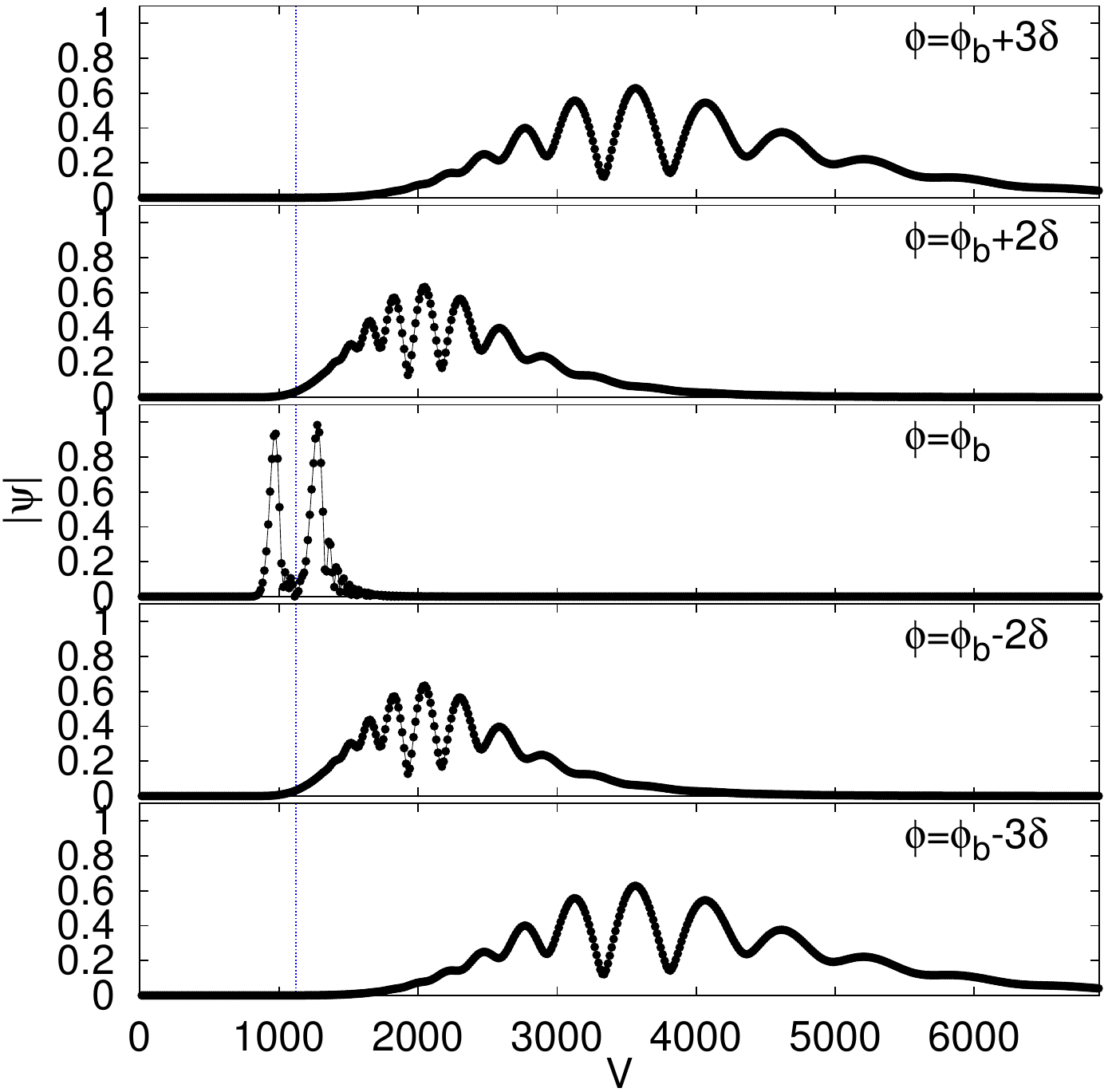}
  \label{f:diener_3Db}
  }
\caption{Evolution of a multipeaked-2 state with $\omega^*=1000\,\sqrt{G}$,
  $\eta=10^{-4}$ and $\delta k=2$. Panel~\subref{f:diener_3Da}: $|\Psi|$
  including the projection onto the $V-\phi$
  plane. Panel~\subref{f:diener_3Db}: $|\Psi|$ at different values of
  $\phi$ close to the bounce, as indicated in the figure, where
  $\phi_{\rm b}=-0.7835$ and $\delta=0.1$.  The blue dotted line indicates the bounce
  volume, $V_{\rm b}=1121.1\,\Vpl$.
}
\label{f:diener_3D}
\end{figure}

%%%%%%%%%%%%%%%%%%%%%%%%%%%%%%%%%%%%%%%%%%%%%%%%%% 
\subsection{Multipeaked-2 states} \label{s:multi2}
\begin{figure}[tbh!]
\subfigure[]
   {
    \includegraphics[angle=0,width=0.62\textwidth,height=!,clip]{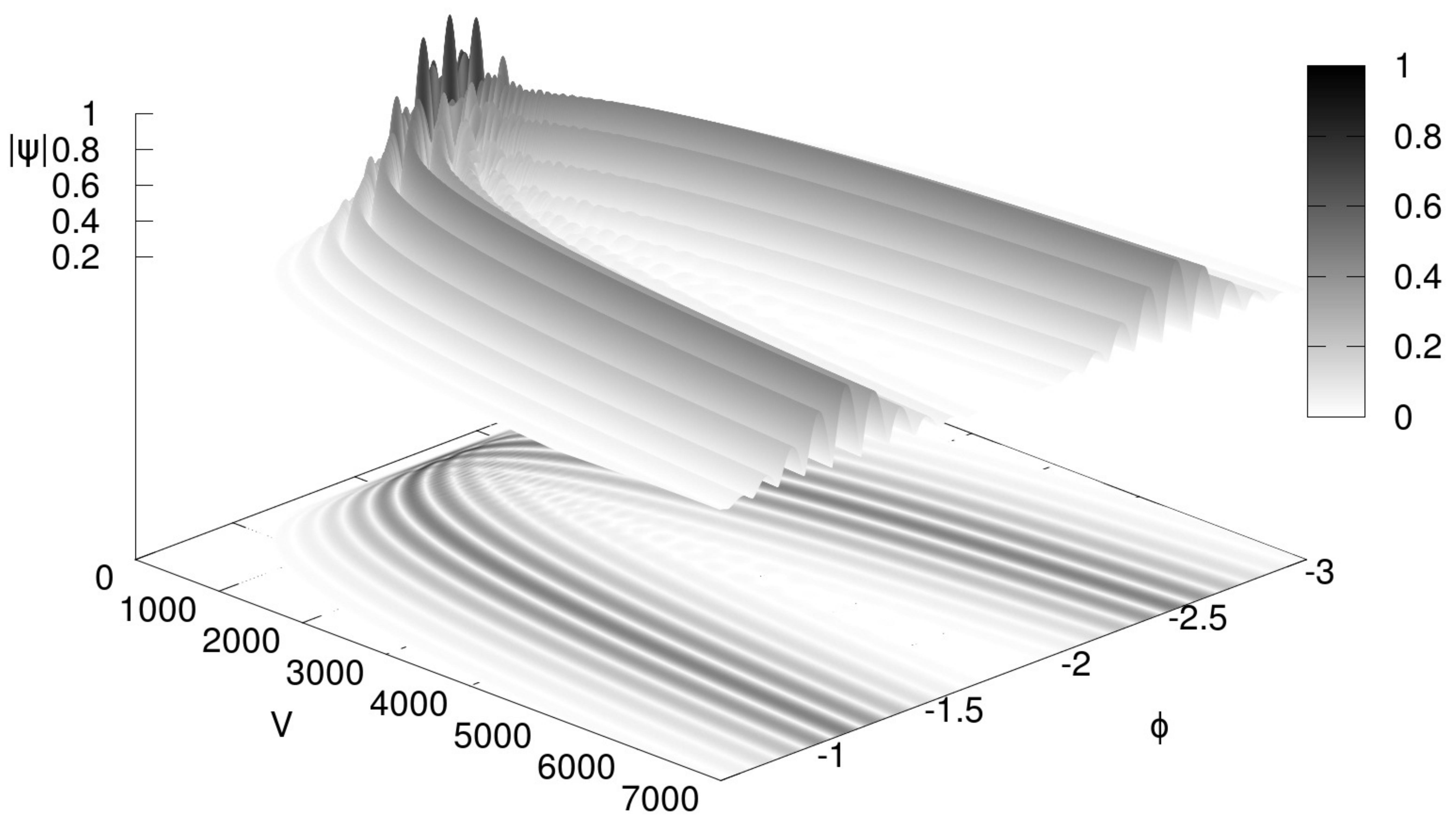}
  \label{f:diener2_3Da}
  }
\subfigure[]
  {
    \includegraphics[angle=0,width=0.35\textwidth,height=!,clip]{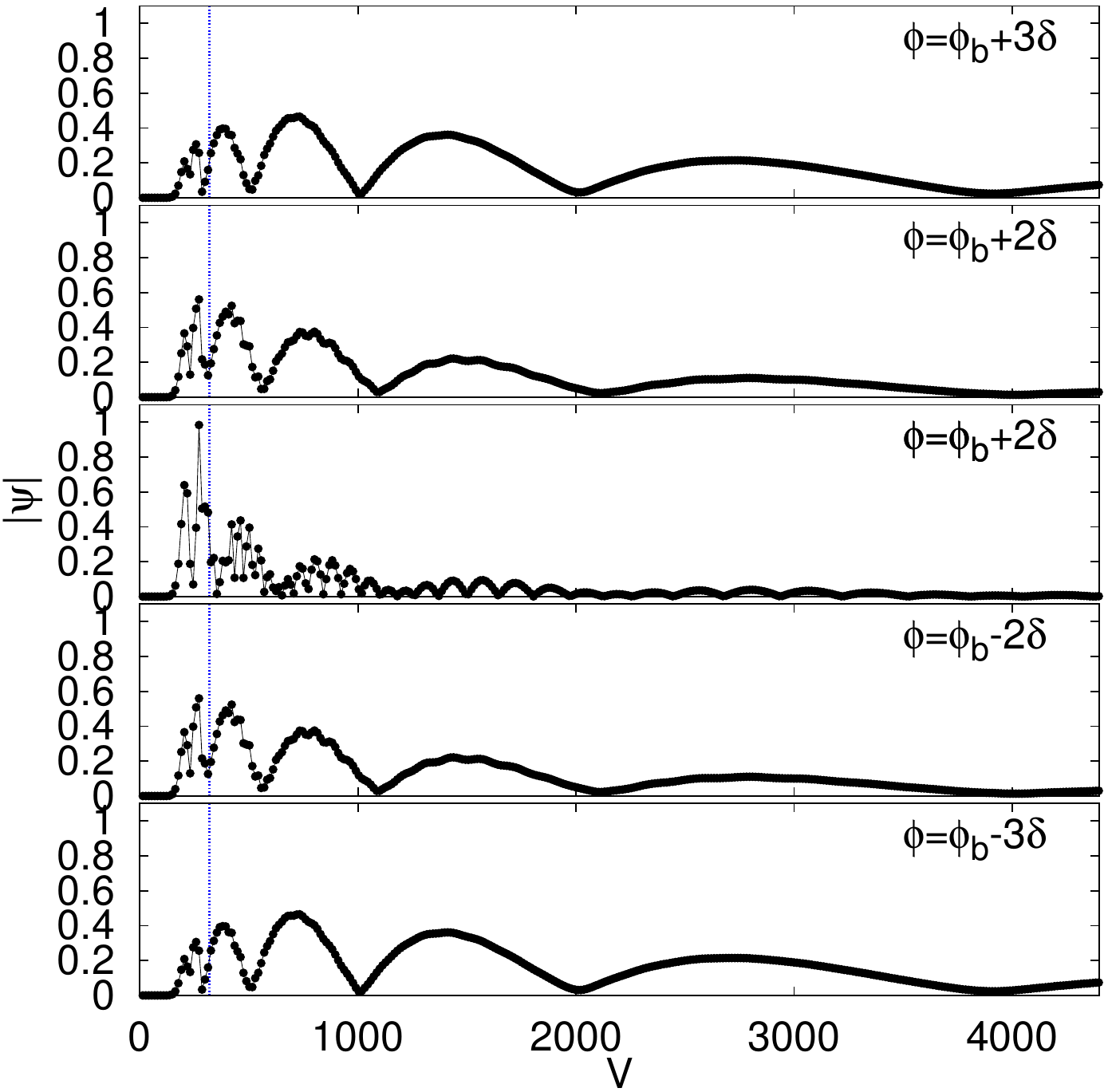}
  \label{f:diener2_3Db}
  }
\caption{Evolution of a multipeaked-2 state with $\omega^*=200\,\sqrt{G}$,
  $\eta=2.5\times10^{-3}$ and $\delta k=2$. Panel~\subref{f:diener_3Da}: $|\Psi|$
  including the projection onto the $V-\phi$
  plane. Panel~\subref{f:diener_3Db}: $|\Psi|$ at different values of
  $\phi$ close to the bounce, as indicated in the figure, where
  $\phi_{\rm b}=-1.796$ and $\delta=0.1$.  The blue dotted line indicates the bounce
  volume, $V_{\rm b}=318.85\,\Vpl$.
}
\label{f:diener2_3D}
\end{figure}

%%%%%%%%%%%%%%%%%%%%%%%%%%%%%%%%%%%%%%%%%%%%%%%%%%
In order to probe the robustness of the quantum bounce even further, we 
consider yet another type of initial state with a non-Gaussian waveform. We call these
multipeaked-2 states, which in comparison to multipeaked-1 states have many more peaks and are 
highly non-Gaussian. 
We show the   wavefunction  evolution for two cases of multipeaked-2 states, one with large and one 
with small $\omega^*$: Figure~\ref{f:diener_3D} corresponds to $\omega^*=1000\,\sqrt{G}$ and $\eta=10^{-4}$ and Figure~\ref{f:diener2_3D} corresponds to $\omega^*=200\,\sqrt{G}$ and 
$\eta=2.5\times10^{-3}$. In both  cases $\delta k=2$. 
The initial data is chosen so that the energy density is much smaller than the Planck 
density and  the expectation value of the volume variable is very large compared to the Planck volume. 
It is evident that, despite the highly non-Gaussian features of the initial 
state, the  wavefunction undergoes a non-singular evolution and 
a quantum bounce takes place. 
It is noteworthy that, like in the case of the Gaussian states, the shape of the 
wavepacket remains the same on the two sides of the bounce, although the 
behavior in the vicinity of the bounce is quite different from the initial state.
This behavior is more noticeable in the smaller $\omega^*$ case, for which 
the bounce occurs at smaller volume.

\begin{figure}[tbh!]
\subfigure[\, $\omega^*=1000\,\sqrt{G}$]
{
    \includegraphics[angle=0,width=0.45\textwidth,height=!,clip]{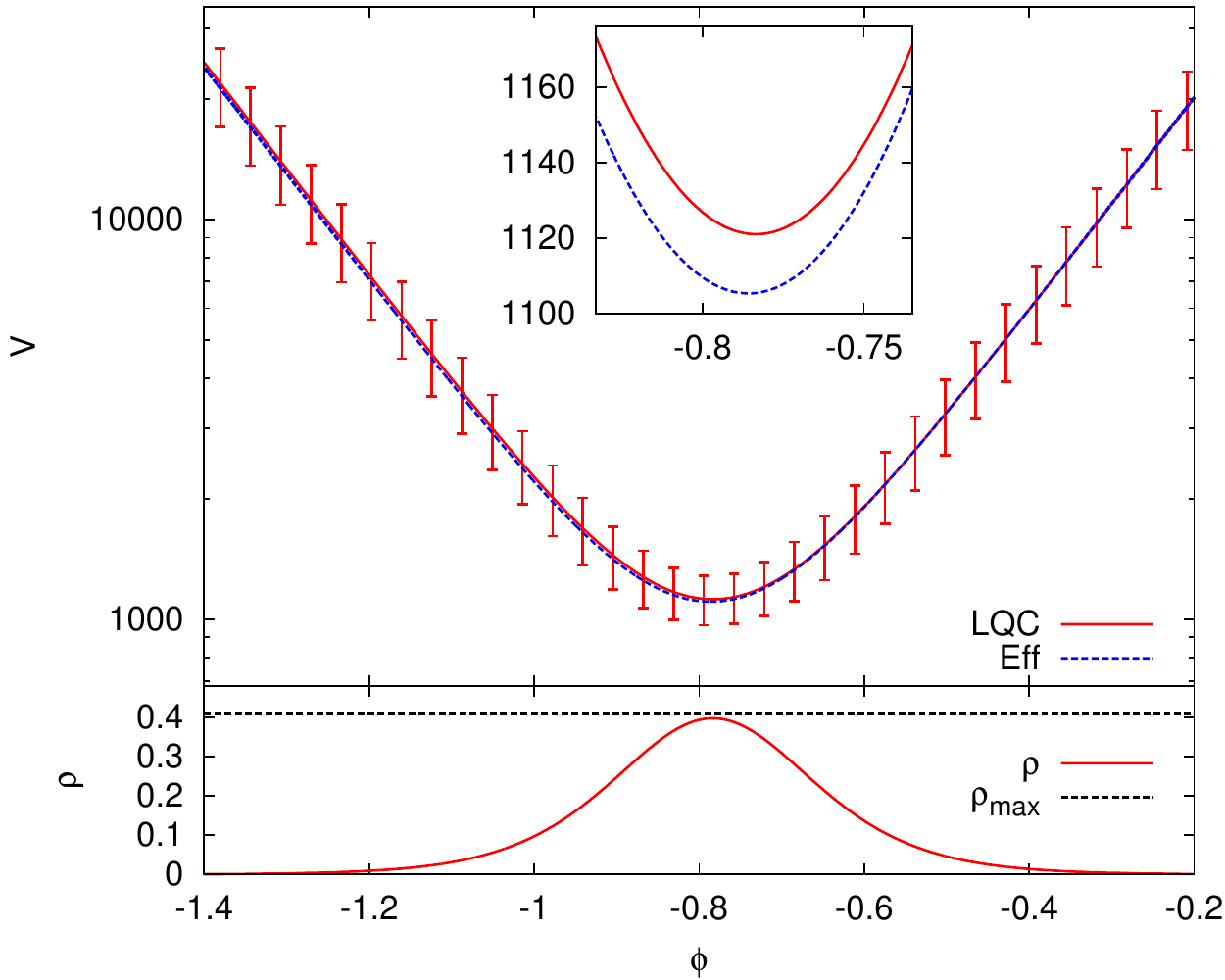}
  \label{f:multi1000}
}
\subfigure[\, $\omega^*=200\,\sqrt{G}$]
{
    \includegraphics[angle=0,width=0.45\textwidth,height=!,clip]{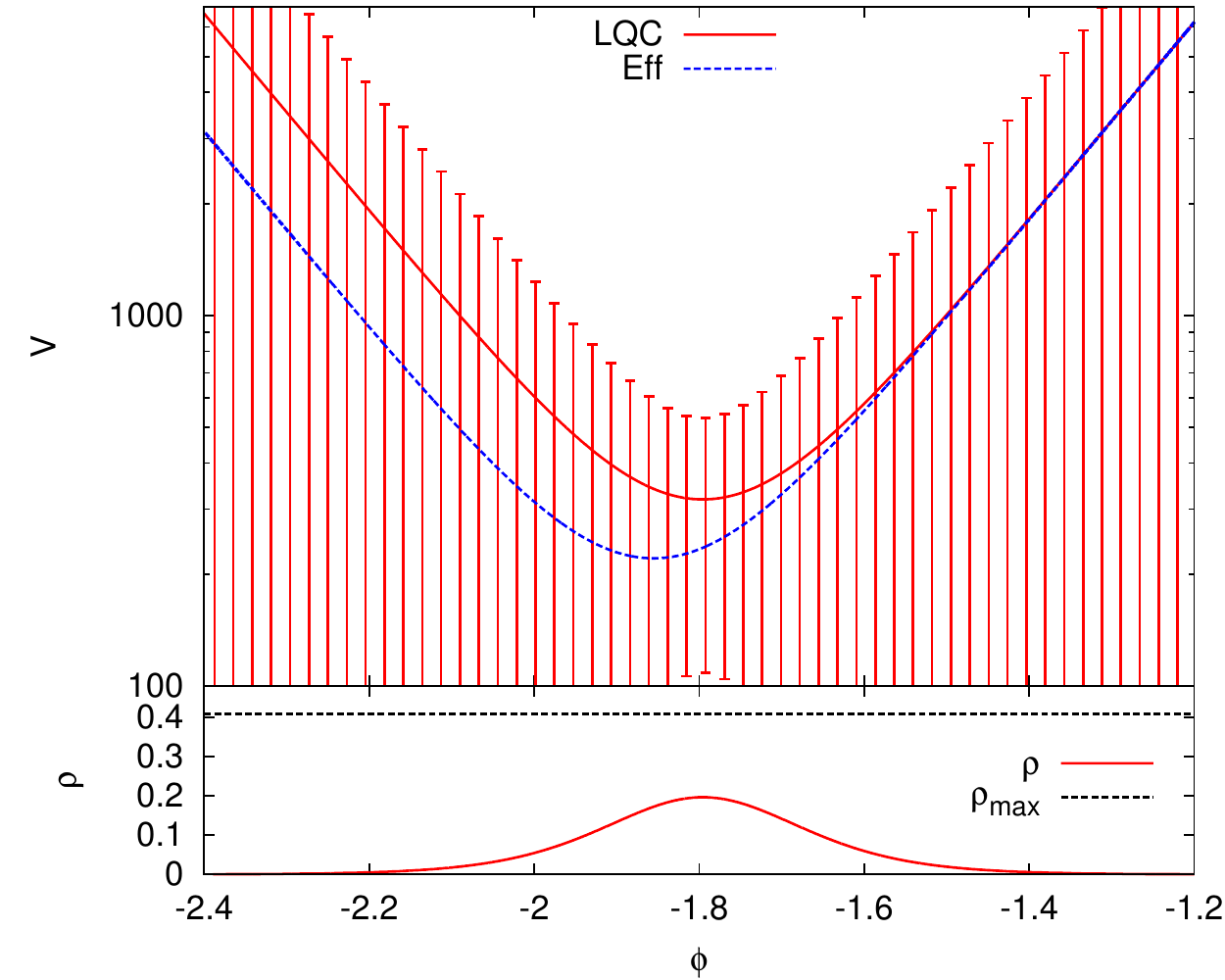}
  \label{f:multi200}
}
\caption{Trajectories and energy densities for  multipeaked-2 states with $\delta
  k=2$ and different values of $\omega^*$ and $\eta$. Panel (a):
  $\omega^*=1000\,\sqrt{G}$ and $\eta=1\times 10^{-4}$.
              Panel (b): $\omega^*=200\,\sqrt{G}$ and $\eta=2.5\times 10^{-3}$. 
              The solid (red) curves show the LQC trajectories, with the error bars representing the 
              volume dispersion,  and the dashed 
              (blue) curves showing the corresponding effective trajectory.
               }
\label{f:multi}
\end{figure}

Let us now compare the trajectories of multi-peaked states in LQC with those of the
effective theory. \fref{f:multi} shows the expectation value of the volume variable as a function 
of $\phi$ for both LQC and the effective theory. 
In the first case, shown in~\fref{f:multi1000},  
%in which $\omega^*$ is large, 
the initial volume dispersion is small ($\Delta V/V=0.22$) and the effective theory is
a good approximation to the full LQC trajectory. 
On the other hand, in the second case 
%for $\omega^*=200\,\sqrt{G}$ 
the initial volume dispersion  is large ($\Delta V/V=1.17$), and there are significant 
differences between the LQC and the effective trajectory
(\fref{f:multi200}).
The bottom panels in~\fref{f:multi} show the evolution of the energy density~$\rho$.
We can see that in both of the cases $\rho$ remains below the absolute maximum
$\rho_{\rm max}$ in sLQC (indicated with a dashed horizontal line in the
figures). However,  $\rho$ gets very close to $\rho_{\rm max}$ at the bounce for
the state with $\omega^* = 1000 \sqrt{G}$ which has smaller relative dispersion in volume. For the state with $\omega^* = 200 \sqrt{G}$, which has larger 
relative dispersion in volume, the density at the bounce is much smaller than $\rho_{\rm max}$.

\begin{figure}[tbh!]
\subfigure[\, $\omega^*=1000\,\sqrt{G}$]
{
    \includegraphics[angle=0,width=0.45\textwidth,height=!,clip]{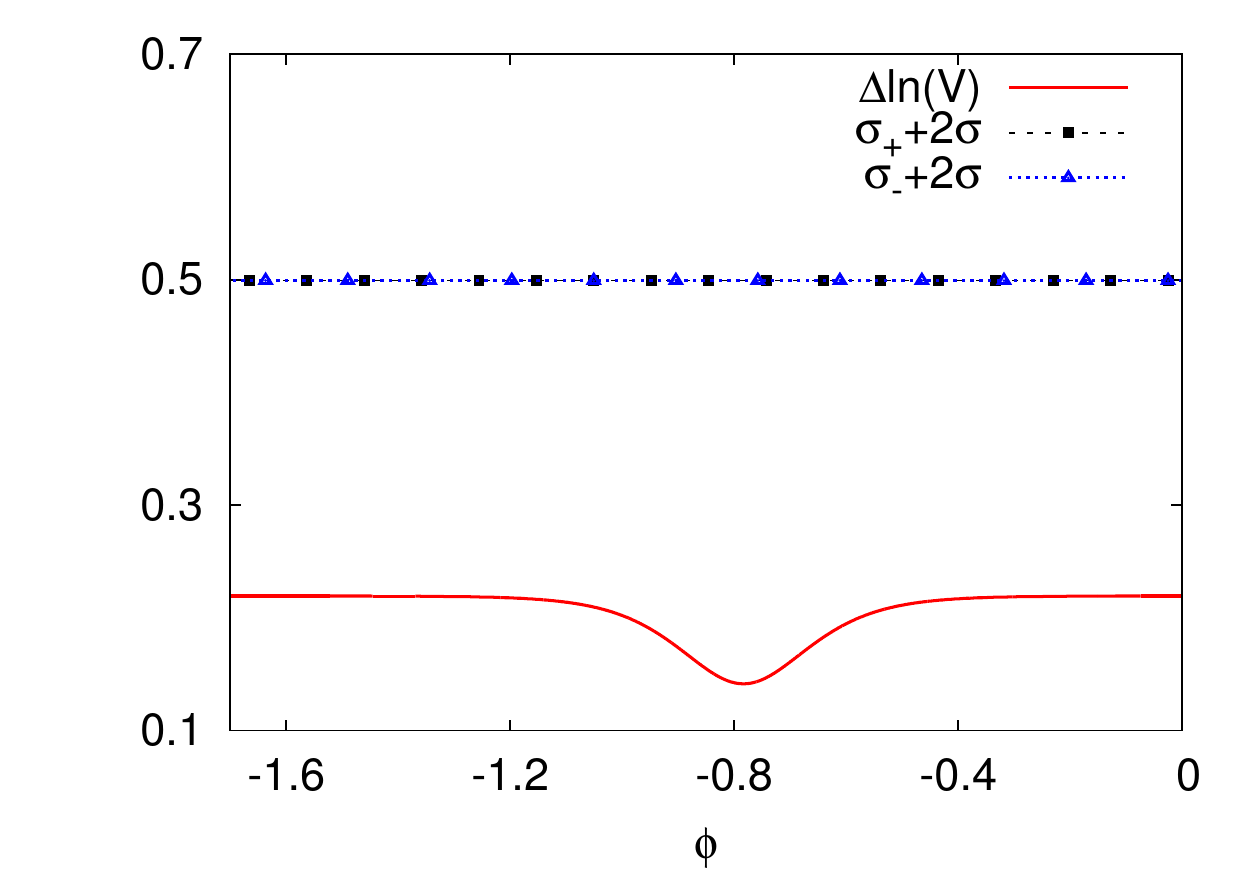}
  \label{f:trianglemulti1000}
}
\subfigure[\, $\omega^*=200\,\sqrt{G}$]
{
    \includegraphics[angle=0,width=0.45\textwidth,height=!,clip]{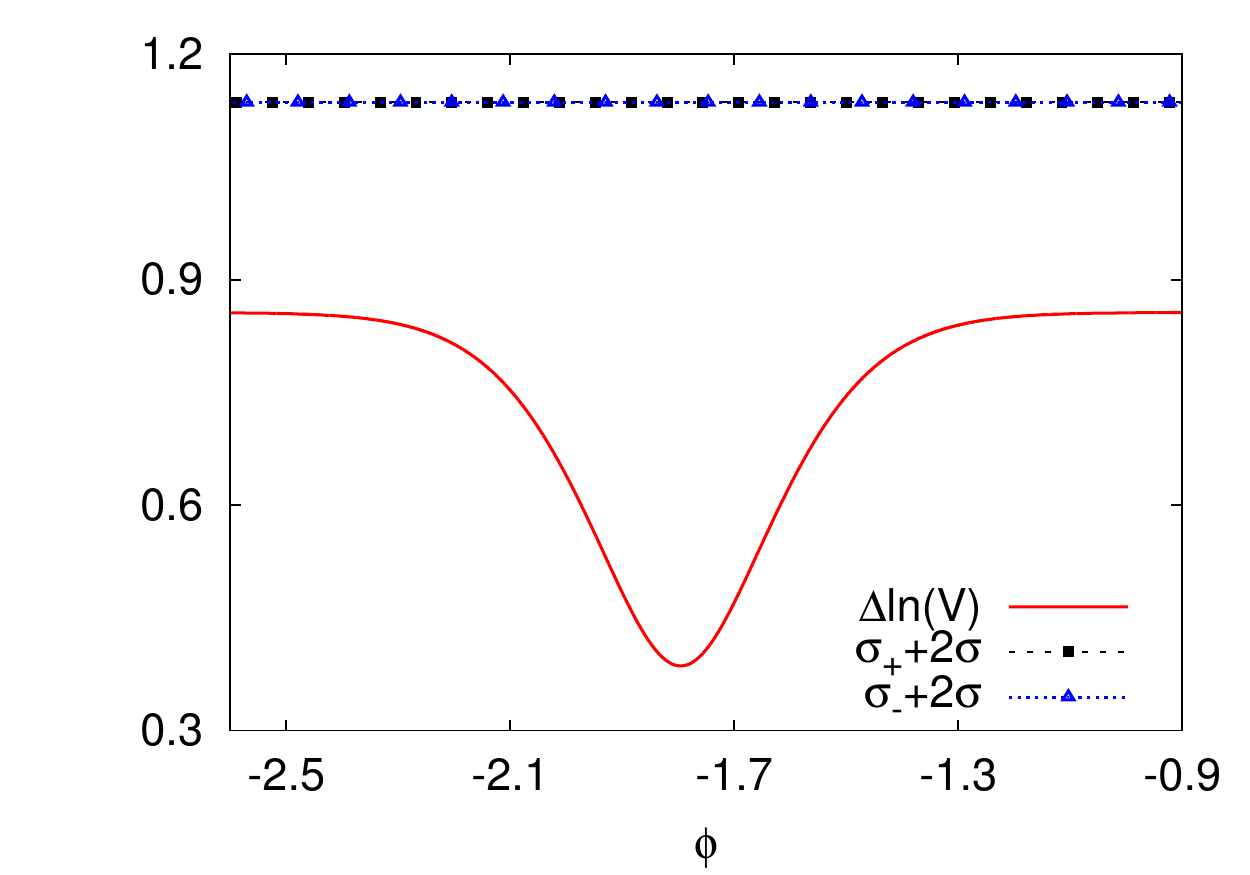}
  \label{f:trianglemulti200}
}
\caption{Validity of the triangle inequality for multipeaked states with 
              $\omega^*=1000\,\sqrt{G}$ and $\eta=1\times10^{-4} $(panel(a)) 
              and with $\omega^*=200\,\sqrt{G}$ and $\eta=2.5\times10^{-3}$ 
	      (panel(b)). In both cases $\delta k=2$.
              The figure clearly demonstrate that $\sigma_\pm+2\sigma$ (shown by horizontal 
              curves) remains greater than $\Delta \ln(\widehat V)$ (shown by 
              the solid red curve) throughout the evolution.
              The asymptotic values of the volume dispersions on both sides of the bounce 
              are practically the same. Therefore, the horizontal curves for 
              $\sigma_++2\sigma$ and  $\sigma_-+2\sigma$ overlap. 
              }
\label{f:trianglemulti}
\end{figure}

We now discuss the validity of
the triangle inequalities for multipeaked-2 states. 
The evolution of the dispersion in volume $\Delta \ln(\widehat V)$ is shown for both
cases in \fref{f:trianglemulti}. It is clear that  $\sigma_\pm+2\sigma$ remains larger than 
$\Delta \ln(\widehat V)$ throughout the evolution in both cases, as predicted by the triangle inequality (\ref{eq:triangle}).  
Similarly to the case of multipeaked-1 states, $2\sigma$ is a constant of motion.  
Asymptotic values of the volume dispersions on both sides of the bounce are 
the same in both cases shown. Therefore, the horizontal curve for $\sigma_++2\sigma$ and 
$\sigma_-+2\sigma$ overlap. 
Another similarity with the multipeaked-1 states is that the difference in the asymptotic values of the relative fluctuations in volume is much smaller than unity for the range of parameters considered here, which, as before, results in 
the satisfaction of the stronger version of the triangle inequality (\ref{eq:corichieps}) for both multipeaked-2 cases (Table \ref{tab:multirelative}). As has been discussed above for the case of multipeaked-1 states, the reason, why highly non-Gaussian multipeaked-2 states satisfy this inequality, is tied to the way they are constructed, i.e.\ by  
using the phase factor $e^{-i \alpha}$ in eq.(\ref{eq:sqstate}) which corresponds to ``method-3'' in Refs. \cite{aps3,dgs2} and the particular choices for parameters of the state. It is important to note that whether (\ref{eq:corichieps}) is satisfied depends on the details of the construction of the state and its parameters, however, all the states irrespective 
of the choice of parameters satisfy the triangle inequality (\ref{eq:triangle}). 

%\clearpage
%%%%%%%%%%%%%%%%%%%%%%%%%%%%%%%%%%%%%%%%%%%%%%%%%%
\section{Discussion}\label{sec:disc}
%%%%%%%%%%%%%%%%%%%%%%%%%%%%%%%%%%%%%%%%%%%%%%%%%%
The goal of our analysis was to understand the quantum evolution of squeezed and
highly non-Gaussian states in LQC. We considered the spatially flat homogeneous
and isotropic spacetime sourced with a massless scalar field in which the
quantum Hamiltonian constraint takes the form of a Klein-Gordon 
equation, with the scalar field playing the role of internal time and where the
spatial Laplacian is a quantum difference operator with uniform 
discreteness in volume. This model has been extensively studied using
rigorous analytical and numerical techniques. 
The first results for this model, which established singularity resolution 
and the quantum bounce, were obtained using numerical simulations with 
Gaussian states initially sharply peaked on a classical 
trajectory \cite{aps2,aps3}. 
These studies brought out many important features of the new physics at the
Planck scale. At curvatures very small compared to the Planck scale,
the  classical GR trajectory is an excellent approximation to the
quantum dynamics. However, near the Planck scale, there are significant
departures
between the quantum and the classical evolution, and quantum geometric effects
provide a non-singular bridge between the classically disjoint expanding 
and contracting branches. The quantum bounce for sharply peaked states occurs
at energy density 
$\rho_{\rm max} \approx 0.409 \rho_{\rm Pl}$ -- the universal 
maximum for the energy density derived using an exactly solvable model in
LQC~\cite{acs}. It was found that the sharply peaked initial states remain
sharply peaked throughout the evolution and the relative fluctuations of
Dirac observables is strongly constrained across the 
bounce~\cite{recall,kp,montoya_corichi1,montoya_corichi2}. Quantum evolution
of sharply peaked states further established that 
the evolution trajectory described by the expectation values of 
the physical observables is in excellent agreement with the one derived from
the effective Hamiltonian. In recent years, this analysis has been
generalized to a 
variety of models and numerical simulations of loop quantum universes with
sharply peaked states for different choices of matter have been performed. 
However, an important issue for the robustness of the new physics in LQC is
to understand the way quantum evolution is affected for states which are
widely spread, are squeezed or are highly non-Gaussian. 
Although these states do not necessarily correspond to large classical universes at late times, 
they do belong to the physical Hilbert space. Therefore, to test the robustness of 
singularity resolution and the occurrence of the quantum bounce it is pertinent to ask the 
following questions: Do states that do
not have any well defined peakedness properties at late times undergo a
quantum bounce? What happens to the growth of fluctuations through the bounce?
And finally, is the effective dynamics derived from the effective 
Hamiltonian (\ref{eq:heff}), still a good approximation to the quantum
evolution for highly squeezed and non-Gaussian states? 
%in the effective 
%description of LQC. 

Numerical simulations with states which are widely spread or are highly
squeezed or non-Guassian is computationally very challenging due to the
tremendously large integration domains involved and the associated stability
constraints. To overcome these challenges, a numerical scheme called Chimera
was recently developed. Its construction is based on an important property of
the loop quantum difference equation -- at large volumes, it is extremely
well approximated by the differential Wheeler-DeWitt equation. Using a hybrid
spatial grid, a sufficiently large inner grid where the evolution is governed
by the quantum difference equation, and a carefully chosen outer grid where the
Wheeler-DeWitt equation is solved, a significant reduction in computational
time and a large increase in efficiency can be achieved. 
The Chimera scheme was 
recently used to study the evolution of widely 
spread Gaussian states \cite{dgs2}. It was found that while the qualitative
features of the quantum bounce remain unaltered, various quantitative details
of the quantum bounce 
depend on the choice of the initial state and their parameters. For these
states it was found that, depending on the dispersion, there may be significant
deviations between the LQC and the corresponding effective trajectory.
Additionally, the energy density at 
the bounce can be much smaller than $\rho_{\rm max}$. As a general result, it 
was found that the effective theory always overestimates the energy density and 
underestimates the volume at the bounce.

%% In this work we used the Chimera scheme to study numerical evolutions of a
%% wide range of states which depart, in some cases greatly, from Gaussianity.
%% Moreover, some of these states have very large dispersion and no well defined
%% peakedness properties. The main result of our analysis is that the quantum
%% bounce occurs for all states, independent of the choice of parameters which
%% control squeezedness and non-Gaussianity. 
%% During the entire evolution the expectation value of the volume observable
%% remains non-zero and the energy density remains bounded irrespective of the
%% initial state.
%% Moreover, despite the non-Gaussian features, the shape of the initial
%% wavefunction is recovered when the energy density after the bounce becomes
%% comparable to the value in the initial data. Our results show that the
%% triangle inequalities on the relative fluctuations are satisfied across the
%% bounce for a wide variety of squeezed and multipeaked states. 
%% The triangle inequality derived in Ref. \cite{kp} is satisfied for all the
%% states considered here, including the ones 
%% with multiple peaks. A version of this triangle inequality obtained under
%% stronger assumptions for sharply peaked states in 
%% Ref. \cite{montoya_corichi1} is found to be violated for highly squeezed 
%% states. 
%% These results demonstrate for the first time that the quantum bounce is a
%% property of far more general states than the Gaussian states and quantum
%% fluctuations remain tightly constrained throughout the evolution, even for
%% highly non-Gaussian states.
%zzz
In this work we used the Chimera scheme to study numerical evolutions of a
wide range of states which depart, in some cases greatly, from Gaussianity.
Moreover, some of these states have very large dispersion and no well defined
peakedness properties. One of the main results of our analysis is that the quantum
bounce occurs for all states, replacing the classical singularity,
independently of the choice of parameters which 
control squeezedness and non-Gaussianity. 
During the entire evolution the expectation value of the volume observable
remains non-zero and the energy density remains bounded irrespective of the
initial state.
Moreover, despite the non-Gaussian features, the shape of the initial
wavefunction is recovered when the energy density after the bounce becomes
comparable to the value in the initial data. Our results show that the
triangle inequalities on the relative fluctuations are satisfied across the
bounce for a wide variety of squeezed and multipeaked states. 
The triangle inequality derived in Ref. \cite{kp} is satisfied for all the
states considered here, including the ones 
with multiple peaks. A version of this triangle inequality obtained under
stronger assumptions for sharply peaked states in 
Ref. \cite{montoya_corichi1} is found to be violated for highly squeezed 
states. 
These results demonstrate for the first time that the quantum bounce is a
property of far more general states than the Gaussian states and quantum
fluctuations remain tightly constrained throughout the evolution, even for
highly non-Gaussian states.

We find that the effective theory captures the main qualitative features of
the quantum bounce quite well, and is a good approximation to LQC for states with
small dispersion, even for states which significantly departs from Gaussianity. 
However, for states with large dispersion 
there are significant differences between an LQC trajectory and the
corresponding effective trajectory. An important contrast between the 
effective theory and LQC is that, in the effective theory, the energy density
at the bounce is always equal to the maximum upper bound, whereas, in the LQC
evolution, the energy density at the bounce can be very small for highly
squeezed and states with no well defined peakedness properties. In agreement
with the recent results obtained in the analysis of widely spread Gaussian
states \cite{dgs2}, we find that the effective theory always overestimates 
the energy density and underestimates the volume at the bounce. This issue
was analyzed in detail using squeezed states by studying the variation of
the energy density at the bounce with the change in spread in the field
momentum, which is directly related to the parameter $\eta$.
We also found excellent agreement between our numerical results and the 
analytical calculations in Ref.~\cite{montoya_corichi2}, which was performed
in the context of sLQC for states which are highly squeezed. 
In summary, these results on one hand indicate the conditions under which the effective 
description may not capture the underlying quantum evolution reasonably well, 
and on the other hand show a synergy between
the numerical simulations of highly non-Gaussian states and the results from sLQC.

Based on the results obtained in this paper, it can be concluded that the
quantum bounce is a robust feature of the spatially flat homogeneous and isotropic LQC with a massless scalar field, irrespective of the choice of the
initial states. The effective theory, although showing deviations for states
with large dispersions, is still able to capture the main qualitative features
of the quantum bounce. The validity of the triangle inequalities
demonstrates that the growth of the dispersion of the states across the
bounce is tightly constrained, which presents very strong numerical evidences
in support of cosmic recall~\cite{recall} and discourages the speculations
related to unbounded growth in the fluctuations of the state. Finally, the
results from our numerical simulations turn out to be in complete synergy with
the predictions of sLQC.

\acknowledgments

We thank Ivan Agullo, Alejandro Corichi, Jorge Pullin and Edward Wilson-Ewing  
for comments.
This work is supported by a grant from John Templeton Foundation and by NSF grant 
PHYS1068743. The opinions expressed in this publication are those of authors and do not 
necessarily reflect the views of John Templeton Foundation. BG  acknowledges  
support from the Coates Scholar Research Award and the Dissertation Year Fellowship of the 
Louisiana State University. This material is based upon work supported by HPC@LSU computing resources.

\section*{References}
%\bibliographystyle{hunsrt}
%\bibliographystyle{h-physrev}
%\bibliographystyle{kp}
%\bibliography{references/lqc}

\begin{thebibliography}{38}
\expandafter\ifx\csname natexlab\endcsname\relax\def\natexlab#1{#1}\fi

\bibitem{as1}
A.~Ashtekar and P.~Singh, ``{Loop Quantum Cosmology: A Status Report}'', {\em
  Class.Quant.Grav.} {\bfseries 28} (2011) 213001,
 \href{http://xxx.lanl.gov/abs/1108.0893}{ arXiv:1108.0893}.
%%CITATION = ARXIV:1108.0893;%%.

\bibitem{aps1}
A.~Ashtekar, T.~Pawlowski, and P.~Singh, ``{Quantum nature of the big bang}'',
  {\em Phys.Rev.Lett.} {\bfseries 96} (2006){\natexlab{a}} 141301,
 \href{http://xxx.lanl.gov/abs/gr-qc/0602086}{ arXiv:gr-qc/0602086}.
%%CITATION = GR-QC/0602086;%%.

\bibitem{aps2}
A.~Ashtekar, T.~Pawlowski, and P.~Singh, ``{Quantum Nature of the Big Bang: An
  Analytical and Numerical Investigation. I.}'', {\em Phys.Rev.} {\bfseries
  D73} (2006){\natexlab{b}} 124038,
 \href{http://xxx.lanl.gov/abs/gr-qc/0604013}{ arXiv:gr-qc/0604013}.
%%CITATION = GR-QC/0604013;%%.

\bibitem{aps3}
A.~Ashtekar, T.~Pawlowski, and P.~Singh, ``{Quantum Nature of the Big Bang:
  Improved dynamics}'', {\em Phys.Rev.} {\bfseries D74} (2006){\natexlab{c}}
  084003,
 \href{http://xxx.lanl.gov/abs/gr-qc/0607039}{ arXiv:gr-qc/0607039}.
%%CITATION = GR-QC/0607039;%%.

\bibitem{szulc_open}
L.~Szulc, ``{Open FRW model in Loop Quantum Cosmology}'', {\em
  Class.Quant.Grav.} {\bfseries 24} (2007) 6191--6200,
 \href{http://xxx.lanl.gov/abs/0707.1816}{ arXiv:0707.1816}.
%%CITATION = ARXIV:0707.1816;%%.

\bibitem{warsaw_flat}
W.~Kaminski and J.~Lewandowski, ``{The Flat FRW model in LQC: The
  Self-adjointness}'', {\em Class.Quant.Grav.} {\bfseries 25} (2008) 035001,
 \href{http://xxx.lanl.gov/abs/0709.3120}{ arXiv:0709.3120}.
%%CITATION = ARXIV:0709.3120;%%.

\bibitem{kv}
K.~Vandersloot, ``{Loop quantum cosmology and the k = - 1 RW model}'', {\em
  Phys.Rev.} {\bfseries D75} (2007) 023523,
 \href{http://xxx.lanl.gov/abs/gr-qc/0612070}{ arXiv:gr-qc/0612070}.
%%CITATION = GR-QC/0612070;%%.

\bibitem{warsaw_closed}
L.~Szulc, W.~Kaminski, and J.~Lewandowski, ``{Closed FRW model in Loop Quantum
  Cosmology}'', {\em Class.Quant.Grav.} {\bfseries 24} (2007) 2621--2636,
 \href{http://xxx.lanl.gov/abs/gr-qc/0612101}{ arXiv:gr-qc/0612101}.
%%CITATION = GR-QC/0612101;%%.

\bibitem{bp}
E.~Bentivegna and T.~Pawlowski, ``{Anti-deSitter universe dynamics in LQC}'',
  {\em Phys.Rev.} {\bfseries D77} (2008) 124025,
 \href{http://xxx.lanl.gov/abs/0803.4446}{ arXiv:0803.4446}.
%%CITATION = ARXIV:0803.4446;%%.

\bibitem{kp1}
W.~Kaminski and T.~Pawlowski, ``{The LQC evolution operator of FRW universe
  with positive cosmological constant}'', {\em Phys.Rev.} {\bfseries D81}
  (2010) 024014,
 \href{http://xxx.lanl.gov/abs/0912.0162}{ arXiv:0912.0162}.
%%CITATION = ARXIV:0912.0162;%%.

\bibitem{ap}
T.~Pawlowski and A.~Ashtekar, ``{Positive cosmological constant in loop quantum
  cosmology}'', {\em Phys.Rev.} {\bfseries D85} (2012) 064001,
 \href{http://xxx.lanl.gov/abs/1112.0360}{ arXiv:1112.0360}.
%%CITATION = ARXIV:1112.0360;%%.

\bibitem{rad}
T.~Pawlowski, R.~Pierini, and E.~Wilson-Ewing, ``{Loop quantum cosmology of a
  radiation-dominated flat FLRW universe}'',
 \href{http://xxx.lanl.gov/abs/1404.4036}{ arXiv:1404.4036}.
%%CITATION = ARXIV:1404.4036;%%.

\bibitem{aps4}
A.~Ashtekar, T.~Pawlowski, and P.~Singh, ``{Pre-inflationary dynamics in loop
  quantum cosmology}'', {\em {(To appear)}}.

\bibitem{cyclic}
P.~Diener, B.~Gupt, M.~Megevand, and P.~Singh {\em {(To appear)}}.

\bibitem{awe2}
A.~Ashtekar and E.~Wilson-Ewing, ``{Loop quantum cosmology of Bianchi I
  models}'', {\em Phys.Rev.} {\bfseries D79} (2009){\natexlab{a}} 083535,
 \href{http://xxx.lanl.gov/abs/0903.3397}{ arXiv:0903.3397}.
%%CITATION = ARXIV:0903.3397;%%.

\bibitem{awe3}
A.~Ashtekar and E.~Wilson-Ewing, ``{Loop quantum cosmology of Bianchi type II
  models}'', {\em Phys.Rev.} {\bfseries D80} (2009){\natexlab{b}} 123532,
 \href{http://xxx.lanl.gov/abs/0910.1278}{ arXiv:0910.1278}.
%%CITATION = ARXIV:0910.1278;%%.

\bibitem{we1}
E.~Wilson-Ewing, ``{Loop quantum cosmology of Bianchi type IX models}'', {\em
  Phys.Rev.} {\bfseries D82} (2010) 043508,
 \href{http://xxx.lanl.gov/abs/1005.5565}{ arXiv:1005.5565}.
%%CITATION = ARXIV:1005.5565;%%.

\bibitem{chioub1}
D.-W. Chiou, ``{Loop Quantum Cosmology in Bianchi Type I Models: Analytical
  Investigation}'', {\em Phys.Rev.} {\bfseries D75} (2007) 024029,
 \href{http://xxx.lanl.gov/abs/gr-qc/0609029}{ arXiv:gr-qc/0609029}.
%%CITATION = GR-QC/0609029;%%.

\bibitem{b1madrid1}
M.~Martin-Benito, G.~Mena~Marugan, and T.~Pawlowski, ``{Loop Quantization of
  Vacuum Bianchi I Cosmology}'', {\em Phys.Rev.} {\bfseries D78} (2008) 064008,
 \href{http://xxx.lanl.gov/abs/0804.3157}{ arXiv:0804.3157}.
%%CITATION = ARXIV:0804.3157;%%.

\bibitem{brizuela_cartin_khanna}
D.~Brizuela, D.~Cartin, and G.~Khanna, ``{Numerical techniques in loop quantum
  cosmology}'', {\em SIGMA} {\bfseries 8} (2012) 001,
 \href{http://xxx.lanl.gov/abs/1110.0646}{ arXiv:1110.0646}.
%%CITATION = ARXIV:1110.0646;%%.

\bibitem{ps12}
P.~Singh, ``{nnaNumerical loop quantum cosmology: an overview}'', {\em
  Class.Quant.Grav.} {\bfseries 29} (2012) 244002,
 \href{http://xxx.lanl.gov/abs/1208.5456}{ arXiv:1208.5456}.
%%CITATION = ARXIV:1208.5456;%%.

\bibitem{acs}
A.~Ashtekar, A.~Corichi, and P.~Singh, ``{Robustness of key features of loop
  quantum cosmology}'', {\em Phys.Rev.} {\bfseries D77} (2008) 024046,
 \href{http://xxx.lanl.gov/abs/0710.3565}{ arXiv:0710.3565}.
%%CITATION = ARXIV:0710.3565;%%.

\bibitem{recall}
A.~Corichi and P.~Singh, ``{Quantum bounce and cosmic recall}'', {\em
  Phys.Rev.Lett.} {\bfseries 100} (2008) 161302,
 \href{http://xxx.lanl.gov/abs/0710.4543}{ arXiv:0710.4543}.
%%CITATION = ARXIV:0710.4543;%%.

\bibitem{kp}
W.~Kaminski and T.~Pawlowski, ``{Cosmic recall and the scattering picture of
  Loop Quantum Cosmology}'', {\em Phys.Rev.} {\bfseries D81} (2010) 084027,
 \href{http://xxx.lanl.gov/abs/1001.2663}{ arXiv:1001.2663}.
%%CITATION = ARXIV:1001.2663;%%.

\bibitem{montoya_corichi1}
A.~Corichi and E.~Montoya, ``{On the Semiclassical Limit of Loop Quantum
  Cosmology}'', {\em Int.J.Mod.Phys.} {\bfseries D21} (2012) 1250076,
 \href{http://xxx.lanl.gov/abs/1105.2804}{ arXiv:1105.2804}.
%%CITATION = ARXIV:1105.2804;%%.

\bibitem{montoya_corichi2}
A.~Corichi and E.~Montoya, ``{Coherent semiclassical states for loop quantum
  cosmology}'', {\em Phys.Rev.} {\bfseries D84} (2011) 044021,
 \href{http://xxx.lanl.gov/abs/1105.5081}{ arXiv:1105.5081}.
%%CITATION = ARXIV:1105.5081;%%.

\bibitem{craig_singh_lqc1}
D.~A. Craig and P.~Singh, ``{Consistent probabilities in loop quantum
  cosmology}'', {\em Class.Quant.Grav.} {\bfseries 30} (2013) 205008,
 \href{http://xxx.lanl.gov/abs/1306.6142}{ arXiv:1306.6142}.
%%CITATION = ARXIV:1306.6142;%%.

\bibitem{vt}
V.~Taveras, ``{Corrections to the Friedmann Equations from LQG for a Universe
  with a Free Scalar Field}'', {\em Phys.Rev.} {\bfseries D78} (2008) 064072,
 \href{http://xxx.lanl.gov/abs/0807.3325}{ arXiv:0807.3325}.
%%CITATION = ARXIV:0807.3325;%%.

\bibitem{ps06}
P.~Singh, ``{Loop cosmological dynamics and dualities with Randall-Sundrum
  braneworlds}'', {\em Phys.Rev.} {\bfseries D73} (2006) 063508,
 \href{http://xxx.lanl.gov/abs/gr-qc/0603043}{ arXiv:gr-qc/0603043}.
%%CITATION = GR-QC/0603043;%%.

\bibitem{dgs2}
P.~Diener, B.~Gupt, and P.~Singh, ``{Numerical simulations of a loop quantum
  cosmos: robustness of the quantum bounce and the validity of effective
  dynamics}'',
 \href{http://xxx.lanl.gov/abs/1402.6613}{ arXiv:1402.6613}.
%%CITATION = ARXIV:1402.6613;%%.

\bibitem{dgs1}
P.~Diener, B.~Gupt, and P.~Singh, ``{Chimera: A hybrid approach to numerical
  loop quantum cosmology}'', {\em Class.Quant.Grav.} {\bfseries 31}
  (2014){\natexlab{b}} 025013,
 \href{http://xxx.lanl.gov/abs/1310.4795}{ arXiv:1310.4795}.
%%CITATION = ARXIV:1310.4795;%%.

\bibitem{MenaMarugan:2011me}
G.~A. Mena~Marugan, J.~Olmedo, and T.~Pawlowski, ``{Prescriptions in Loop
  Quantum Cosmology: A comparative analysis}'', {\em Phys.Rev.} {\bfseries D84}
  (2011) 064012,
 \href{http://xxx.lanl.gov/abs/1108.0829}{ arXiv:1108.0829}.
%%CITATION = ARXIV:1108.0829;%%.

\bibitem{Ashtekar:1995zh}
A.~Ashtekar, J.~Lewandowski, D.~Marolf, J.~Mourao, and T.~Thiemann,
  ``{Quantization of diffeomorphism invariant theories of connections with
  local degrees of freedom}'', {\em J.Math.Phys.} {\bfseries 36} (1995)
  6456--6493,
 \href{http://xxx.lanl.gov/abs/gr-qc/9504018}{ arXiv:gr-qc/9504018}.
%%CITATION = GR-QC/9504018;%%.

\bibitem{Marolf:1995cn}
D.~Marolf, ``{Refined algebraic quantization: Systems with a single
  constraint}'',
 \href{http://xxx.lanl.gov/abs/gr-qc/9508015}{ arXiv:gr-qc/9508015}.
%%CITATION = GR-QC/9508015;%%.

\bibitem{josh}
J.~Willis, ``{On the low energy ramifications and a mathematical extension of
  loop quantum gravity}'', {\em {PhD, The Pennsylvania State University}},
  2004.

\bibitem{craig_singh_wdw1}
D.~Craig and P.~Singh, ``{Consistent Histories in Quantum Cosmology}'', {\em
  Found.Phys.} {\bfseries 41} (2011) 371--379,
 \href{http://xxx.lanl.gov/abs/1001.4311}{ arXiv:1001.4311}.
%%CITATION = ARXIV:1001.4311;%%.

\bibitem{craig_singh_wdw2}
D.~A. Craig and P.~Singh, ``{Consistent Probabilities in Wheeler-DeWitt Quantum
  Cosmology}'', {\em Phys.Rev.} {\bfseries D82} (2010) 123526,
 \href{http://xxx.lanl.gov/abs/1006.3837}{ arXiv:1006.3837}.
%%CITATION = ARXIV:1006.3837;%%.

\bibitem{Craig:2012gw}
D.~A. Craig, ``{Dynamical eigenfunctions and critical density in loop quantum
  cosmology}'', {\em Class.Quant.Grav.} {\bfseries 30} (2013) 035010,
 \href{http://xxx.lanl.gov/abs/1207.5601}{ arXiv:1207.5601}.
%%CITATION = ARXIV:1207.5601;%%.

\end{thebibliography}

\end{document}